\documentclass[letterpaper]{emulateapj}
\shorttitle{The Occurrence Rate of Potentially Habitable Planets Orbiting M dwarfs}
\shortauthors{Dressing \& Charbonneau}
\usepackage{epsfig}
\usepackage{amsmath}
\usepackage{rotating}
\usepackage{natbib}
\usepackage{enumerate}
\usepackage{hyperref}
\usepackage{url}
\usepackage{longtable}
\usepackage{xcolor}
\usepackage{xspace}
\def\mearth{{\rm\,M_\oplus}}                                                    
                                                        
\def\rsun{{\rm\,R_\odot}}                                                       
\def\rearth{{\rm\,R_\oplus}} 
\def\fearth{{\rm\,F_\oplus}} 
                                                          
\def\kepler {{\emph{Kepler}\,}}

\def\nstars{2543} 

\def\ntcefirst{3111\xspace} 
\def\ntcestarsfirst{534\xspace} 
\def\nvisrej{2600\xspace} 
\def\nvisfirst{511\xspace}
\def\nvisstarsfirst{246\xspace}
\def\nseqfirst{323\xspace}
\def\nseqstarsfirst{246\xspace}
\def\naccfirst{143\xspace}
\def\naccstarsfirst{97\xspace}
\def\nnotaccfirst{180\xspace}
\def\nfpfirst{9\xspace}

\def\ntcesecond{104\xspace} 
\def\ntcestarssecond{33\xspace} 
\def\naccsecond{15\xspace} 
\def\naccstarssecond{12\xspace} 

\def\ncois{156\xspace} 
\def\nnew{one\xspace} 
\def\nmiss{7\xspace} 
\def\nknownkoi{155\xspace} 

\def\ncoisbox{149\xspace} 

\def\nfakeperstar{2000\xspace}
\def\ninj{5086000\xspace} 
\def\nfullbls{83699\xspace}
\def\nfailprebls{604278\xspace}
\def\nextrasim{4398023\xspace}
\def\fracfailideal{8\%\xspace} 
\def\fracfailreal{4\%\xspace}
\def\maxdf{91.2\%\xspace}

\def\nlitcois{34}

\def\occearthfifty{$0.56^{+0.06}_{-0.05}$\xspace}
\def\occsefifty{$0.46^{+0.07}_{-0.05}$\xspace}
\def\occearthhundred{$0.65^{+0.07}_{-0.05}$\xspace}
\def\occsehundred{$0.57^{+0.08}_{-0.06}$\xspace}
\def\totalsmall{$2.5 \pm 0.2$\xspace}

\def\occhzearth{$0.16^{+0.17}_{-0.07}$\xspace} 
\def\occhzse{$0.12^{+0.10}_{-0.05}$\xspace} 
\def\occhzneps{$0.15^{+0.10}_{-0.05}$\xspace} 
\def\occhzonetwo{$0.27^{+0.16}_{-0.08}$\xspace} 
\def\occvenusmars{$0.24^{+0.18}_{-0.08}$\xspace} 
\def\occvenusmarsse{$0.21^{+0.11}_{-0.06}$\xspace} 

\def\occhzcloudyonetwo{$0.43^{+0.14}_{-0.09}$\xspace} 

\def\occpetigura{$0.83^{+0.09}_{-0.05}$\xspace} 
\def\occzsom{$0.86^{+0.08}_{-0.04}$\xspace} 

\def\dnearhz{$2.6 \pm 0.4$} 
\def\dnearhzuplim{3.5} 
\def\dtranshz{$10.6^{+1.6}_{-1.8}$} 
\def\dtranshzuplim{14.6} 
\def\knearhz{2.9} 
\def\ktranshz{6.0} 

\def\dnearhzsmall{$2.1 \pm 0.2$} 
\def\dtranshzsmall{$8.6^{+0.7}_{-0.8}$} 

\begin{document}
\title{The Occurrence of Potentially Habitable Planets Orbiting M Dwarfs Estimated from the Full Kepler Dataset and an Empirical Measurement of the Detection Sensitivity}
\author{Courtney D. Dressing\altaffilmark{1,2}}
\author{David Charbonneau\altaffilmark{1}}
\altaffiltext{1}{Harvard-Smithsonian Center for Astrophysics, 60 Garden St., Cambridge, MA 02138}
\altaffiltext{2}{\tt cdressing@cfa.harvard.edu}
\vspace{0.5\baselineskip}
\date{\today}

\begin{abstract}
We present an improved estimate of the occurrence rate of small planets orbiting small stars by searching the full four-year \kepler data set for transiting planets using our own planet detection pipeline and conducting transit injection and recovery simulations to empirically measure the search completeness of our pipeline.  We identified \ncois~planet candidates, including \nnew~object that was not previously identified as a \kepler Object of Interest. We inspected all publicly available follow-up images, observing notes, and centroid analyses, and corrected for the likelihood of false positives. We evaluated the sensitivity of our detection pipeline on a star-by-star basis by injecting 2000 transit signals into the light curve of each target star. For periods shorter than 50~days, we find \occearthfifty~Earth-size planets ($1-1.5\rearth$) and \occsefifty~super-Earths ($1.5-2\rearth$) per M~dwarf. In total, we estimate a cumulative planet occurrence rate of \totalsmall planets per M dwarf with radii $1-4\rearth$ and periods shorter than 200~days. Within a conservatively defined habitable zone based on the moist greenhouse inner limit and maximum greenhouse outer limit, we estimate an occurrence rate of \occhzearth~Earth-size planets and \occhzse~super-Earths per M~dwarf habitable zone. Adopting the broader insolation boundaries of the recent Venus and early Mars limits yields a higher estimate of \occvenusmars~Earth-size planets and \occvenusmarsse~super-Earths per M~dwarf habitable zone. This suggests that the nearest potentially habitable non-transiting and transiting Earth-size planets are \dnearhz~pc and \dtranshz~pc away, respectively. If we include super-Earths, these distances diminish to \dnearhzsmall~pc and \dtranshzsmall~pc.
\end{abstract}


\maketitle

\section{Introduction}
In this paper, we focus on the population of planets orbiting small stars. Such planets are important for constraining the galactic census of exoplanets because the majority of stars in the galaxy are low-mass stars \citep{henry_et_al2006, winters_et_al2015}. In addition to understanding the overall occurrence rate of planets orbiting low-mass stars, we would like to know how planet occurrence depends on factors such as planet radius, orbital period, stellar insolation, and host star properties. Furthermore, small stars afford the best near-future opportunities for detailed characterization studies of small planets and their atmospheres \citep{charbonneau+deming2007}. In order to prepare for these observations, we would like to know the likely distance to the closest such targets.

Computing the occurrence rate of small planets around small stars is complicated by the fact that the parameters of low-mass stars are more difficult to measure than the parameters of Sun-like stars. The main emphasis of the \kepler mission was the detection of planets around Sun-like stars, so the assumptions made in the construction of the \kepler Input Catalog (KIC) were tailored to be appropriate for Sun-like stars. Accordingly, \citet{brown_et_al2011} cautioned against relying on KIC classifications for stars cooler than 3750K. 

Several previous studies have attempted to improve the KIC parameters for the coolest \kepler target stars. In our previous paper \citep{dressing+charbonneau2013}, we used the photometry provided in the KIC to reclassify all stars cooler than 4000K using models \citep{dotter_et_al2008, feiden_et_al2011} and assumptions more appropriate for low-mass stars. \citet{gaidos2013} conducted a similar analysis for the population of planet candidate host stars. Other authors further constrained the properties of particular low-mass stars and associated planet candidates by acquiring follow-up spectroscopic and high resolution imaging observations \citep{johnson_et_al2012, muirhead_et_al2012b, muirhead_et_al2012a, ballard_et_al2013, muirhead_et_al2013, swift_et_al2013}. Recognizing the importance of characterizing the full target sample as well as the planet host stars in order to constrain the planet occurrence rate, \citet{mann_et_al2012} acquired spectra for a subset of non-planet host stars. They found that the majority of bright ($Kp < 14$) \kepler target stars are giant stars (several hundred of which were classified as dwarfs in the KIC) but that $93\%$ of fainter stars are correctly classified as dwarfs.

In this paper, we combine the current best estimates of the properties of small \kepler target stars in order to estimate the frequency of small planets around small stars. Our analysis was preceded by several studies of the planet occurrence rate based on \kepler data and we adopt some techniques from the earlier studies. In particular, we draw upon the framework established by \citet{howard_et_al2012}, \citet{fressin_et_al2013}, \citet{dressing+charbonneau2013}, \citet{petigura_et_al2013a}, and \citet{petigura_et_al2013b}.

Working with the first three quarters of \kepler data, \citet{howard_et_al2012} estimated the frequency of planets around main-sequence GK stars. They found that the occurrence rate of planets increased sharply with decreasing planet size and moderately with increasing orbital period. They also found evidence for a cutoff period below which the planet occurrence rate falls off more quickly with decreasing period. The position of the cutoff period appeared to move outward from near 2~days for larger planets ($R_p > 8\rearth$) to roughly 7~days for $2-4\rearth$~planets. 

\citet{youdin2011} used the search completeness estimates from \citet{howard_et_al2012} to model the planet occurrence rate around Sun-like stars by a joint powerlaw in orbital period and planet radius. He found an occurrence rate of $0.19$~planets per star with periods shorter than 50~days and radii larger than $2\rearth$. He also extrapolated outward to predict an occurrence rate of roughly three Earth-like planets per star with periods shorter than a year. 

Due to the low false positive rate expected for the \kepler planet candidate sample \citep{morton+johnson2011}, the \citet{howard_et_al2012} and \citet{youdin2011} analyses assumed that all of the candidates were bona fide transiting planets. They further assumed that \kepler would have been able to detect all transiting planets with cumulative SNR above a set threshold of $10\sigma$. The first assumption biased their occurrence rate estimates toward higher values while the second assumption would have resulted in an underestimate if the actual search completeness were lower. 

\citet{fressin_et_al2013} conducted a follow-up study of the \kepler planet occurrence rate incorporating both contamination from false positives and a more sophisticated model of pipeline sensitivity. In particular, they used a hierarchical approach in which they first estimated the population of Jupiter-size planet candidates that might be astrophysical false positives. They then iteratively determined the occurrence rate of small planets by modeling the fraction of larger planet candidates that might be masquerading as smaller planet candidates in diluted transit events. \citet{fressin_et_al2013} found a global false positive rate of $9.4\pm 0.9\%$ and noted that considering false positives is particular important when calculating the occurrence rate of giant planets ($6-22\rearth$, FP rate = $17.7\%\pm 2.9$) and Earth-size planets ($0.8-1.25\rearth$, FP rate = $12.3\%\pm 3.0$). 

\citet{fressin_et_al2013} also used their hierarchical model to estimate the completeness threshold of the \kepler pipeline. They found that a linear ramp model in which $0\%$ of signals with SNR $< 6$ and $100\%$ of signals with SNR $> 16$ were detected provided a better fit to the observed planet candidate population than an abrupt step function. Accounting for both a non-zero false positive rate and a ramp sensitivity model, \citet{fressin_et_al2013} estimated that $14.9\pm2.4\%$ of FGK stars host an Earth-size planet ($0.8-1.25\rearth$) with a period between 0.8 and 50~days. In a similar study of \kepler data, \citet{dong+zhu2013} found that roughly $20\%$ of main sequence stars with $5000\rm{K}< T_{\rm{eff}} < 6500\rm{K}$ host $1-2\rearth$ planets in periods less than 50~days. 

More recently, \citet{petigura_et_al2013a} developed their own planet search pipeline in order to search for additional planet candidates around \kepler stars. Their TERRA pipeline uses a custom light curve detrending algorithm based on principal component analysis \citep{petigura+marcy2012}. After searching for planets around 42,557 relatively quiet GK stars, \citet{petigura_et_al2013b} found that $7.7\pm1.3\%$ of GK stars host small planets ($1-2\rearth$) in periods between 25 and 50~days. They also extrapolated to predict that $22\% \pm 8\%$ of GK stars host $1-2\rearth$ planets receiving between $1/4$ and $4$~times the insolation received by the Earth. Their calculation incorporated a $10\%$ correction for false positives. As a benefit of writing their own pipeline, \citet{petigura_et_al2013b} were able to explicitly measure the completeness of their planet sample by injecting and attempting to recover transiting planets. 

In a follow-up study, \citet{foreman-mackey_et_al2014} used the reported search completeness and planet candidates from \citet{petigura_et_al2013b} to rederive the planet occurrence rate using a hierarchical Bayesian model. The \citet{foreman-mackey_et_al2014} analysis differed from the \citet{petigura_et_al2013b} analysis in two key aspects: (1) \citet{foreman-mackey_et_al2014} considered measurement errors in the stellar and transit parameters and (2) they did not assume that the planet occurrence rate was flat in log period, instead using a flexible Gaussian process to model the occurrence rate assuming a smooth functional form. As a result, \citet{foreman-mackey_et_al2014} found an occurrence rate of potentially habitable Earth-size planets three times lower than the \citet{petigura_et_al2013b} estimate.

\citet{silburt_et_al2015} considered a sample of 76,711~\kepler target stars with radii of $0.8-1.2\rsun$ and estimated the search completeness using the reported Combined Differential Photometric Precision \citep[CDPP,][]{christiansen_et_al2012}. They employed an iterative simulation to investigate the dependence of the planet occurrence rate on planet radius without subdividing the data into bins and accounted for errors in the planet radii. For planets with periods of 20--200~days, \citet{silburt_et_al2015} reported that the occurrence rate is higher for planets with radii of $2-2.8\rearth$ than for smaller or larger planets. In total, they estimated that a typical Sun-like star hosts $0.46\pm0.03$ planets with periods of 20--200~days and radii of $1-4\rearth$. In agreement with \citet{petigura_et_al2013b}, \citet{silburt_et_al2015} noted that the planet occurrence rate is flat in log period. Within a broad habitable zone extending from 0.99--1.7~AU, \citet{silburt_et_al2015} estimated an occurrence rate of  $0.064^{+0.034}_{-0.011}$~small ($1-2\rearth$) planets per Sun-like star.

Focusing specifically on \kepler's smallest target stars, we \citep{dressing+charbonneau2013} estimated an occurrence rate of $0.90^{+0.04}_{-0.03}$ planets per star for $0.5-4\rearth$ planets with periods shorter than 50~days. We based our previous analysis on Q1--Q6 \kepler planet candidate list and assumed that \kepler detected all planets with cumulative SNR$ > 7.1\sigma$. Using conservative habitable zone limits from \citet{kasting_et_al1993}, we estimated an occurrence rate of $0.15^{+0.13}_{-0.06}$ potentially habitable Earth-size ($0.5-1.4\rearth$) planets per small star. \citet{kopparapu2013} then revised this estimate to $0.48^{+0.12}_{-0.24}$~planets per star using the broader updated habitable zone boundaries from \citet{kopparapu_et_al2013}. His result agreed well with the estimate of $0.46^{+0.18}_{-0.15}$ potentially habitable $0.8-2\rearth$ planets per star from \citet{gaidos2013}. Unlike \citet{kopparapu2013}, \citet{gaidos2013} adopted habitable zone boundaries corresponding to the $50\%$ cloud cover case from \citet{selsis_et_al2007}. 

\citet{morton+swift2014} adopted a slightly different technique to estimate the frequency of small planets around small stars. They assumed that the planet radius distribution is independent of orbital period and modeled each planet using a weighted kernel density estimator when computing the occurrence rate. They found that the occurrence rate estimates  from \citet{dressing+charbonneau2013}, \citet{kopparapu2013}, and \citet{gaidos2013} for planets smaller than $1.4\rearth$ should be increased by an additional incompleteness factor of 1.6 if the assumption made by \citet{morton+swift2014} about the period-independence of the planet radius distribution is correct. 

Like \citet{silburt_et_al2015}, \citet{gaidos_et_al2014b} used an iterative simulation to estimate the planet occurrence rate, but they elected to focus on stars cooler than 4200K. For orbital periods of $1-180$~days and radii of $0.5-6\rearth$, \citet{gaidos_et_al2014b} calculated a cumulative occurrence rate of $2.01\pm0.36$ planets per M~dwarf. \citet{gaidos_et_al2014b} also remarked that the planet occurrence rate is highest for planets with radii of approximately $1\rearth$ and lower for larger and smaller planets.

The frequency of potentially habitable planets around small stars has also been estimated from radial velocity surveys. Based on six years of observations with the HARPS spectrograph, \citet{bonfils_et_al2013} estimated an occurrence rate of $0.41^{+0.54}_{-0.13}$ potentially habitable planets per M dwarf. Their definition of ``potentially habitable'' encompassed planets with $1 \leq M\sin i \leq 10$ within the ``early Mars'' and ``recent Venus'' boundaries of the habitable zone presented in \citet{selsis_et_al2007}. A subsequent study by \citet{robertson_et_al2014} revealed that GJ~581d, one of the two planets upon which \citet{bonfils_et_al2013} based their occurrence rate estimate, is likely a manifestation of stellar activity. \citet{robertson_et_al2014} reported a revised occurrence rate of 0.33~potentially habitable planets per M~dwarf. The updated RV-based estimate is more similar to the estimates based on \kepler data, but accurately determining an occurrence rate with only a single planet \citep[GJ~667Cc,][]{anglada-escude_et_al2012, bonfils_et_al2013, delfosse_et_al2013} is challenging. Additionally, direct comparison of planet occurrence estimates from RV and transit surveys is complicated by the need to employ a compositional model to  translate planet masses into radii.

In this paper, we implement the following improvements to refine our 2013 estimate of the frequency of small planets around small stars:
\begin{itemize}
\item{We use the full Q0-Q17 \kepler data set.}
\item{We utilize archival spectroscopic and photometric observations to refine the stellar sample.}  
\item{We explicitly measure the pipeline completeness.}
\item{We inspect follow-up observations of planet host stars to properly account for transit depth dilution due to light from nearby stars.}
 \item{We apply a correction for false positives in the planet candidate sample.}
 \item{We incorporate a more sophisticated treatment of the habitable zone.}
\end{itemize}
In Section~\ref{sec:stars} we describe the selection of our stellar sample, which includes some stars whose parameters have been characterized spectroscopically via follow-up observations. We explain our planet detection pipeline in Section~\ref{sec:detpipe} and our procedure for vetting candidates in Section~\ref{sec:vet}. We present light curve fits for the accepted planet candidates in Section~\ref{sec:mcpipe}. In Section~\ref{sec:injpipe}, we  assess the completeness of our pipeline. We then estimate and discuss the planet occurrence rate in Section~\ref{sec:occrate} before concluding in Section~\ref{sec:conc}.

\section{Stellar Sample Selection}
\label{sec:stars}

We selected our stellar sample by first downloading a table of all 4915~stars with $T_{\rm eff} < 4000$ and $\log g > 3$ from the Q1--16 \kepler Stellar Catalog on the NASA Exoplanet Archive\footnote{\url{http://exoplanetarchive.ipac.caltech.edu}} \citep{akeson_et_al2013}. This catalog is described in \citet{huber_et_al2014} and combines the best estimates available for each star from a variety of photometric, spectroscopic, and asteroseismic analyses. The properties for the stars in the downloaded sample were primarily determined from photometry \citep{brown_et_al2011, dressing+charbonneau2013, gaidos2013, huber_et_al2014}, but 2\%~of the sample had spectroscopically-derived parameters  \citep{mann_et_al2012, muirhead_et_al2012b, mann_et_al2013b, martin_et_al2013}. For the majority of the stars in the sample (79\%), the stellar parameters were drawn from our 2013 analysis \citep{dressing+charbonneau2013}. 

Some of the stars in the downloaded subset had light curves indicative of binary stars, variable stars, or enhanced spot activity. In addition, some of the stars were observed only for a small number of days. In order to accurately estimate the planet occurrence rate for small stars, we wanted to select the subset of stars with the highest search completeness. We therefore performed the following series of cuts on the sample. 

First, we counted the number of timestamps for which each star had ``good'' data (i.e., not flagged). We rejected all 2101~stars with fewer than 48940 unflagged long cadence data points. Since \kepler obtained long cadence data using 29.4~minute integration times, this cut requires 1000~days of data. One of the main goals of this paper is to measure the occurrence rate of potentially habitable planets and we wanted to ensure that \kepler would have been able to observe multiple transits of planets within the habitable zones (HZ; see Section~\ref{ssec:hzocc}) of the stars in our final sample. For reference, the median orbital period at the outer edge of the \citet{kopparapu_et_al2013} HZ for the stars in our final sample is 131~days and the longest period at the outer HZ is 207~days. 

Second, we removed 63~stars that \citet{mcquillan_et_al2013} categorized as likely giants based on their stochastic photometric variability. The affected stars have red colors (median $J-H = 0.83$) consistent with their revised classification as giants. Although two of the stars had revised classifications from \citet{dressing+charbonneau2013}, the remaining 61 had parameters from the Kepler Input Catalog.

For reference, we checked whether any of our target stars were known eclipsing binaries by consulting the Kepler Eclipsing Binary Catalog. We examined both the published Version 2 \citep{slawson_et_al2011} and the online beta version of the Third Revision.\footnote{\url{keplerebs.villanova.edu}} Eleven of the target stars were listed in both versions of the catalog, six stars were listed in Version~2 only and three stars were listed in Version~3 only. 

Six of the twenty targets with matches in the Eclipsing Binary~Catalogs were listed in the NASA Exoplanet Archive as false positive systems (KID~5820218 = KOI~1048, KID~6620003 = KOI~1225, KID~8823426 = KOI~1259, KID~9761199 = KOI~1459, KID~9772531 = KOI~950, and KID~10002261 = KOI~959). Two were listed as planet candidate host stars (KID~5384713 = KOI~3444 and KID~11853130 = KOI~3263). Confirmed giant planet KOI~254.01 \citep[KID 5794240, ][]{johnson_et_al2012} was also included as an EB match because of the very large transit depth. As evidenced by the presence of KOI~254.01 in the EB catalogs, the catalogs contain both actual EBs and likely planets. Accordingly, we did not remove the twenty targets with matches in the EB catalogs from our target sample. 
\begin{figure}[tbp] 
\begin{center}
\centering
\includegraphics[width=0.5\textwidth]{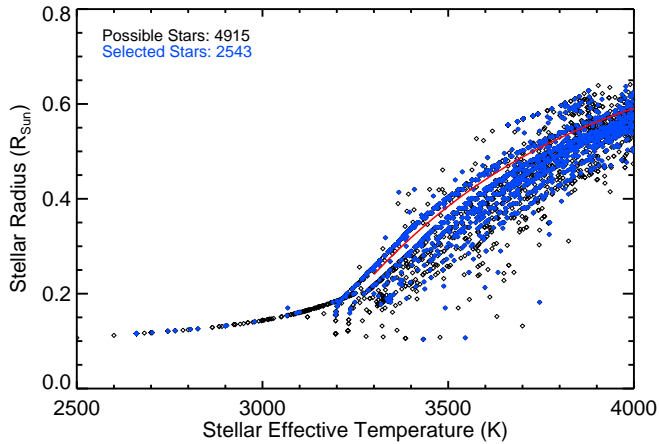}
\end{center}
\caption{Radii and stellar effective temperatures of the stars in our final selected subsample (blue) compared to the full sample initially downloaded from the NASA Exoplanet Archive (black). The red line is an empirical relation between effective temperature and radius \citep[][see Section~\ref{ssec:starbias}]{mann_et_al2013c}.}
\label{fig:stars}
\end{figure}

We then detrended all of the light curves as described in detail in Section~\ref{ssec:detrending} using smoothing lengths of 500, 1000, and 2000~minutes. We constructed a histogram of the flux distributions for each of the detrended light curves and measured the $\chi^2$ of a fit to a Gaussian flux distribution. The flux distribution of a well-behaved single star should be Gaussian after detrending, but the flux distribution of an eclipsing binary can appear bimodal. We therefore flagged for visual inspection all 353~stars for which reduced $\chi^2 > 3$ for any of the detrended light curves. We also measured the standard deviations $\sigma_{500}, \sigma_{1000}, \sigma_{2000}$ of the three detrended light curves for each star and took the ratios of the standard deviations of light curves detrended using different median filters. We flagged 42~stars for which any of the ratios $\sigma_{500}/\sigma_{1000}$, $\sigma_{500}/\sigma_{2000}$, or $\sigma_{1000}/\sigma_{2000}$ were below 0.8. This cut was designed to pick out light curves for which the detrending algorithm failed to remove longer timescale variability. 

Finally, we visually inspected the detrended and PDC-MAP photometry \citep{smith_et_al2012, stumpe_et_al2012} for the 395~flagged stars with \mbox{$>1000$~days} of data and large $\chi^2$ or standard deviation ratios. We rejected 207~stars with highly variable detrended light curves or strong indications of classification as an eclipsing binary. We also rejected KID~12207013 because of the unusual light curve morphology displayed in certain quarters. We accepted the remaining 187~flagged stars, increasing our final selected sample to \nstars~stars. We note that the 208~stars that were rejected during the visual inspection stage are unlikely to harbor detectable planet candidates exactly because their light curves are highly variable. Similarly, our injection tests would likely recover only a small fraction of any planets injected into their light curves.  Accordingly, the exclusion of these stars from our stellar sample has negligible effect on our estimated rates of planet occurrence.

Our final selected sample of \nstars~stars is compared to the initial downloaded sample in Figure~\ref{fig:stars}. The temperature range of the sample extends from 2661K to 3999K, with a median stellar effective temperature of 3746K. The median stellar radius is $0.47\rsun$ and the stars have radii spanning from $0.10\rsun$ to $0.64\rsun$. The metallicity range is [Fe/H] = -2.5 to [Fe/H] = 0.56, with a slightly sub-solar median metallicity of [Fe/H] $=-0.1$. However, most of the metallicity estimates were derived from photometry \citep{dressing+charbonneau2013} and are not well-constrained. The brightest star in the sample has a \kepler magnitude $Kp=10.07$, but the median brightness is $Kp=15.5.$ The faintest star has $Kp=16.3$. The sample contains 100~known planet (candidate) host stars with 83~planet candidates and 80~confirmed planets. 

\section{Planet Detection Pipeline}
\label{sec:detpipe}
The first step in our planet detection pipeline was to clean the light curves to prepare them for the transit search. Next, we searched each light curve sequentially for planets, allowing the code to detect multiple planets per star when warranted by the data. We then accepted or rejected putative detections using the vetting procedure described in Section~\ref{sec:vet} and checked that the automatically accepted transits were not ephemeris matches with other KOIs. Finally, we visually inspected all surviving candidates and reviewed all available follow-up analyses produced by the \kepler team and the community. We discuss each step of the planet detection pipeline in more detail in the following sections. 

\subsection{Preparing the light curves}
\label{ssec:detrending}
\begin{figure}[tbp] 
\begin{center}
\centering
\includegraphics[width=0.5\textwidth]{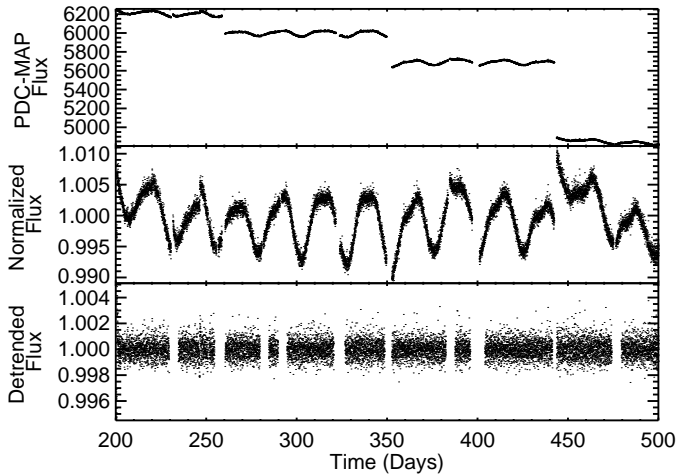}
\end{center}
\caption{Illustration of the detrending process using a section of the light curve of KID~5531953 (KOI~1681). \emph{Top:} PDC-MAP flux versus time. \emph{Middle: } Normalized flux versus time. \emph{Bottom: } Flux detrended using a 2000~minute filter versus time.}
\label{fig:detrend}
\end{figure}

We obtained all available long cadence data for each target via anonymous ftp from the MAST.\footnote{\url{http://archive.stsci.edu/kepler/publiclightcurves.html}}  We then excluded all data points flagged as low quality and detrended each quarter of data independently. We produced three detrended versions of each light curve using a running sigma-clipped mean filter with widths of 500, 1000, or 2000~minutes. When calculating the mean, we excluded all points more than $3\sigma$ away from the median value of the light curve within the filtered region.  We then divided the flux data by the smoothed light curve to obtain a detrended, normalized light curve for that quarter. Figure~\ref{fig:detrend} provides an illustration of our light curve detrending process. 

Next, we searched for data gaps and anomalies within the detrended light curves. We defined a data gap as $\geq 0.75$ days of missing photometry. Data gaps are frequently accompanied by sharp increases or decreases in flux that can confound searches for planets. We removed these events by excising all data points within 1~day of the start of a data gap or less than 3~days after the end of a data gap. 

\subsection{Searching for Transiting Planets}
\label{ssec:search}
We designed our planet search algorithm to take advantage of the known stellar properties for each target star. First, we predicted the expected transit duration as a function of orbital period based on the mass and radius of the target star \citep{winn2010}. We initially computed the transit duration for a central transit of a planet in a circular orbit, but we reduced the minimum duration considered by a factor of $15$ to account for grazing transits and eccentric orbits. We then constructed a box-fitting least squares (BLS) periodogram for each of twelve logarithmically spaced intervals between 0.5 and 200~days. Our main planet search program was written in IDL, but the BLS portion was implemented using the Fortran package and IDL wrapper provided by Scott Fleming.\footnote{\url{http://www.personal.psu.edu/users/s/w/swf13/SGE/clio.html}}

We used different boundaries for the transit ``duty cycle'' (the ratio of the transit duration to the orbital period) for each period range based on our predicted transit durations. For each search, we used a light curve detrended with a smoothing filter of 500, 1000, or 2000~minutes. The choice of light curve was set by the expected transit duration. For predicted transit durations shorter than 200~minutes, we selected the shortest smoothing filter such that the expected transit duration was less than one tenth of the smoothing window. For transit durations longer than 200~minutes we used the 2000~minute filter. 

We then determined the signal detection efficiency (SDE) for each possible signal in the composite periodogram using Equation~6 in \citet{kovacs_et_al2002}. We checked whether any peaks had $SDE > 6$ and stopped searching if no peaks were above the threshold. If peaks were detected, we ranked the peaks in order of decreasing SDE.

Starting with the most significant peak, we re-ran the BLS algorithm considering only periods close to the period of the identified peak. In these high-resolution runs, we considered transit durations ($\tau_{\rm dur}$) as short as $1/30$th of the expected transit duration of a planet in a circular orbit to increase the chance that our code would be able to recover planets in grazing or eccentric orbits. We used these higher resolution BLS periodograms to determine the epochs of the putative transit events. We then ran a Monte Carlo analysis to find the preliminary transit model \citep{mandel+agol2002} that best described the candidate event. 

In our Monte Carlo analysis, we allowed the transit center to shift by 2 hours (up to a maximum of 1/200th of the orbital period for short-period events). For each choice of transit center we generated a new version of the detrended light curve by dividing the raw PDC-MAP \emph{Kepler} photometry by a straight line fit to the photometry immediately preceding and following each putative transit. Specifically, we considered data points more than one and less than 3.5 expected full transit durations away from the putative transit center. We assumed circular orbits and estimated quadratic limb darkening parameters from the $T_{\rm eff}$ and $\log g$ of the target star by interpolating between the coefficients determined by \citet{claret+bloemen2011}. We considered $a/R_*$ between 50\% and 200\% of the expected value for the trigger orbital period, impact parameters between 0 and 1, and $R_P/R_*$ as large as the square root of the depth of the trigger event. 

In some cases, the highest peak in the periodogram was actually at a harmonic of the true planet period, so we repeated the Monte Carlo transit fitting analysis at $\frac{1}{n}$ and $n$ times the trigger period for $n=2-7$. We then selected the period for which the $\Delta \chi^2$ compared to a straight line fit was maximized. We rejected all putative transit events for which none of the models were preferred at $5\sigma$ and recorded the parameters of the best-fit model in other cases. We further refined the planet parameters during the vetting stage as described in Section~\ref{sec:vet}.

We then repeated the process described in the previous three paragraphs to fit the next highest peak in the periodogram. If the preceding transit fit had been accepted, then we excised the data near transit prior to fitting the next peak. When all peaks above the threshold level were exhausted, we generated a new periodogram using only the out-of-transit data and reran the peak identification and transit model fitting process with the new periodogram. The code automatically stopped searching for planets when no peaks with SDE $ > 6$ were found, when none of the transit models for the identified peaks were accepted, or when the code had completed three~iterations of searching for planets. (Note that multiple planets could be detected in a single round of searching.)

\section{Vetting}
\label{sec:vet}
Our transit detection pipeline identified \ntcefirst putative transit events associated with \ntcestarsfirst stars. Some of those signals might have been systematics or astrophysical false positives instead of bona fide transiting planet candidates. Accordingly, we performed a series of cuts to select the events consistent with transiting planets. First, we visually inspected the candidate transit events to identify signals that were not clearly associated with spacecraft systematics or stellar activity. Of the \ntcefirst candidate events, \nvisfirst~events survived initial visual inspection. The \nvisrej~events rejected at the visual inspection stage displayed morphologies consistent with classification as spacecraft systematics or sinusoidal brightness variations indicative of starspots rather than transiting planets. The \nvisfirst~signals surviving visual inspection were associated with \nvisstarsfirst unique host stars. 

Some of the candidate events were harmonics of signals detected at integer multiples of the true period. We then ranked the accepted signals for each host star in order of decreasing $\Delta \chi^2$ as calculated during the detection phase and iteratively fit and excised the transits of each signal in order to ensure that the lower $\Delta \chi^2$ signals were not simply harmonics of the strongest signals. We rejected all signals with resulting $\Delta \chi^2$ below $5\sigma$. This ``sequential vetting'' step reduced the number of candidate events to \nseqfirst possible transits for \nseqstarsfirst unique stars. 

Next, we conducted a second, more intensive round of visual examination for the remaining candidate events. We compared the shapes and depths of odd and even transits, checked for the appearance of secondary eclipses, considered the depth of the putative transit relative to other possible features at the same orbital period, and investigated whether the putative transit events were dominated by a small number of deep events. After visually vetting the candidates, we checked whether putative signals had been previously identified as false positives and excluded all such events. We rejected \nnotaccfirst signals (including \nfpfirst known false positives) and accepted \naccfirst signals associated with \naccstarsfirst unique stars. 

During the vetting stage, we noticed that the phase-folded light curve for KOI~2283.01 exhibited two transit-like events with markedly different depths when folded to the 17.402~day orbital period listed in the Q1--16 KOI catalog. We therefore rejected KOI~2283.01 as a blend containing an eclipsing binary. This interpretation is consistent with the large observed centroid source offset shift of $5.3\sigma$.  

We then performed a final search for additional planets in the systems in which a previously detected planet had survived the vetting process. We executed this search by phase-folding the detrended light curve on the orbital periods of all accepted signals and removing all data points within plus or minus one best-fit duration of transit center prior to re-running the search process detailed in Section~\ref{ssec:search}. In all cases we used the light curve that had been detrended using a 2000~minute filter. The motivation for repeating this search after the first round of vetting was that uncertainties in the initial periods, durations, and transit centers of the accepted signals might have limited the effectiveness of the clipping performed in the initial search. 

Our second round search revealed \ntcesecond~candidate signals for \ntcestarssecond stars. We vetted the signals using the same vetting pipeline as in the first round search and accepted \naccsecond~additional candidate transiting planets associated with \naccstarssecond stars. Next, we excised the transits of the signals accepted in the second round and performed a third round of transit searches. No additional signals were accepted during the third round. The full sample of \ncois~planet candidates included \naccfirst~signals associated with \naccstarsfirst stars revealed in the first round of searching and \naccsecond~signals associated with \naccstarssecond stars revealed in the second round. 

We compared the periods, $P$, and epochs, $t_0$, of the accepted planet candidates to the catalog of known eclipsing binaries, periodic variable stars, and KOIs compiled by \citet{coughlin_et_al2014}. We excluded the host star from the match process in order to avoid matching a signal to itself. We did not find any corresponding signals within our specified match tolerances of 
\begin{align}
\left | P_{\rm match}-P\right | &\leq \rm{min}\left(2~\rm{hr}, 0.001\times P\right)\nonumber \\
& and  \\
 \left | t_{0,\rm match}-t_0\right | &\leq \rm{min}\left(4~\rm{hr}, 0.001\times P\right)\nonumber
\end{align}

\subsection{New Planet Candidate}
\label{ssec:newcan} 
The majority of the accepted planet candidates corresponded to signals previously identified as KOIs. We found that \nknownkoi~putative planet candidates had periods and epochs matching those of known planet candidates or confirmed planets \citep{borucki_et_al2010, borucki_et_al2011a, borucki_et_al2011b, batalha_et_al2013, burke_et_al2014}. We accepted one new signal in a system with previously known KOIs.
 \begin{deluxetable}{ccccccc}
\tablecolumns{7}
\tabletypesize{\footnotesize}
\tablecaption{New Candidate Accepted By Our Pipeline}
\tablehead{
\colhead{KID} &
\colhead{KOI} &
\colhead{P} &
\colhead{t0} &
\colhead{$R_p$} &
\colhead{$a$} &
\colhead{$\Delta \chi^2$} \\
\colhead{} &
\colhead{} &
\colhead{(days)} &
\colhead{(days)} &
\colhead{($\rearth$)} &
\colhead{(AU)} &
\colhead{}\\
}
5531953 & 1681 & 21.913843 & 17.036402 & 1.03 & 0.16 & 41.4
\label{tab:new}
\end{deluxetable}

\begin{figure*}[htbp] 
\begin{center}
\centering
\includegraphics[width=0.9\textwidth]{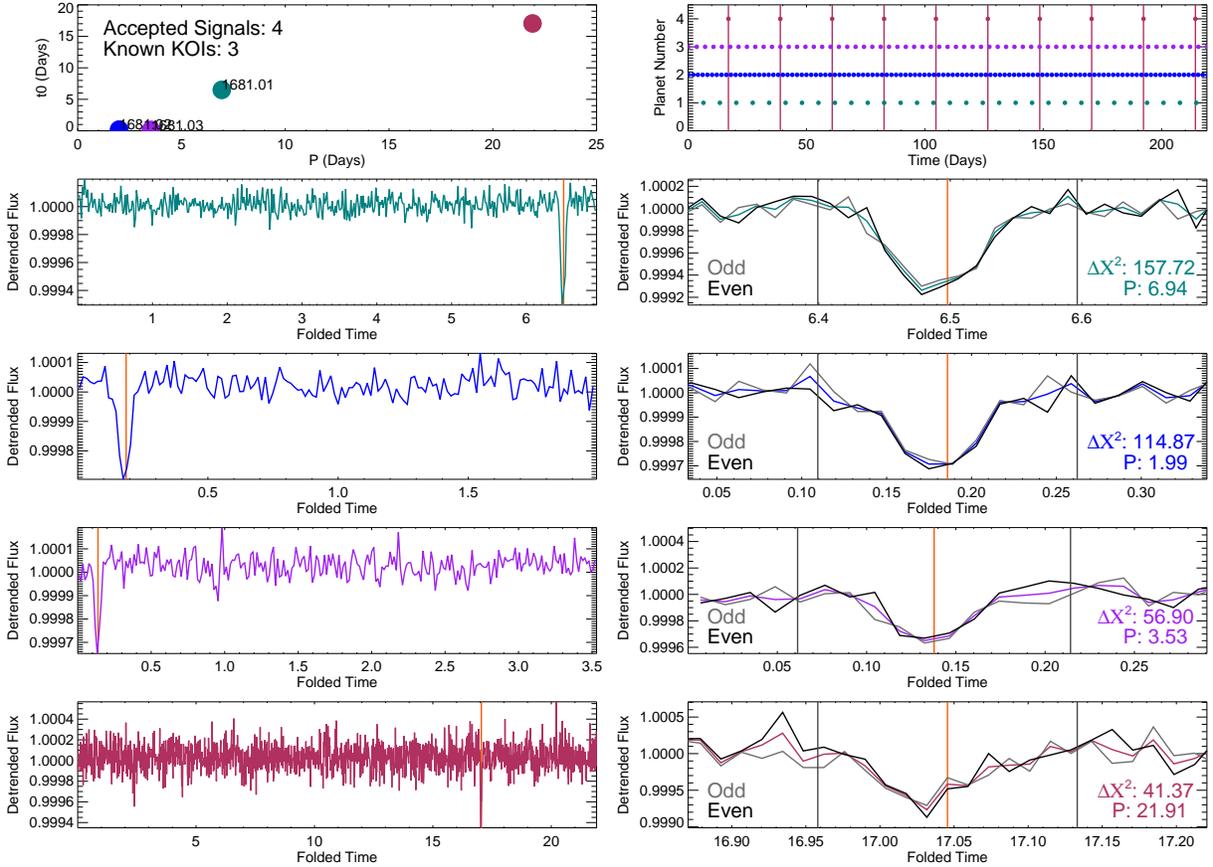}
\end{center}
\caption{Transit signals detected in the KID~5531953 system. \emph{Top left: } Period and epochs of all four signals identified by our pipeline (circles) and the known planet candidates in the system (marked by text). \emph{Top right: } Transit times for each of the four planet candidates. \emph{Second row from top: } Detrended flux versus time folded to the 6.9~day period of KOI~1681.01. The left panel displays the full binned phase-folded light curve and the right panel shows a zoomed-in view near transit center. The orange line marks the transit center. The light gray and dark gray lines show the binned phase-folded light curve for only the odd and even transits, respectively. We excised data points between the vertical gray lines before folding the data to the period of the next planet. \emph{Third row from top: } Same as second row but for the 1.99~day period of KOI~1681.02. The transits of KOI~1681.01 are not included. \emph{Fourth row from top: } Same as previous row but for the 3.53~day period of KOI~1681.03. The transits of KOI~1681.01 and KOI~1681.02 are not included. \emph{Bottom row: } Same as second row but for the 21.9~day period of the new signal detected by our pipeline. The transits of KOI~1681.01, 1681.02, and 1681.03 are not included.}
\label{fig:sys1681}
\end{figure*}

The KOI~1681 (KID~5531953) system contains three known planet candidates with periods of 6.51, 1.99, and 3.53~days. Our pipeline detected a planet candidate in the system with a radius of $1\rearth$ and a period of $21.9$ days, roughly 11 times the 1.99~day orbital period of KOI~1681.02. We describe additional planet properties in Table~\ref{tab:new}.
As shown in Figure~\ref{fig:sys1681}, the transit signal is still visible in the light curve after the transits of the other three planets have been removed, suggesting that this is a new planet candidate rather than an alias of KOI~1681.02. The signal was later identified in the Q1-17 DR 24 Kepler pipeline run\footnote{\url{http://tinyurl.com/kepler-dr24}} and is now listed as planet candidate 1681.04.

\subsection{Accounting for Transit Depth Dilution}
\label{ssec:dilution}
For the \nknownkoi~known KOIs in our planet candidate sample, we inspected the DV reports prepared by the \kepler team and all publicly available follow-up data to check for signs that the KOIs were false positives. As described below, we learned that several of the planet host stars in our sample have stellar companions at separations within $1''$. Accordingly, the measured transit depths for those planet candidates would have been diluted by the additional light in the aperture. 

Two of those systems (KOI~1422 and KOI~2626) were well characterized by \citet{cartier_et_al2014}. In their analysis, \citet{cartier_et_al2014} determined stellar parameters for the double star system KOI~1422 and the triple star system KOI~2626 using HST WFC3/UVIS photometry. They were unable to constrain which of the stars hosted the associated planet candidates, but they were able to provide revised estimates for the radii and orbital parameters of the associated planet candidates for each choice of host star. In our analysis, we therefore chose to represent the planet candidates in these systems using ``fractional planets'' orbiting each of the possible host stars rather than assuming that the planet candidates orbit the system primaries or excluding them from the analysis.  

For the remaining systems with close stellar companions we did not have sufficient information about the companion star to model planet candidates orbiting each star in the system. Instead, we corrected for the transit depth dilution by multiplying the estimated planet radius by the correction factor $c = \sqrt{2.512^{-\Delta K}+1}$, where $\Delta K$ is the difference in \kepler magnitudes between the apparent magnitudes of the target star and companion star. In the worse case scenario of an equal-brightness binary, the correction increases the estimated radius of the planet candidate by roughly $40\%$.

We applied correction factors for three systems: KOI~605 (41\%), KOI~3010 (41\%), and KOI~3284 (2\%). 
In the KOI-605 system, which contains two candidates, D. Ciardi's Keck/NIRC2 adaptive optics images\footnote{\url{https://cfop.ipac.caltech.edu/edit_obsnotes.php?id=605}} revealed that the system consists of two stars separated by less than $0\farcs1$ with nearly equal brightness in the \kepler bandpass. 

D. Ciardi also acquired Keck/NIRC2 observations\footnote{\url{https://cfop.ipac.caltech.edu/edit_obsnotes.php?id=3010}} of KOI~3010 showing that the system is a close binary with a separation of $0\farcs3$. The two stars appear to have nearly equal brightnesses. For KOI~3284, Keck/NIRC2 and Gemini/DSSI images\footnote{\url{https://cfop.ipac.caltech.edu/edit_obsnotes.php?id=3284}} revealed a companion 3.56~magnitudes fainter than the target star at a separation of $0\farcs4$. 

Our correction procedure explicitly assumed that any associated planet candidates orbit the target star, but they might actually orbit the companion star. If the planet candidates do indeed orbit the companion star, then the radii of the planet candidates will need to be reevaluated once the properties of the companion star are established (see \citealt{ciardi_et_al2015} for a detailed discussion). In most cases, the available photometry was insufficient to determine whether the nearby companion is physically associated with the target star or to constrain the properties of the companion star.

\subsection{False Positive Correction}
\label{ssec:fp}
In addition to correcting for transit depth dilution in systems with nearby companions, we incorporated a general false positive correction to account for the possibility that some of the smaller transiting planets in our sample might be diluted eclipses or transits of larger planets. We therefore consulted Table~1 of \citet{fressin_et_al2013} to determine the false positive probability ($FPP$) for a planet with a given radius. We apply this radius-dependent correction to the planet occurrence map derived in Section~\ref{sec:occrate}. 

\subsection{Known Planet Candidates Missed by Our Pipeline}
Our sample of accepted candidate events included all but \nmiss~of the 161~known planet candidates and confirmed planets meeting our sample cuts of planet radii larger than $0.5 \rearth$ and orbital periods shorter than 200~days. We list the missed candidates in Table~\ref{tab:missed}. We note that \citet{swift_et_al2015} rejected one of the missed candidates (KOI~1686.01) as a possible planet because the phase folded light curve did not display a convincing transit event and that the reported disposition in the NASA Exoplanet Database was later changed to False Positive. Three of the missed candidates are in the same system (KOIs~3444.01, 3444.03, and 3444.04) and none of the \nmiss missed candidates produced accepted peaks in the BLS periodograms. Although we were reassured that our pipeline recovered most of the previously detected planet candidates, our goal was not to reproduce the \kepler planet candidate list but to design a single pipeline that could be used to both search for planets and characterize pipeline completeness. Thus we do not consider these additional \nmiss~KOIs in our analysis below.
\begin{deluxetable}{ccccrcr}
\tablecolumns{7}
\tabletypesize{\footnotesize}
\tablecaption{Known KOIs Missed By Our Pipeline}
\tablehead{
\colhead{} &
\colhead{} &
\colhead{Kepler} &
\colhead{mag} &
\colhead{P} &
\colhead{$R_p$}  &
\colhead{Kepler} \\
\colhead{KID} &
\colhead{KOI} &
\colhead{name} &
\colhead{($Kp$)} &
\colhead{(days)} &
\colhead{($\rearth$)} &
\colhead{SNR\tablenotemark{a}}
}
6149553 & 1686.01\tablenotemark{b} & ... & 15.89 & 56.87 & 1.3 & 5.2\tablenotemark{c}\\
8890150 & 2650.02 & 395b & 15.99 & 7.05 & 1.1 & 12.4\\
9605552 & 3102.01 & ... & 15.98 & 9.32 & 1.0 & 6.1\\
5384713 & 3444.01 & ... & 13.69 & 12.67 & 1.0 & 14.7\\
5384713 & 3444.03 & ... & 13.69 & 2.64 & 0.6 & 12.2 \\
5384713 & 3444.04 & ... & 13.69 & 14.15 & 0.8  & 9.3\\
2986833 & 4875.01 & ... & 15.78 & 0.91 & 1.0 & 10.2
\enddata
\tablenotetext{a}{As reported in the Cumulative KOI table at the NASA Exoplanet Archive on 22 November 2014.}
\tablenotetext{b}{As of 22 March 2015, the reported disposition of KOI~1686.01 at the NASA Exoplanet Archive has been changed to False Positive.}
\tablenotetext{c}{The transit SNR for KOI~1686.01 was not reported in the Cumulative KOI table, the Q1--Q16 table, or the Q1--Q12 table. This is the value from the Q1--Q8 table. The value in the Q1--Q6 table was 7.6.}
\label{tab:missed}
\end{deluxetable}

\begin{figure*}[htbp] 
\begin{center}
\centering
\includegraphics[width=0.48\textwidth]{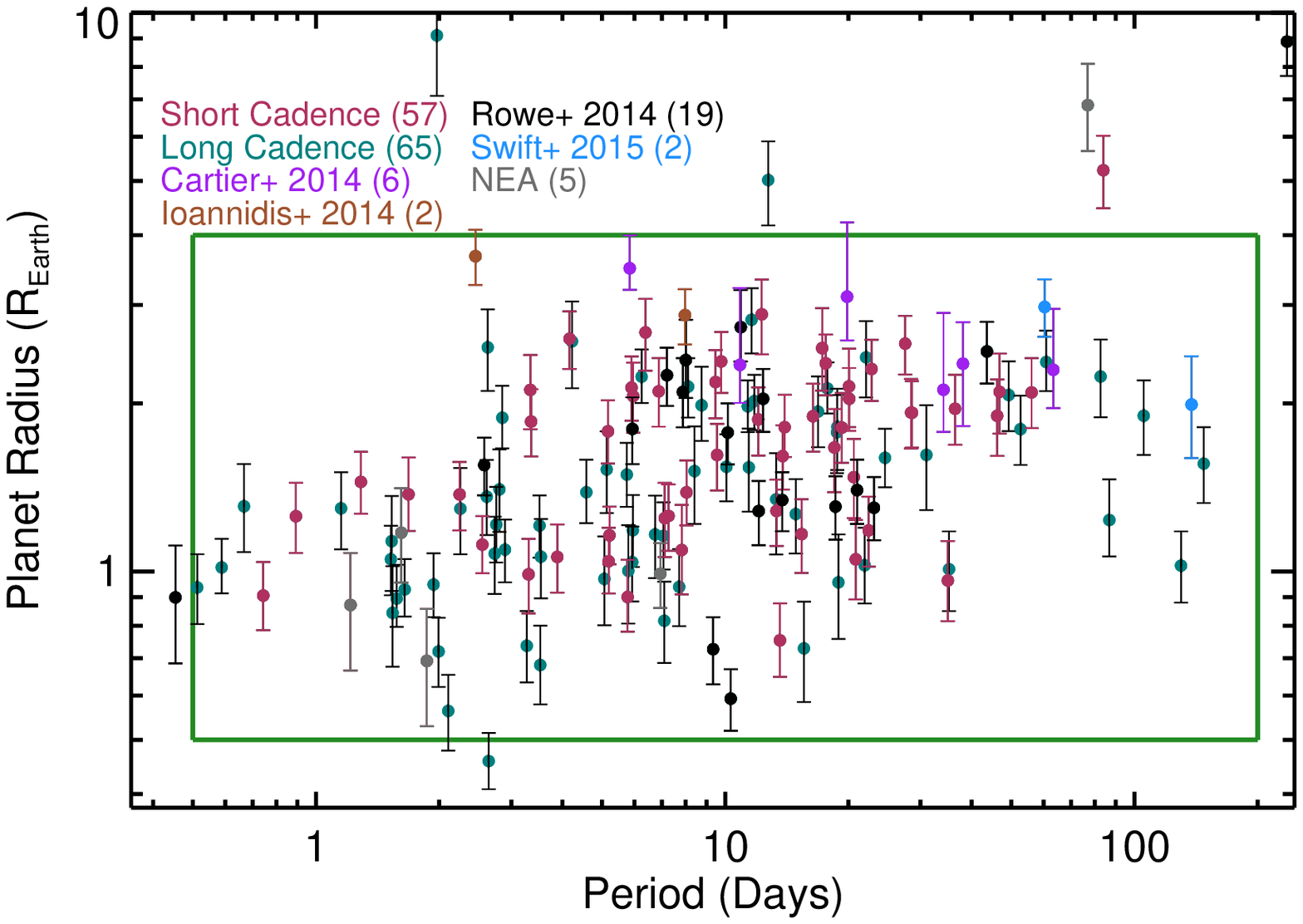}
\includegraphics[width=0.48\textwidth]{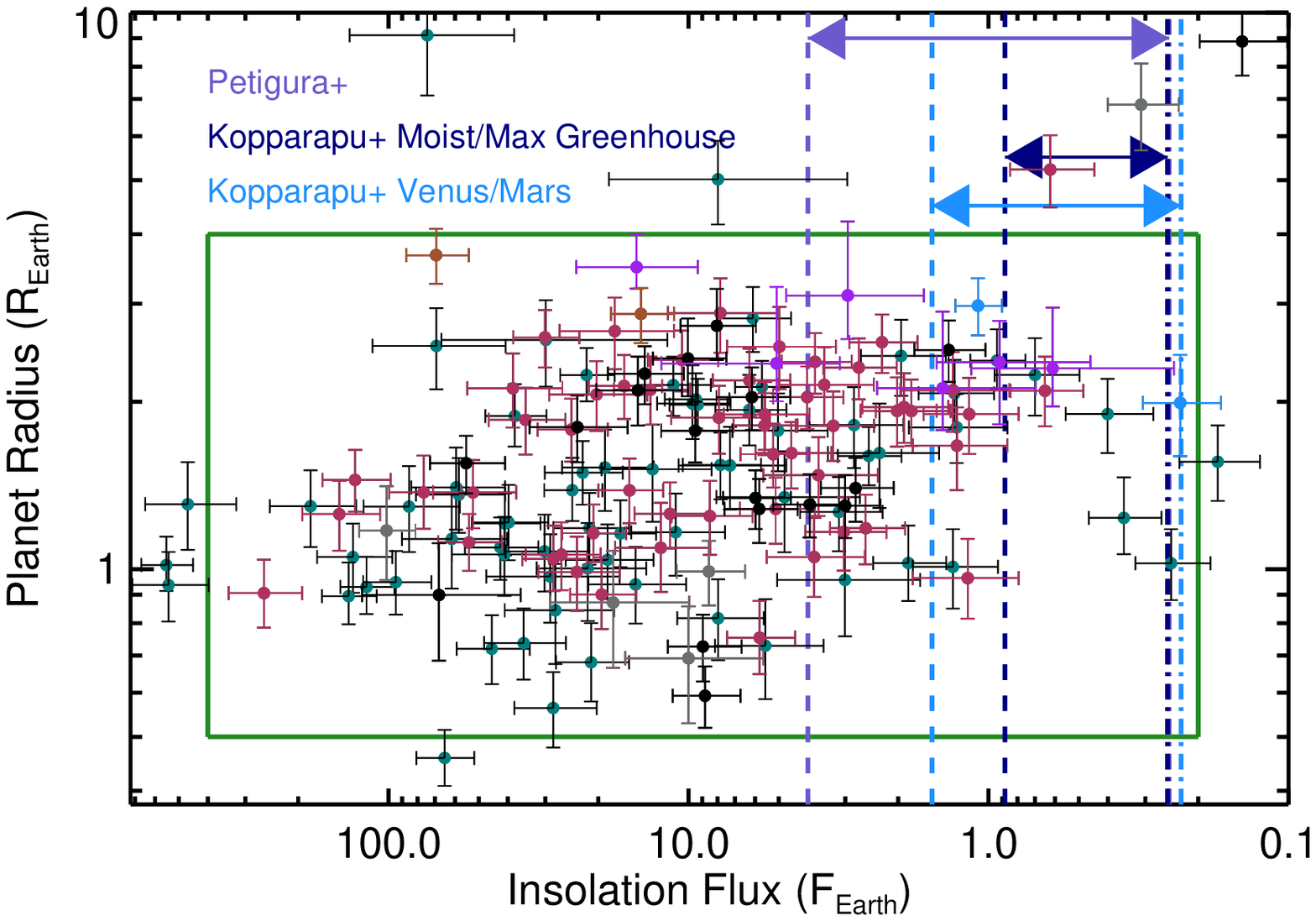}
\end{center}
\caption{Radii of the planet candidates detected by our pipeline versus orbital period (\emph{Left}) or insolation flux (\emph{Right}) with $1\sigma$ errors. In most cases, we refit the planet parameters by conducting an MCMC analysis fitting transit models to the short cadence (crimson points) or long cadence (teal points) \kepler data. For the remaining \nlitcois~planet candidates, we adopted transit parameters from \citet[][purple points]{cartier_et_al2014}, \citet[][brown points]{ioannidis_et_al2014}, \citet[][black points]{rowe_et_al2014}, \citet[][blue points]{swift_et_al2015}, or the NASA Exoplanet Archive (gray points). In all cases, the errors on the planet properties incorporate uncertainties in both stellar and transit parameters. The green boxes indicate the boundaries within which we report planet occurrence rates and the arrows in the right panel mark several variations of the habitable zone as explained in the legend. The errors for KOIs~1422.01, 1422.02, 1422.03, 1422.04, 1422.05, and 2626.01 appear particularly large because we accounted for the possibility that any of the three stars in the KOI~2626 system or either of the two stars in the KOI~1422 system harbor the transiting planets (see Section~\ref{ssec:dilution} for details). KOI~254.01 has an estimated radius of $11\rearth$ and therefore does not appear on these plots.}
\label{fig:planetmc}
\end{figure*}

\section{Planet Properties}
\label{sec:mcpipe}
Accurate planet radius estimates are a key ingredient in the planet occurrence calculation. We therefore refined the preliminary transit parameters found in Section~\ref{ssec:search} by conducting a Bayesian Markov Chain Monte Carlo (MCMC) analysis with a Metropolis-Hastings acceptance criterion \citep{metropolis_et_al1953}. We varied the orbital period $P$, epoch of transit center $t_0$, planet-to-star radius ratio $R_p/R_*$, semimajor axis to stellar radius ratio $a/R_*$ and impact parameter $b$. 
We assumed that all of the orbits were circular and fixed quadratic limb darkening parameters to the values predicted from the stellar temperatures and surface gravities \citep{claret+bloemen2011}. We conducted some of the planet fits using short cadence \kepler data to better constrain the shape of transit during ingress and egress. 

For each planet candidate, we ran $N$ chains starting at initial positions set by perturbing the initial solution found during the detection and validation process by up to $5\sigma$ in each parameter. We manually adjusted the step sizes for each parameter such that the acceptance fractions were between 10--30\%. We ran each chain for at least $10^4$ steps before initiating periodic convergence tests by calculating the Gelman-Rubin potential scale reduction factor $\hat{R}$ for each parameter \citep{gelman_et_al2004}. We terminated the chains when $\hat{R}< 1.05$ for all parameters and then accounted for ``burn-in'' by removing all steps taken prior to the point at which the likelihood first became higher than the median likelihood of the chain. After merging the chains, we adopted the median values of each parameter as the best-fit value and assigned errors encompassing the 68\% of values nearest to the chosen best-fit value. We provide the best-fit parameters for each detected planet candidate in Table~\ref{tab:cois} and display them in Figure~\ref{fig:planetmc}. 

Several of the planet candidates in our sample exhibit large transit timing variations and were poorly fit by the method described above. In those cases, we adopted the transit parameters found in previous studies using combined fits to the individual transit times and the planet properties. For the candidates displaying TTVs, we incorporated fits from \citet{rowe_et_al2014} for KOIs~248.01, 248.02, 248.03, 248.04, 314.01, 314.02, 314.03, 448.01, 448.02, 886.01, 886.02, and 886.03 and from \citet{ioannidis_et_al2014} for KOIs~676.01 and 676.02. 

In addition, we adopted the light curve parameters from the 2 January 2015 version of the NASA Exoplanet Archive for 5~candidates: 961.01, 961.03, 1681.01, 2329.01, and 3263.01. We also adopted the fits from \citet{rowe_et_al2014} for KOIs 254.01, 430.01, 430.02, 438.01, 775.02, 868.01, and 961.02, from \citet{swift_et_al2015} for KOIs~1902.01 and 3444.02 and from \citet{cartier_et_al2014} for KOIs 1422.01, 1422.02, 1422.03, 1422.04, 1422.05, and 2626.01. The KOIs with parameters from \citet{cartier_et_al2014} were detected in multi-star systems in which the identities of the host stars are unknown. As described in Section~\ref{ssec:dilution}, we accounted for all possible system configurations by using fractional planets distributed around each of the possible host stars.

As depicted in Figure~\ref{fig:planetmc}, we found that \ncoisbox of the accepted planet candidates had revised radii  \mbox{$0.5 < R_P < 4 \rearth$} and orbital periods \mbox{$0.5 < P < 200$~days.} Of the remaining candidates, one had a shorter period (KOI 961.02, $P$ = 0.45~days), one was too small (KOI~5692.01), and six were too large (KOI~254.01, 868.01, 901.01, 902.01, 1176.01, and 3263.01). KOI~868.01 also has an orbital period longer than 200~days. 

\section{Planet Injection Pipeline}
\label{sec:injpipe}
In order to accurately measure the planet occurrence rate based on the results of our planet search, we needed to know the completeness of our planet candidate list. We measured the completeness of our planet detection pipeline by injecting transiting planets into the \mbox{PDC-MAP} light curves, detrending them, and running the detrended light curves through our detection algorithm. We did not introduce the signals at the pixel level and we are therefore unable to comment on how the initial light curve extraction process affects transiting planets. We refer instead to \citet{christiansen_et_al2013} for a discussion of pixel-level effects. They found that transits injected at the pixel level are usually recovered with high fidelity (final SNR = $96\% - 98\%$ expected SNR).

The transit detection process as modeled in this paper consists of two distinct stages: (1) the putative event is identified as a peak in the BLS periodogram and (2) the signal is accepted because a transit model provides a $5\sigma$ improvement to a straight-line fit. We took advantage of the two-step nature of the search process when determining the search completeness for each star in our survey. 

For each star, we generated a set of \nfakeperstar~trial planets with orbital periods drawn from a log uniform distribution extending from 0.5 to 200~days and uniformly distributed epochs of transit, radii ($0.5-4\rearth$), and impact parameters ($0-1$). We then constructed transit models \citep{mandel+agol2002} for the trial planets using the assigned planetary parameters and limb darkening parameters estimated from the coefficients in \citet{claret+bloemen2011} based on the stellar temperatures and surface gravities. We resampled the transit models to the 29.4~minute long cadence integration time.

The first question we wished to address was whether the transits of the trial planets would be accepted as $5\sigma$ detections in the ideal scenario in which the light curve was perfectly detrended and the orbital parameters were determined exactly. Accordingly, we first multiplied the detrended light curves by the transit models \citep{mandel+agol2002} generated using the assigned trial planet parameters. We then checked whether the difference in the $\chi^2$ between the best-fit transit model (which we guessed perfectly) and a no-transit model exceeded the $5\sigma$ detection threshold of $\Delta \chi^2 = 30.863$. If the transit model was not preferred, then we recorded the trial planet as a non-detection. For a typical star, \fracfailideal~of the trial planets were rejected at this stage.

For the trial planets that would be accepted in the ideal scenario, we then conducted a more realistic test by multiplying the transit model by the raw PDC-MAP photometry before detrending using the straight line fit method described in Section~\ref{ssec:search}. We then re-checked whether the transit model was preferred at $5\sigma$. Trial planets that were not accepted at this stage were also recorded as non-detections. Overall, \fracfailreal~of the trial planets that were accepted in the ideal case of a perfectly detrended light curve were not accepted in this more realistic test.

Finally, we tested whether the remaining trial planets would have been identified as peaks in the BLS periodograms by running a full test for at least 25~trial planets for each star. We selected the trial planets for the full test by ranking the signals detected in the second round of testing in order of increasing $\Delta \chi^2$, where $\Delta \chi^2$ was the value at which they were preferred to a non-transiting model. We chose the first 25~trial planets in the ranked list for which a random number draw yielded a result greater than 0.5. In other words, we thinned the sample of trial planets by 50\% and selected those closest to the expected sensitivity threshold. If fewer than ten trial planets were recovered, we conducted up to 25 additional runs (for a total of 50 simulations) until at least ten planets were recovered.

For the selected trial planets, we multiplied the corresponding transit model by the raw \mbox{PDC-MAP} photometry for the assigned host star and detrended the light curve using a 2000~minute median filter as explained in Section~\ref{ssec:detrending}. We then fed the injected light curves into the detection pipeline described in Section~\ref{sec:detpipe}.  Although we considered the full range of planet periods (0.5--200 days), we halted the search process as soon as the injected signal was detected. If the signal was not detected, we terminated the search using the usual conditions discussed in Section~\ref{ssec:search}.

\begin{figure}[htbp] 
\begin{center}
\centering
\includegraphics[width=0.48\textwidth]{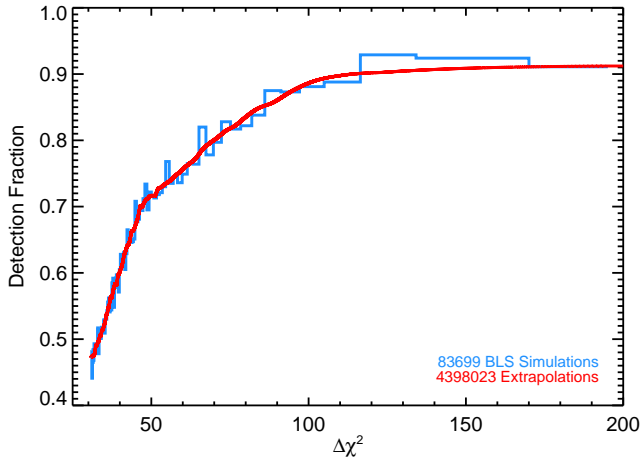}
\end{center}
\caption{Empirical detection sensitivity versus the $\Delta \chi^2$ between fitting the detrended light curve with the injected transit model or with a no-transit model. The light blue histogram depicts the recovery fraction for the \nfullbls~BLS trial planets in bins of 1000~planets. The solid red line marks the estimated likelihood of detection for the \nextrasim~non-BLS injected planets predicted from the smoothed histogram of BLS results.}
\label{fig:ramp}
\end{figure}

\subsection{Predicting Transit Detectability}
\label{ssec:injextrap}
In total, we ran \nfullbls~complete BLS injection simulations for the \nstars~stars in our sample. We also injected \nfailprebls~planets that had $\Delta \chi^2$ below our $5\sigma$ detection threshold. The remaining \nextrasim~injected planets had $\Delta \chi^2$ above the detection threshold but were not tested in the full BLS simulation. We predicted the detectability of these trial planets by finding the fraction of BLS trial planets recovered as a function of the $\Delta \chi^2$ computed in the second round of transit model tests. We ranked the BLS trial planets by $\Delta \chi^2$ and computed the recovery fraction for each consecutive group of 500~planets. Next, we smoothed the resulting histogram and predicted the likelihood of detection for the \nextrasim~non-BLS runs using a cubic spline interpolation based on the smoothed histogram. We limited the maximum detection likelihood to \maxdf, which was the maximum value of the cubic spline  in the histogram of the recovery rate for the full BLS runs. Figure~\ref{fig:ramp} displays the histogram of the recovery rate for the BLS trial planets and the extrapolated likelihoods of detection for the non-BLS runs. 

\subsection{Assessing Pipeline Performance}
\label{ssec:recovery}
In total, we injected \ninj~transiting planets into the light curves of the \nstars~stars in our sample. We provide a catalog of injected planet parameters and recovery results in Table~\ref{tab:injplanets}. Our pipeline successfully recovered $86\%$ of signals injected with an expected SNR between 15 and 20.  For lower SNR, the pipeline performance decreased roughly linearly with anticipated SNR until reaching $52\%$ recovery for signals with anticipated SNR between 5 and 7. For the purpose of assessing pipeline performance as a function of SNR, we modeled the anticipated SNR of a transiting planet as:
\begin{equation}
\rm{SNR} = \frac{\delta}{\rm{CDPP}_{\rm{transit}}}\sqrt{n_{\rm{transit}}}
\label{eq:snr}
\end{equation}

where $\delta$ is the median decrease in brightness during the injected transit, $\rm{CDPP}_{\rm{transit}}$ is the Combined Differential Photometric Precision (CDPP) on the timescale of a full transit of a planet on a circular orbit, and $n_{\rm{transit}}$ is the number of transits expected given the orbital period of the planet and the number of days the star was observed. As in \citet{dressing+charbonneau2013}, we estimated the CDPP on the timescale of a planetary transit by interpolating over the provided CDPP measured on 3-, 6-, and 12-hour timescales.

\begin{deluxetable}{crrrr}
\tablecolumns{5}
\tabletypesize{\footnotesize}
\tablecaption{Injected Planets}
\tablehead{
\colhead{} & 
\colhead{Period} &
\colhead{Radius} &
\colhead{Insolation} &
\colhead{Recovery} \\
\colhead{KID} & 
\colhead{(Days)} &
\colhead{($\rearth$)} &
\colhead{($\fearth$)} &
\colhead{Status\tablenotemark{a}}
}
\startdata
003835071 & 0.686 & 0.712 & 409.813 & 1.000 \\
005693298 & 0.876 & 0.531 & 250.636 & 1.000 \\
011560326 & 1.900 & 2.700 & 48.871 & 0.912 \\
008229458 & 2.700 & 0.788 & 61.212 & 0.520 \\
004931385 & 20.420 & 1.717 & 3.056 & 0.650 \\
009691776 & 55.352 & 2.555 & 0.568 & 0.904 \\
010032631 & 85.665 & 1.978 & 0.664 & 0.791 \\
002441562 & 112.865 & 0.925 & 0.195 & 0.000 \\
011013096 & 155.207 & 1.795 & 0.123 & 0.908 \\
002692704 & 184.735 & 1.181 & 0.225 & 0.000 \\
... & ... & ... & ... & ... 
\enddata
\tablecomments{Table~\ref{tab:injplanets} is published in its entirety in the electronic edition of the Astrophysical Journal. A portion is shown here for guidance regarding its form and content.}
\tablenotetext{a}{For the \nfullbls~injected planets that were tested in the complete pipeline and the \nfailprebls~planets with $\Delta \chi^2$ below our $5\sigma$ detection threshold, the recovery status indicates whether the planet was detected (1 for recovered planets, 0 for unrecovered planets). For the remaining \nextrasim~injected planets that had $\Delta \chi^2$ above the detection threshold but were not tested in the full BLS simulation, the recovery status indicates the estimated likelihood of detection (see \mbox{Section~\ref{sec:injpipe}}).}
\label{tab:injplanets}
\end{deluxetable}

\begin{figure*}[htbp] 
\begin{center}
\centering
\includegraphics[width=0.48\textwidth]{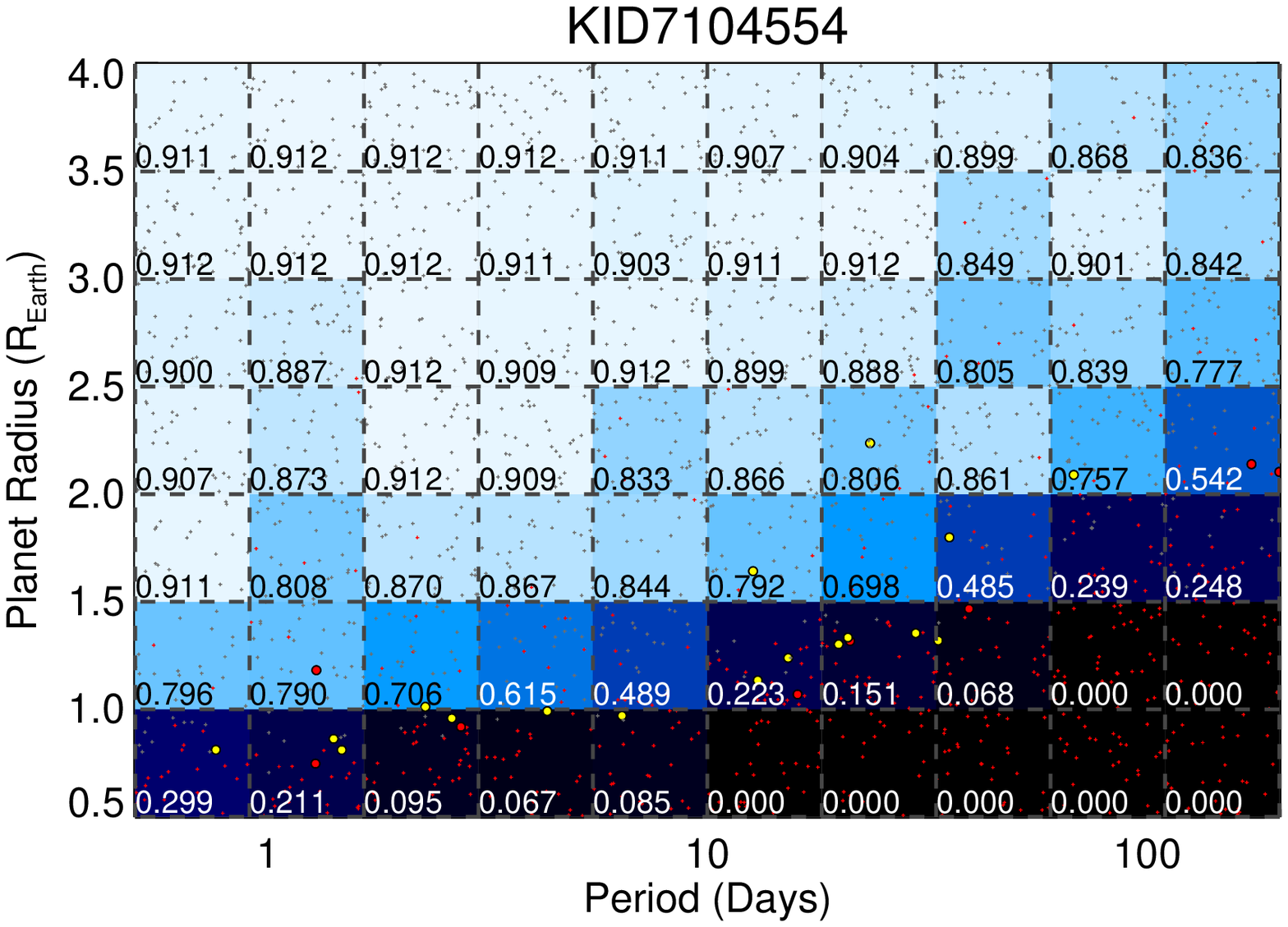}
\includegraphics[width=0.48\textwidth]{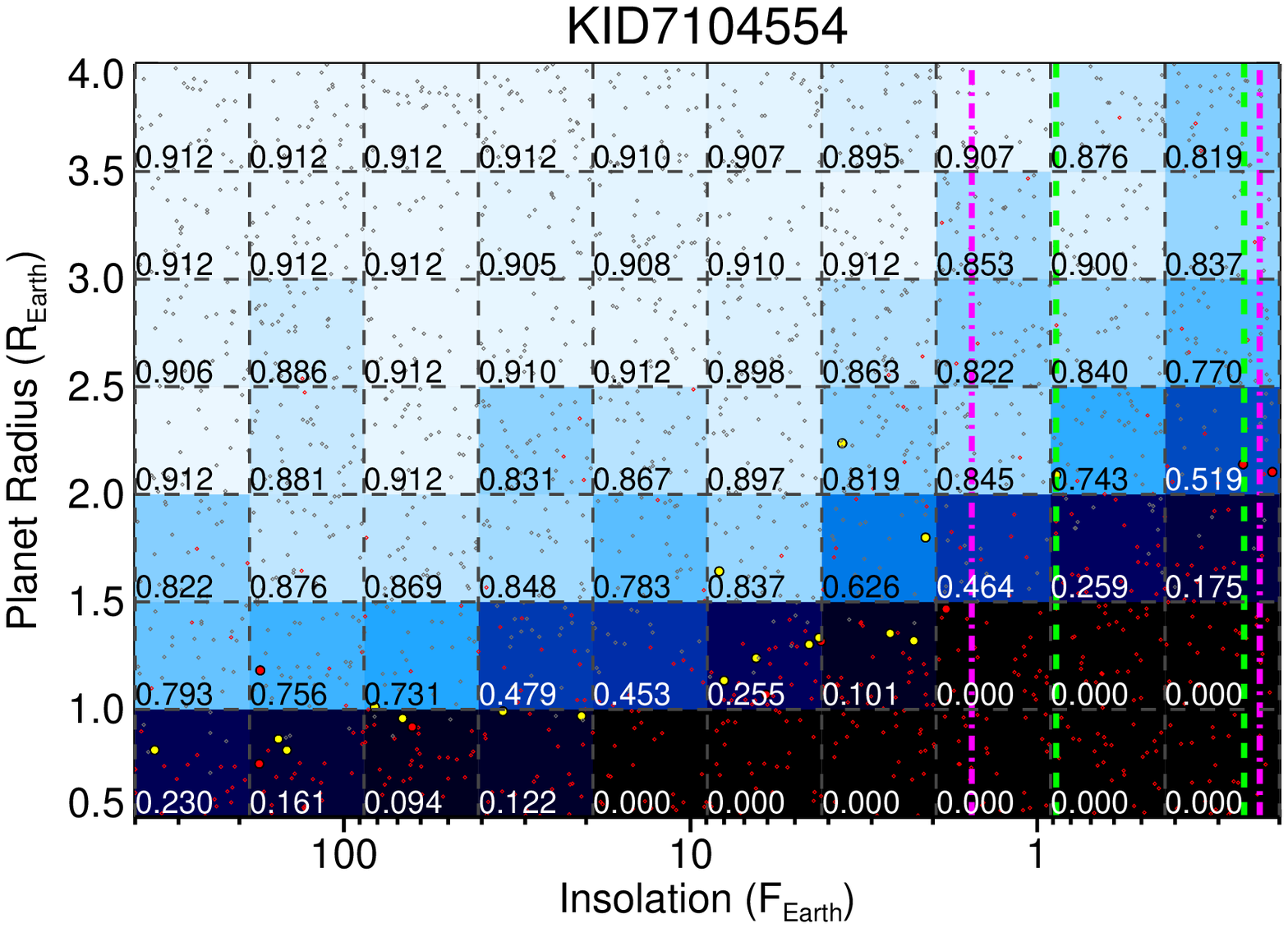}
\end{center}
\caption{Transit detectability maps for KID~7104554 as a function of planet radius and orbital period (\emph{Left}) or insolation flux (\emph{Right}) based on the results of our injection simulation. The small red plus symbols mark the 498 injected planets with $\Delta \chi^2$ below the $5\sigma$ detection threshold. The 25~large circles indicate injected planets with $\Delta \chi^2$ above the detection threshold that were recovered (yellow, 17 planets) or undetected (red, 8 planets) during the full BLS test phase. The small gray plus symbols are the remaining 1477~injected planets with $\Delta \chi^2$ above the detection threshold that were not selected for the full BLS test. The numbers within each cell denote the recovery fraction within the cell boundaries and the cells are color-coded so that darker colors correspond to lower detectability. The green dashed lines mark the maximum greenhouse (Max GH) and moist greenhouse (Moist GH) insolation limits from \citet{kopparapu_et_al2013} and the magenta dot-dashed lines mark the less conservative Recent Venus and Early Mars limits, also from \citet{kopparapu_et_al2013}.}
\label{fig:kidsearchmap}
\end{figure*}

As a benefit of injecting multiple trial planets per star, we generated unique transit detectability maps for each star in our sample. For example, Figure~\ref{fig:kidsearchmap} displays the transit detectability maps for KID~7104554, a $Kp=15.3$, $T_{\rm eff} = 3957K$ star with lower search completeness than the larger sample. We created the star-by-star transit detectability maps by gridding the injected planets in radius/period and radius/insolation space and calculating the fraction of detectable planets within each grid cell. For the subset of trial planets that passed both $\Delta \chi^2$ tests yet were not selected for the full BLS search, we estimated the recovery fraction as explained in Section~\ref{ssec:injextrap}.

\begin{figure*}[htbp] 
\begin{center}
\centering
\includegraphics[width=0.48\textwidth]{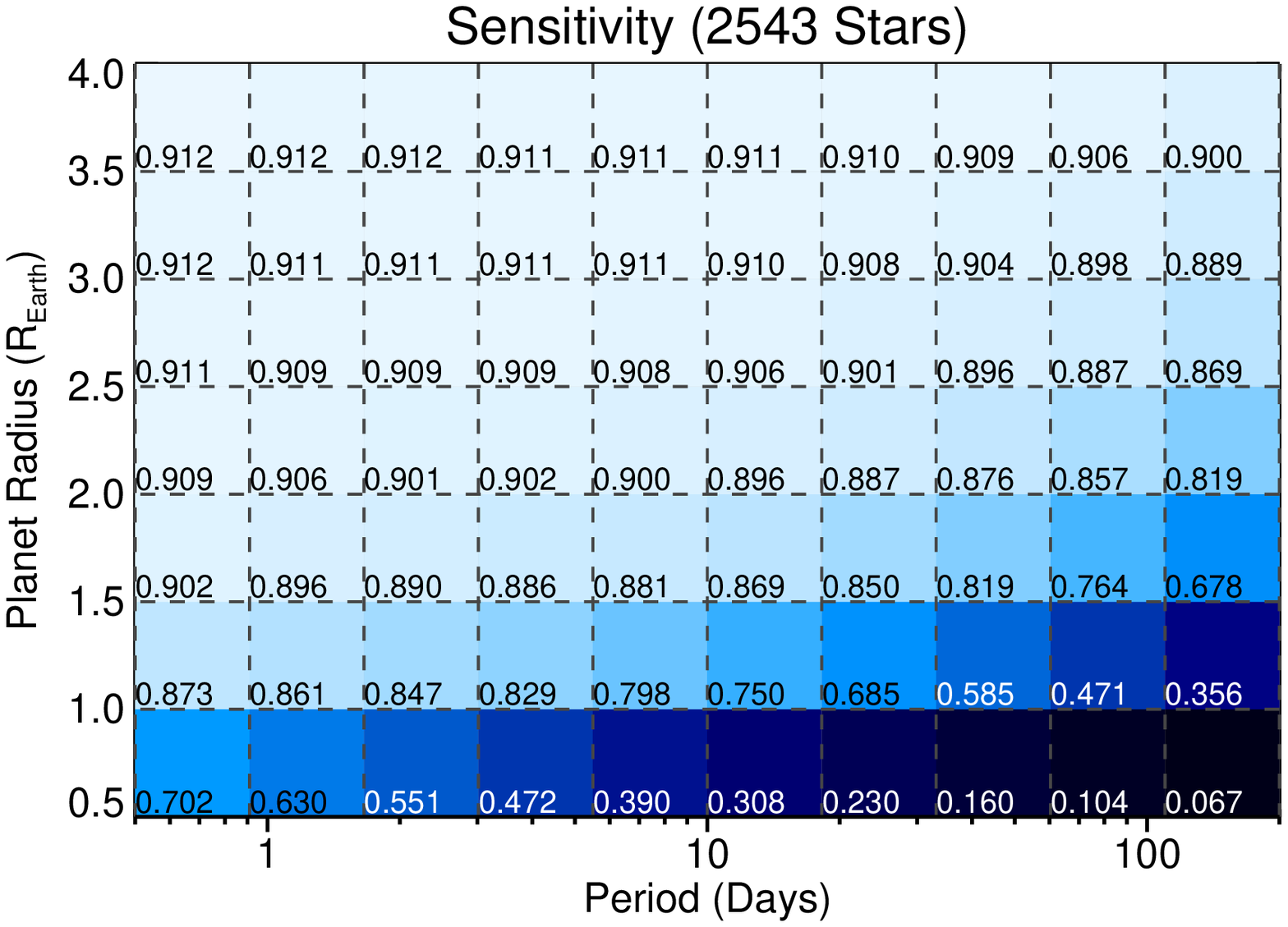}
\includegraphics[width=0.48\textwidth]{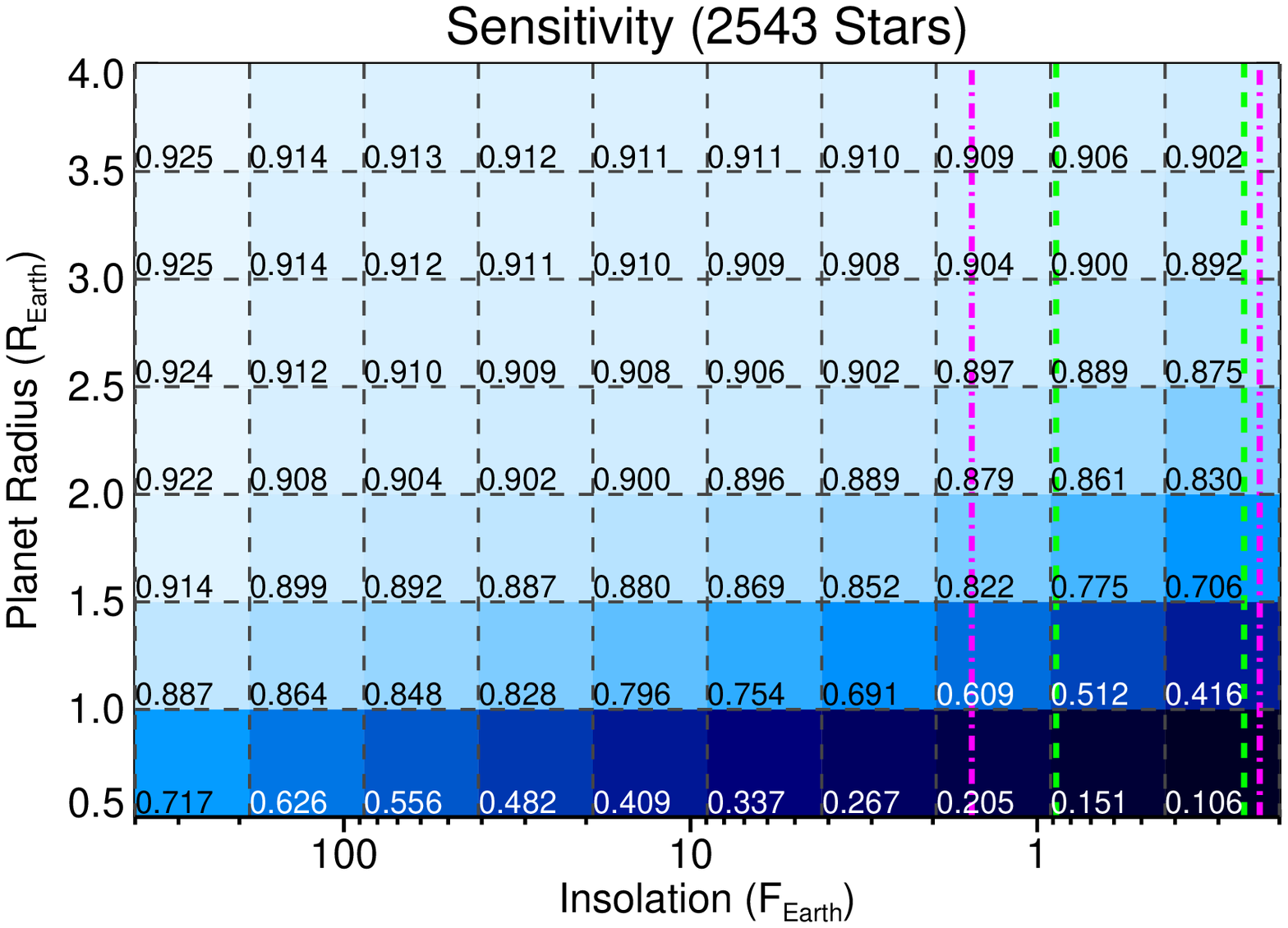}
\end{center}
\caption{Combined transit detectability maps for the full stellar sample as a function of planet radius and orbital period (\emph{Left}) or insolation flux (\emph{Right}) based on the results of our injection simulation. The numbers within each cell denote the recovery fraction within the cell boundaries and the cells are color-coded so that darker colors correspond to lower detectability. These figures were produced by combining individual completeness maps for each star such as those displayed in Figure~\ref{fig:kidsearchmap}. As in Figure~\ref{fig:kidsearchmap}, the vertical lines in the right panel mark two definitions of the habitable zone.}
\label{fig:searchmap}
\end{figure*}

\begin{figure*}[htbp] 
\begin{center}
\centering
\includegraphics[width=0.48\textwidth]{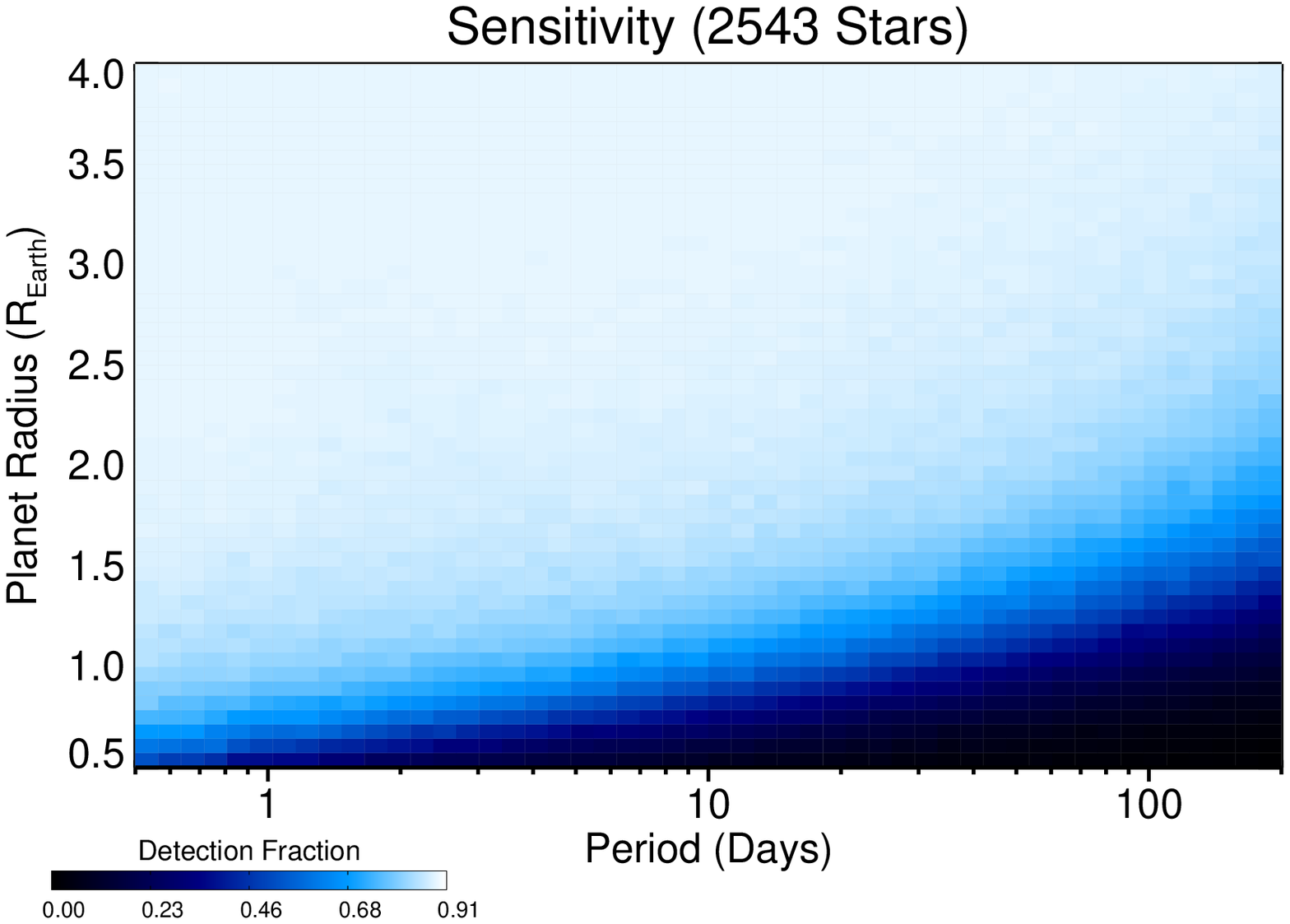}
\includegraphics[width=0.48\textwidth]{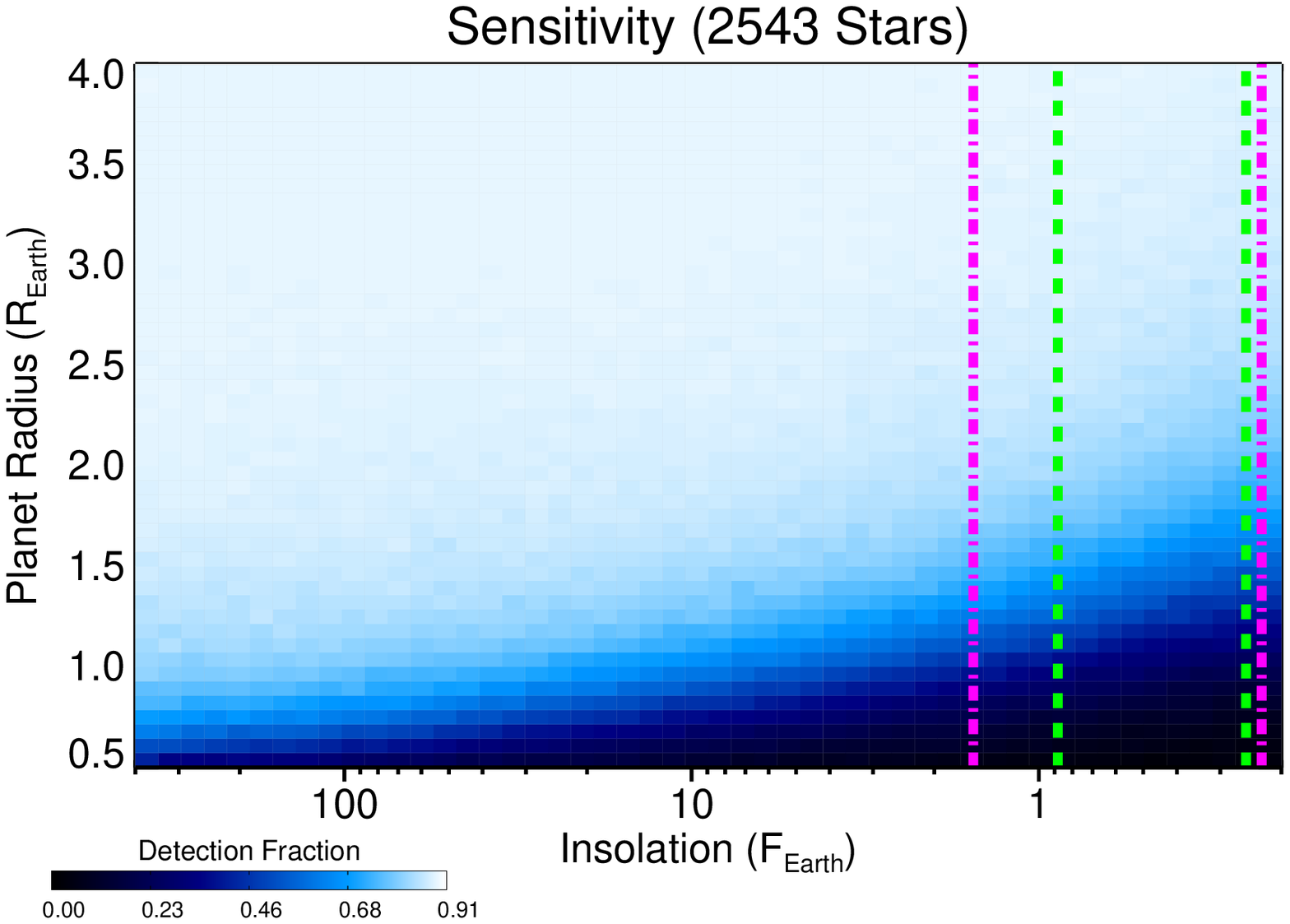}
\end{center}
\caption{Smoothed maps of the fraction of injected planets that were detected by our pipeline. As indicated in the color bars, darker points correspond to lower detection fractions. \emph{Left:} Planet radius versus period. \emph{Right:} Planet radius versus insolation.  As in Figure~\ref{fig:kidsearchmap}, the vertical lines mark two definitions of the habitable zone.}
\label{fig:detfrac}
\end{figure*}

After generating transit detectability maps in radius-period and radius-insolation space for each star independently, we created transit detectability maps for the full sample by summing the individual maps. Although the combined maps displayed in Figure~\ref{fig:searchmap} are useful for comparing the sensitivity of any individual star to the sensitivity of the larger sample, the binning is rather coarse. We therefore generated a second set of combined sensitivity maps by sorting the full set of \ninj~injected planets into smaller grid cells in radius/period and radius/insolation space. We then calculated the recovery fractions within each of the cells to produce the smoother sensitivity maps displayed in Figure~\ref{fig:detfrac}. 

As shown in Figure~\ref{fig:detfrac}, we found that our pipeline is very sensitive to injected planets with radii larger than \emph{$2.5\rearth$}. Such planets were detected with nearly 90\% efficiency out to the maximum injected orbital period of 200~days. Our pipeline had a significantly harder time detecting $1.0-1.5\rearth$ planets with periods longer than 100~days (recovery fraction = 36\%) and $0.5-1.0\rearth$ planets with periods longer than 5~days (recovery fraction = 22\%). Planets smaller than $1.0\rearth$ were nearly undetectable (recovery rate approximately $6\%$) at orbital periods longer than 150~days.

Inspecting the transit recovery map as a function of insolation revealed that the transit detectability changes sharply across the habitable zone (HZ). At the inner edge of the HZ (median orbital period of 50~days for the stars in our sample), we recovered 84\% of $2.0\rearth$ planets and 34\% of $1.0\rearth$ planets. At the outer edge of the habitable zone (median orbital period of 130~days), the sensitivity decreased to 80\% for $2.0\rearth$ planets and 25\% for $1.0\rearth$ planets. This large change in sensitivity in a very interesting region of planet radius and insolation space reveals that the pipeline sensitivity within the habitable zone is not well described by a single number.

\subsection{Calculating Search Completeness}
\label{ssec:maps}
The overall planet search completeness depends both on the detectability of a particular transiting planet and the likelihood that a particular planet will be observed to transit. We accounted for the latter factor by determining the mean geometric probability of transit for planets orbiting the stars in our sample at particular periods or insolation levels. For a given orbital period, we computed the corresponding semimajor axis for a planet orbiting each of the stars in our sample. Next, we divided the stellar radii by the calculated semimajor axes to find the transit probability for a planet in a circular orbit. 

\begin{figure*}[htbp] 
\begin{center}
\centering
\includegraphics[width=0.48\textwidth]{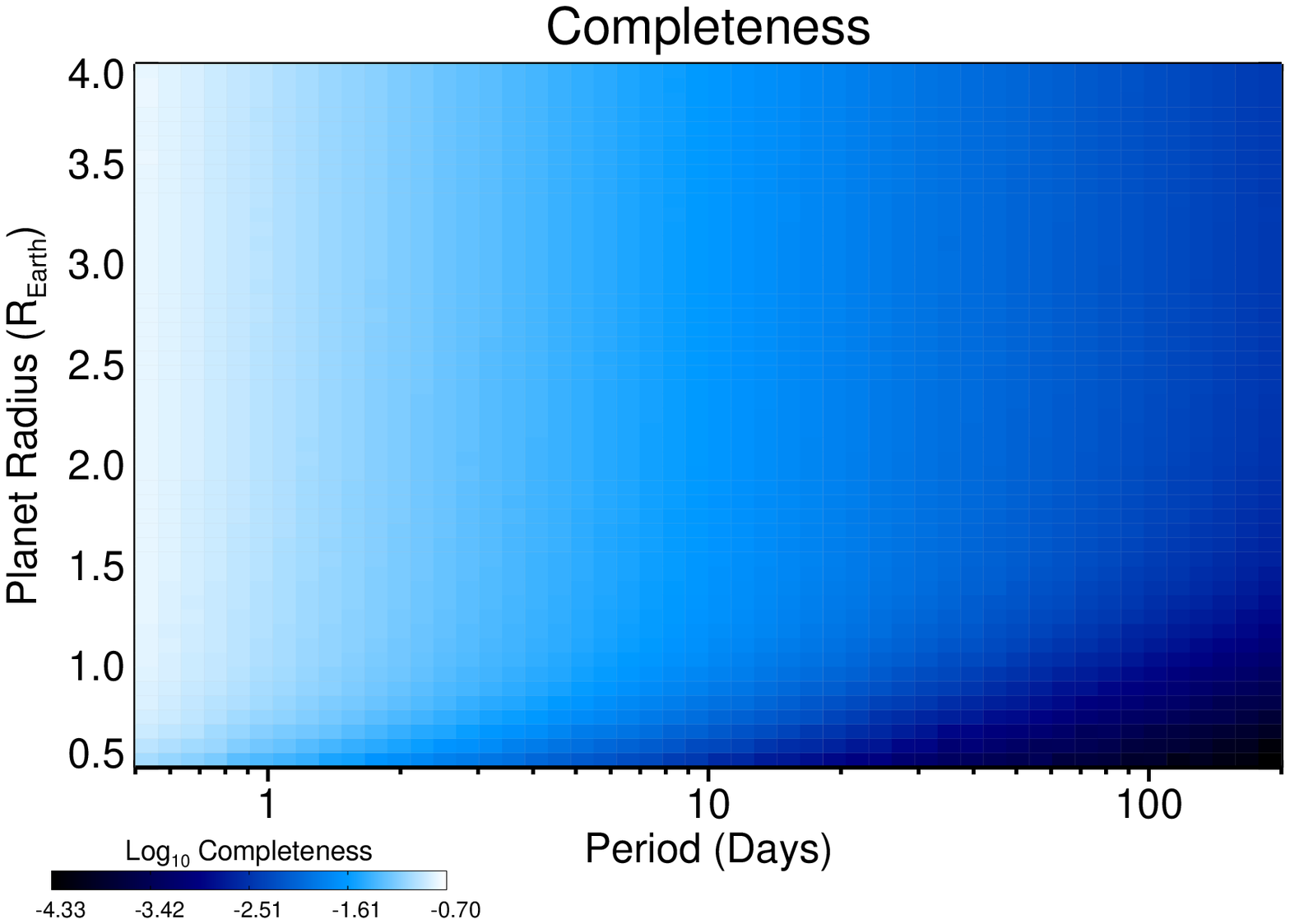}
\includegraphics[width=0.48\textwidth]{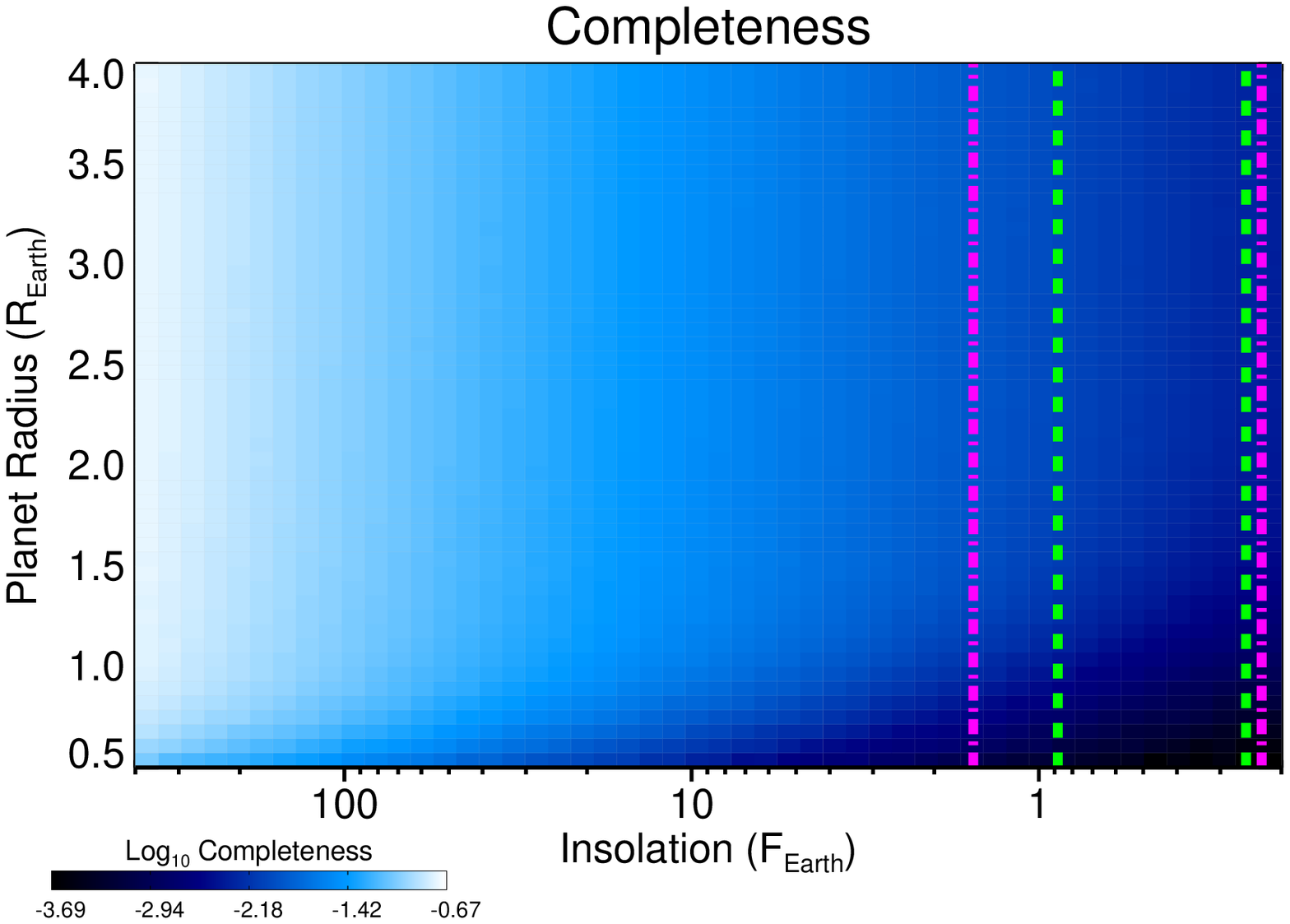}
\end{center}
\caption{Smoothed maps of the search completeness accounting for both pipeline sensitivity and the geometric probability of transit. As indicated in the color bars, darker points correspond to lower search completeness. \emph{Left:} Planet radius versus period. \emph{Right:} Planet radius versus insolation.  As in Figure~\ref{fig:kidsearchmap}, the vertical lines mark two definitions of the habitable zone.}
\label{fig:comp}
\end{figure*}

We then multiplied the transit probability by a correction factor to account for the fact that the planets in our sample are more likely to be eccentric and were accordingly more likely to transit \citep{barnes2007, kipping2014}. We adopted a correction factor of $1.08$ based on a beta distribution fit by \citet{kipping2013b} to transiting planets with periods shorter than 382.3~days. Neglecting this correction factor would lead to an underestimate of search completeness and an overestimate of the planet occurrence rate by roughly $8\%$ \citep{kipping2014}. The resulting search completeness plots are displayed in Figure~\ref{fig:comp}.

\section{The Planet Occurrence Rate}
\label{sec:occrate}
In order to estimate the planet occurrence rate, we first generated smoothed maps of the detected planet population. For each planet candidate, we counted the number of links from the MCMC posteriors that fell within each grid cell in radius/period and radius/insolation space.\footnote{We directly incorporated the posteriors for the KOIs fit by \citet{rowe_et_al2014}. For the other 15~KOIs with parameters drawn from previously published papers, we modeled the radii and periods by constructing Gaussian distributions using the reported values and errors.} When converting each link of the chains from light curve parameters to physical values, we accounted for uncertainties in the stellar parameters by drawing new stellar parameters from Gaussians centered at the reported values with widths set by the reported errors. (In cases where the reported errors were asymmetric, we adopted the larger value.) We weighted each link so that the total weight equaled one minus the false positive correction (see Section~\ref{ssec:fp}) for a planet with the given radius. 

The errors on the planet radii and insolation flux were large enough that the posteriors from multiple candidates overlapped to produce smoothed distributions. For the orbital periods, however, the errors were small enough that each planet appeared isolated. For the purpose of calculating the planet occurrence rate, we artificially inflated the spread of the period values so that the standard deviation of $\log_{10}P$ distribution was equal to 0.1. We then constructed smoothed distributions of the insolation flux  received by each planet by converting the periods into semimajor axes using Kepler's third law and stellar masses drawn from gaussian distributions centered on the reported value with widths set by the reported errors. 

\begin{figure*}[htbp] 
\begin{center}
\centering
\includegraphics[width=0.48\textwidth]{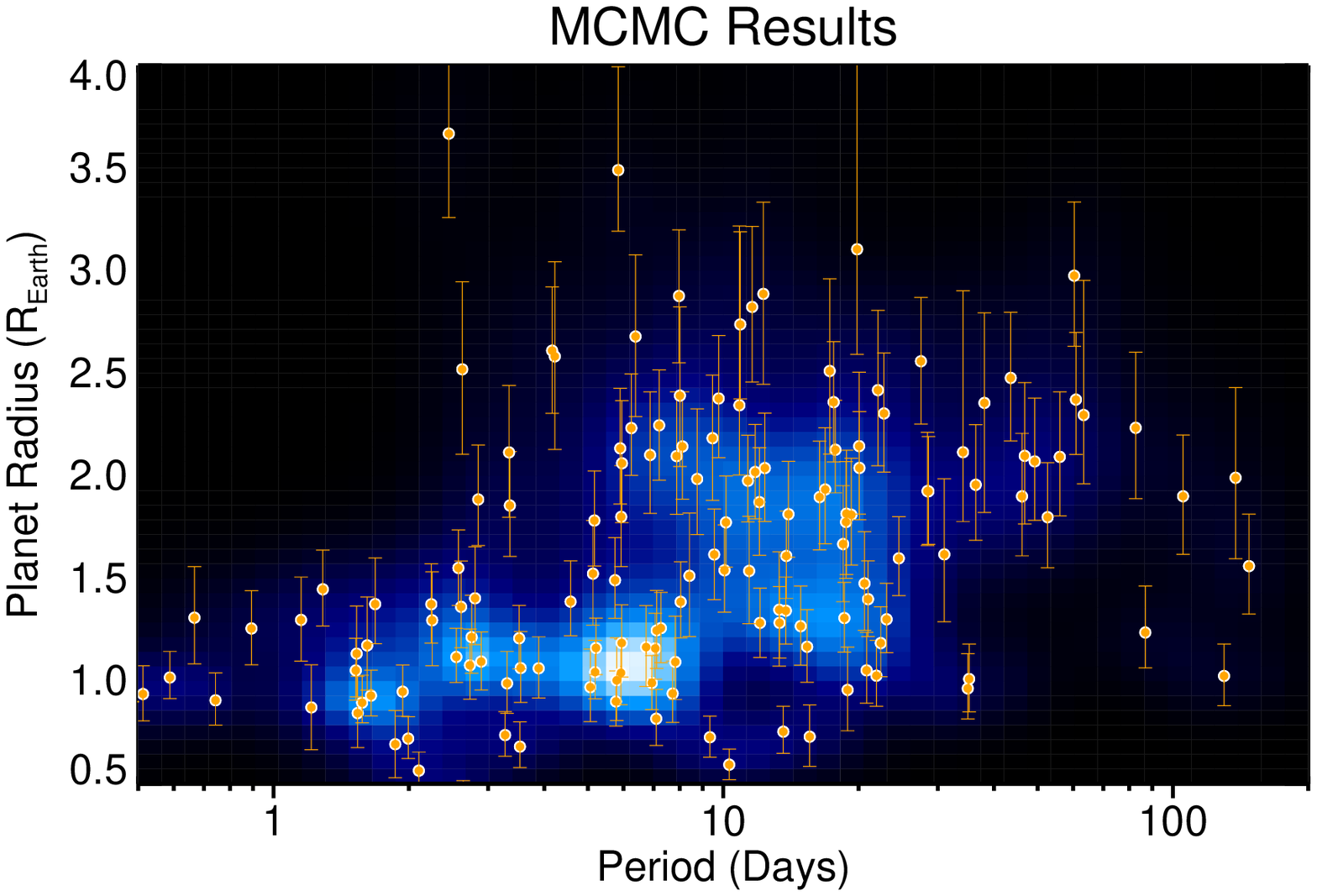}
\includegraphics[width=0.48\textwidth]{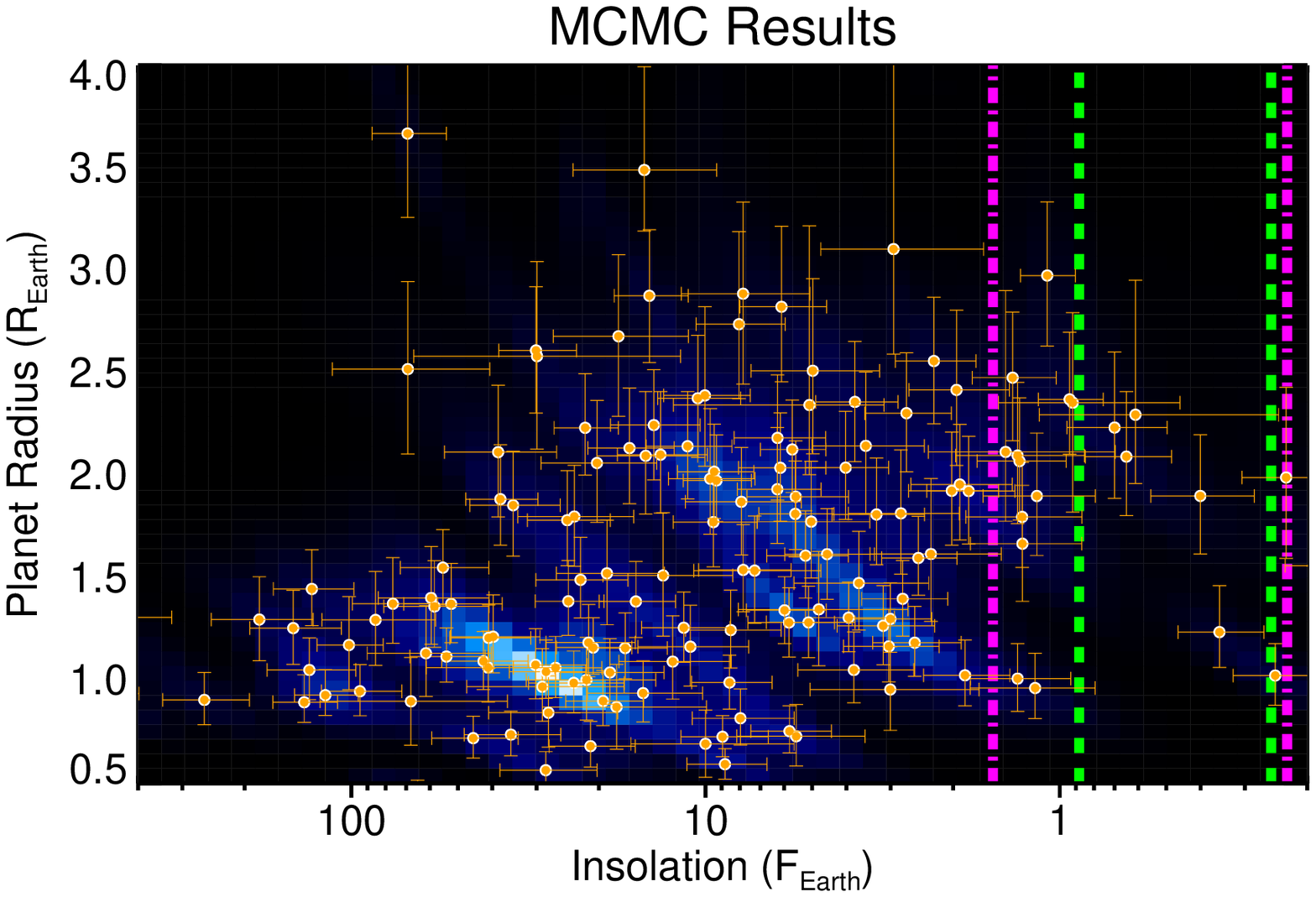}
\end{center}
\caption{Smoothed distribution of planet candidates detected by our pipeline. The color scale is linear with lighter colors indicating a higher number of planets. \emph{Left: }Planet radius versus period. \emph{Right: } Planet radius versus insolation. The orange points with error bars are the planet candidates detected by our pipeline. As in Figure~\ref{fig:kidsearchmap}, the vertical lines mark two definitions of the habitable zone.}
\label{fig:koimap}
\end{figure*}

As shown in Figure~\ref{fig:koimap}, the detected planet population has peaks within the region $P=1-20$~days and \mbox{$R_p=0.7-2.5\rearth$}. There is also a noticeable lack of large planets ($R_p \geq 1.7\rearth$) and shorter periods ($P \leq 2$~days). In radius-insolation space (right panel of Figure~\ref{fig:koimap}), the highest peaks of the smoothed candidate distribution are located at insolations of $20-50$ and $2.5-10$~times the insolation received by the Earth. 

\begin{figure*}[htbp] 
\begin{center}
\centering
\includegraphics[width=0.48\textwidth]{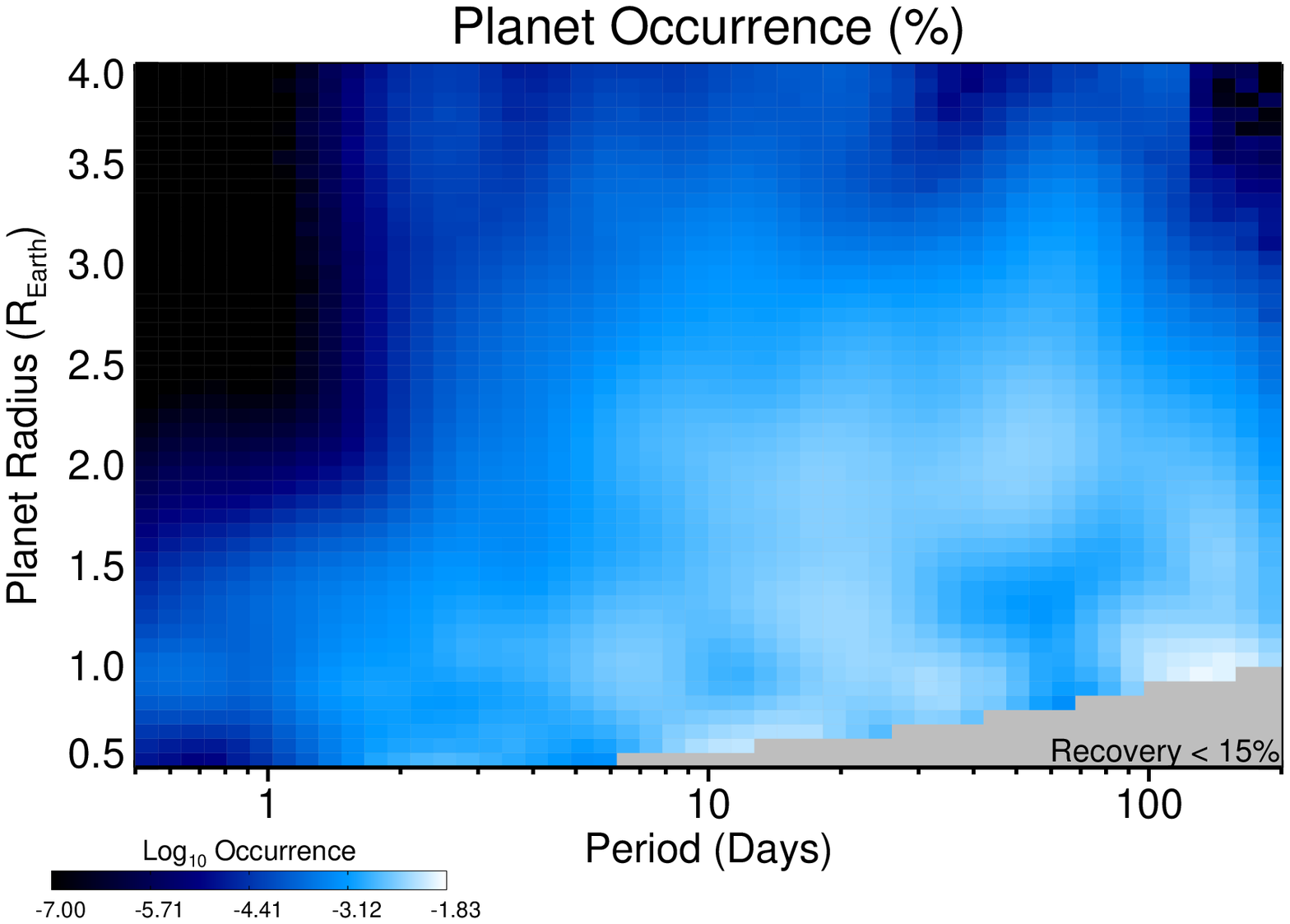}
\includegraphics[width=0.48\textwidth]{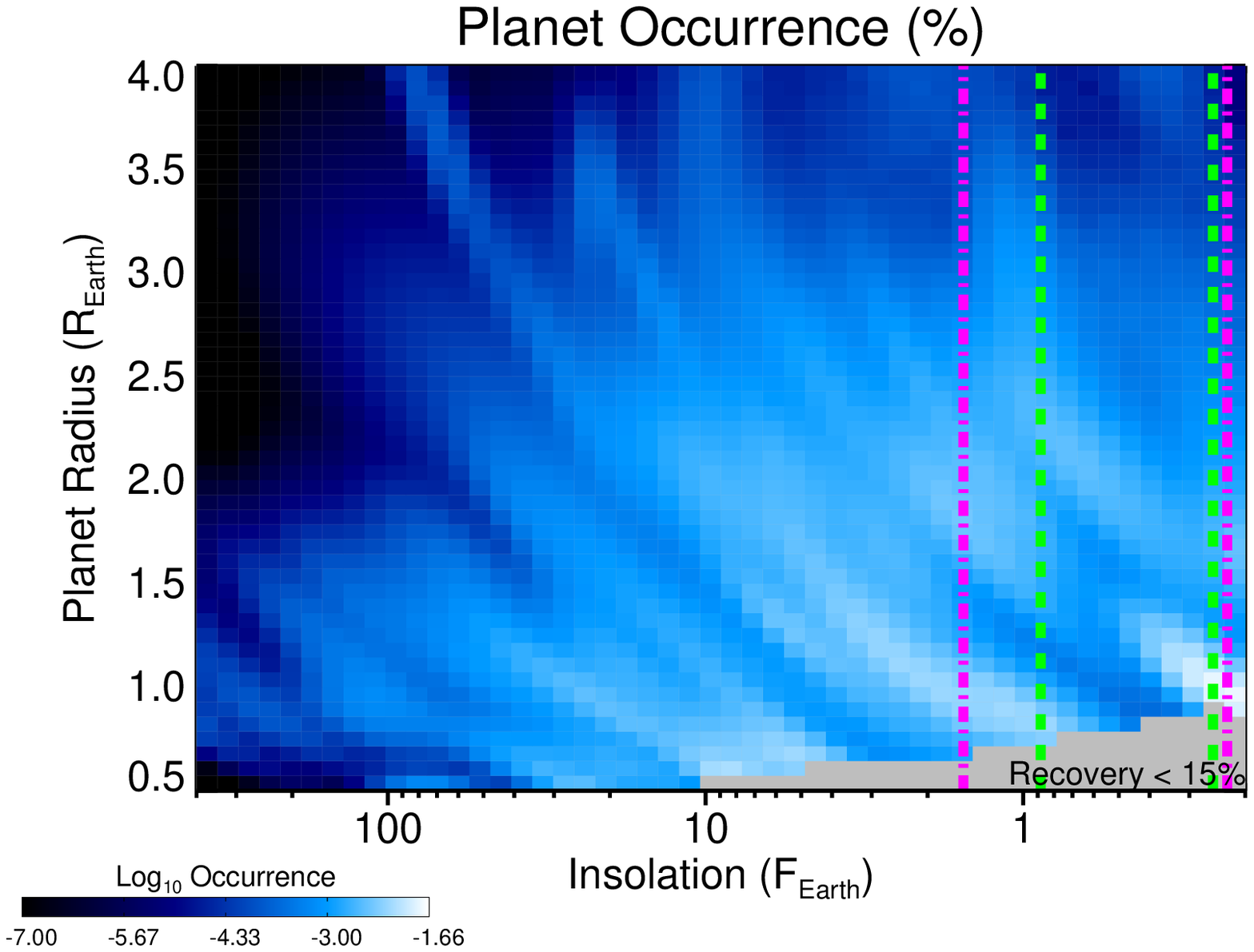}
\end{center}
\caption{Smoothed plot of the derived planet occurrence rate as a function of planet radius versus orbital period (\emph{Left}) or insolation (\emph{Right}). Lighter colors indicate higher planet occurrence per grid cell and regions in which our pipeline detected $<15\%$ of injected signals are marked in gray. As in Figure~\ref{fig:kidsearchmap}, the vertical lines in the right panel mark two definitions of the habitable zone.}
\label{fig:occmap}
\end{figure*}

\begin{figure*}[tbp] 
\begin{center}
\centering
\includegraphics[width=0.48\textwidth]{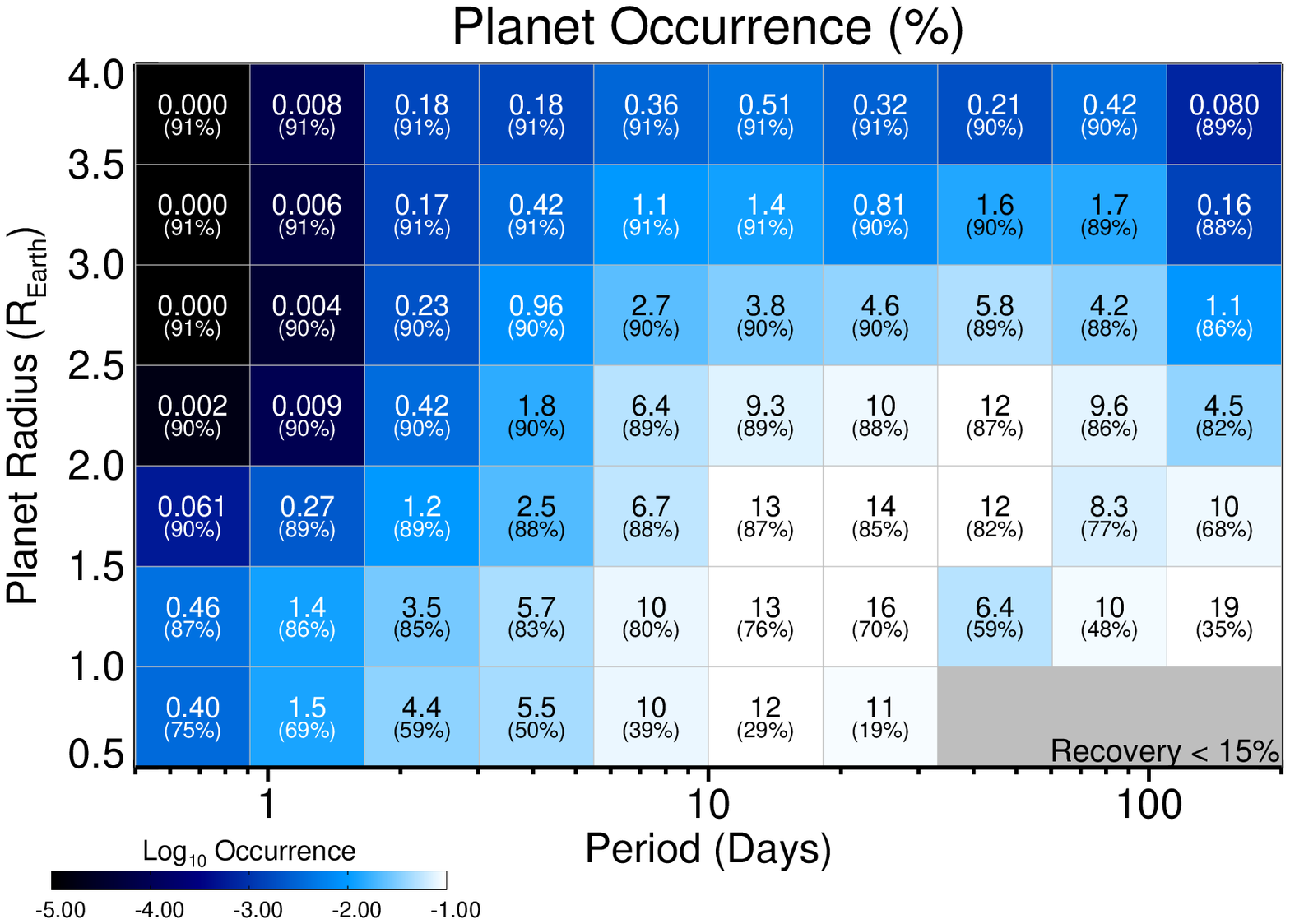}
\includegraphics[width=0.48\textwidth]{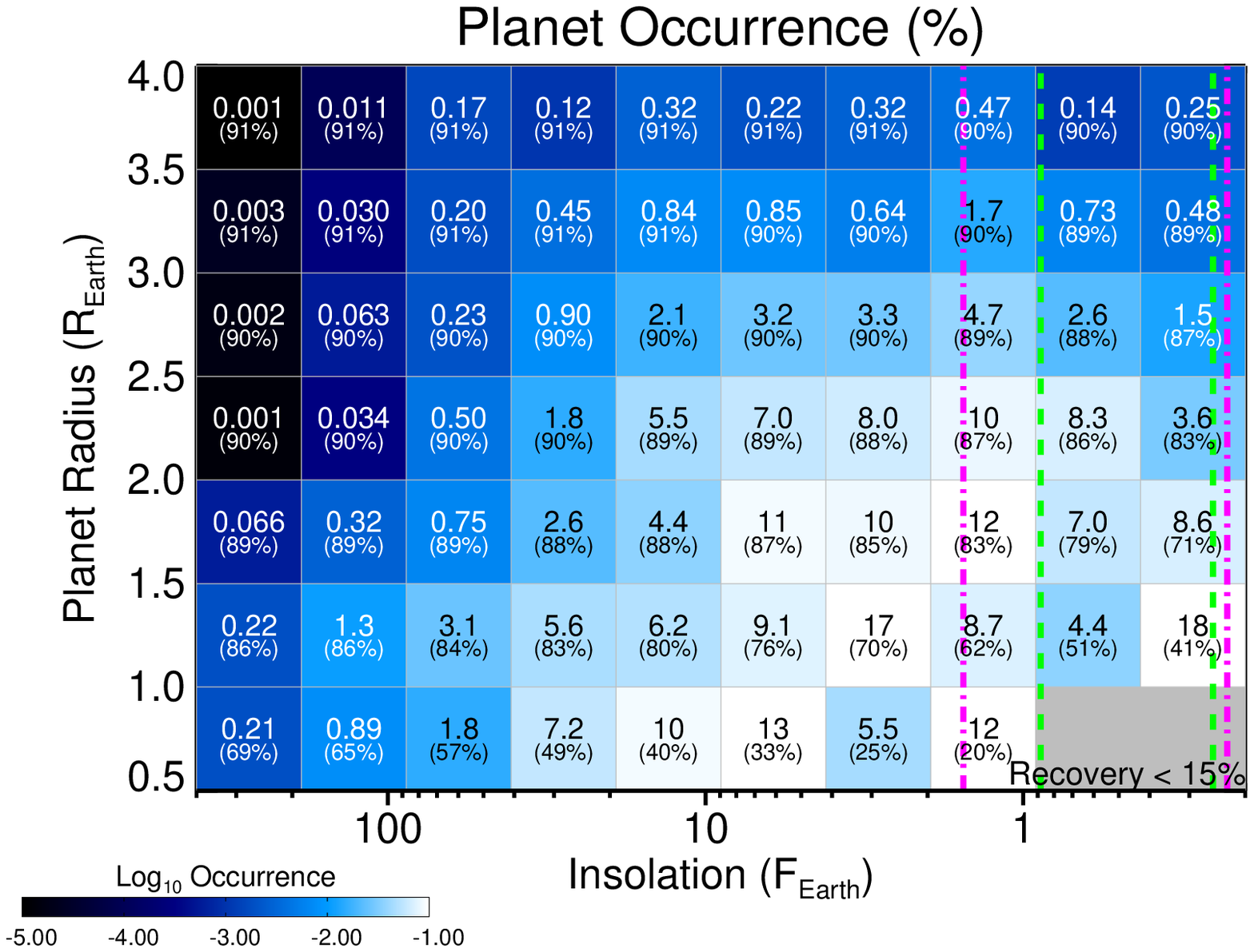}
\end{center}
\caption{Binned planet occurrence rate in period/planet radius space (\emph{Left}) and insolation/planet radius space (\emph{Right}). The numbers within each grid cell indicate the planet occurrence rate as a percentage (top) and the percentage of injected planets that were recovered by our pipeline (bottom). The gray regions have injected planet recovery rates below 15\%. Some boxes have large Poisson errors; please see Tables \ref{tab:perocc} --\ref{tab:hzocc}.  As in Figure~\ref{fig:kidsearchmap}, the vertical lines in the right panel mark two definitions of the habitable zone.}
\label{fig:occgrid}
\end{figure*}

We estimated the planet occurrence rate by dividing the smoothed maps of the detected planet population in Figure~\ref{fig:koimap} by the smoothed maps of the search completeness in Figure~\ref{fig:comp}. The resulting maps of the planet occurrence rate are displayed in Figure~\ref{fig:occmap}. The division reveals that the lack of detected planets in the upper left corner of the left panel of Figure~\ref{fig:koimap} is quite meaningful. That region has very high search completeness, so the lack of detected planets in that region of parameter space implies that hot mini-Neptunes and Neptunes are rare around low-mass stars. At the opposite corner of the diagram in the small-planet, long-period regime, the relatively small number of detected planets does not indicate a low occurrence rate. On the contrary, those planets were detected despite relatively low search completeness, so the underlying occurrence rate of such planets is predicted to be high. 

Consulting the right panel of Figure~\ref{fig:occmap}, the estimated occurrence rate of small planets is highest at insolations below roughly $0.5\fearth$, but that is a region of low search completeness and the occurrence rate for such planets is not well constrained. In order to more readily see trends in the planet occurrence rate as a function of planetary properties, we binned the smoothed occurrence distributions shown in Figure~\ref{fig:occmap} to produce the gridded diagrams shown in Figure~\ref{fig:occgrid}. The gridded version of the radius-period diagram (left panel of Figure~\ref{fig:occgrid}) clearly demonstrates that planet occurrence increases with decreasing planet radius and increasing $\log_{10}P$. 

\subsection{Dependence on Planet Radius \& Period}
\label{ssec:occrpper}

\begin{figure}[tbp] 
\begin{center}
\centering
\includegraphics[width=0.5\textwidth]{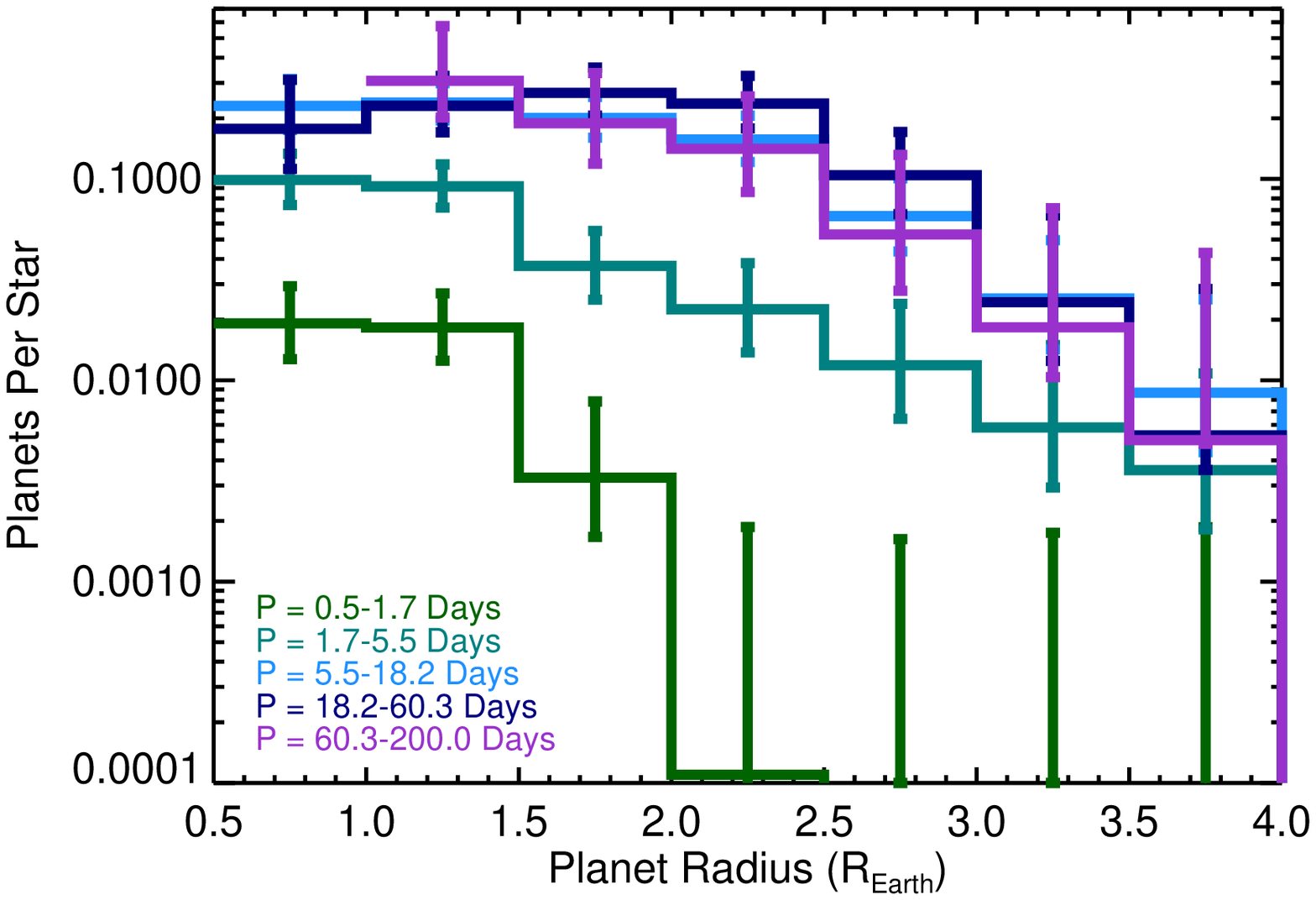}
\includegraphics[width=0.5\textwidth]{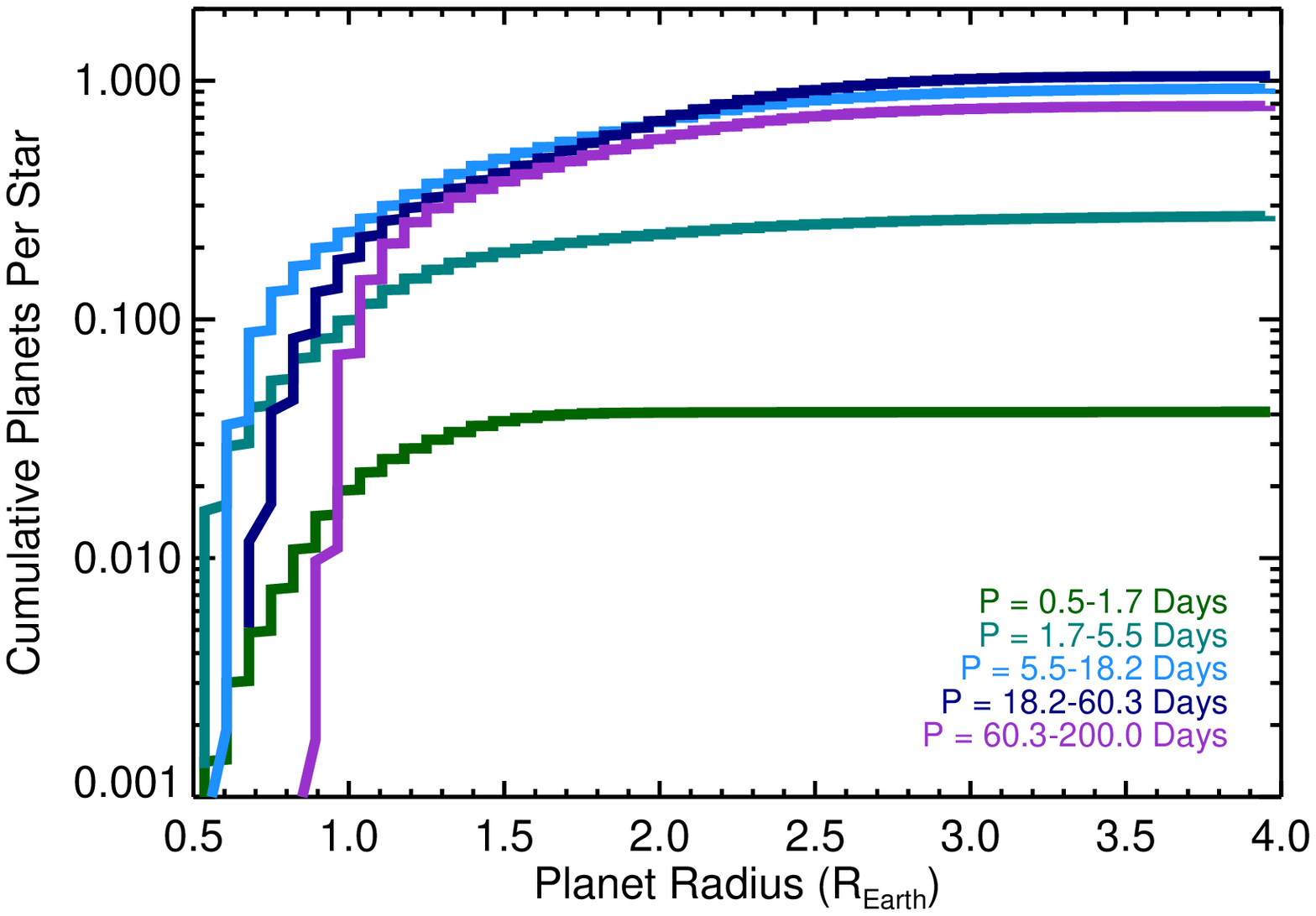}
\end{center}
\caption{Planet occurrence (top) and cumulative planet \mbox{occurrence} (bottom) versus planet radius for planets with periods of $0.5-1.7$~days (dark green), $1.7-5.5$~days (teal), $5.5-18.2$~days (light blue), $18.2-60.3$~days (navy), and $60.3-200$~days (purple). The error bars are based on binomial statistics and the assumed smoothing of the planet population. In this figure and \mbox{Figures~\ref{fig:perdep}--\ref{fig:fluxdep}} we do not present occurrence rates for regions with pipeline sensitivity below $15\%$. }
\label{fig:rpdep}
\end{figure}

\begin{figure}[tbp] 
\begin{center}
\centering
\includegraphics[width=0.5\textwidth]{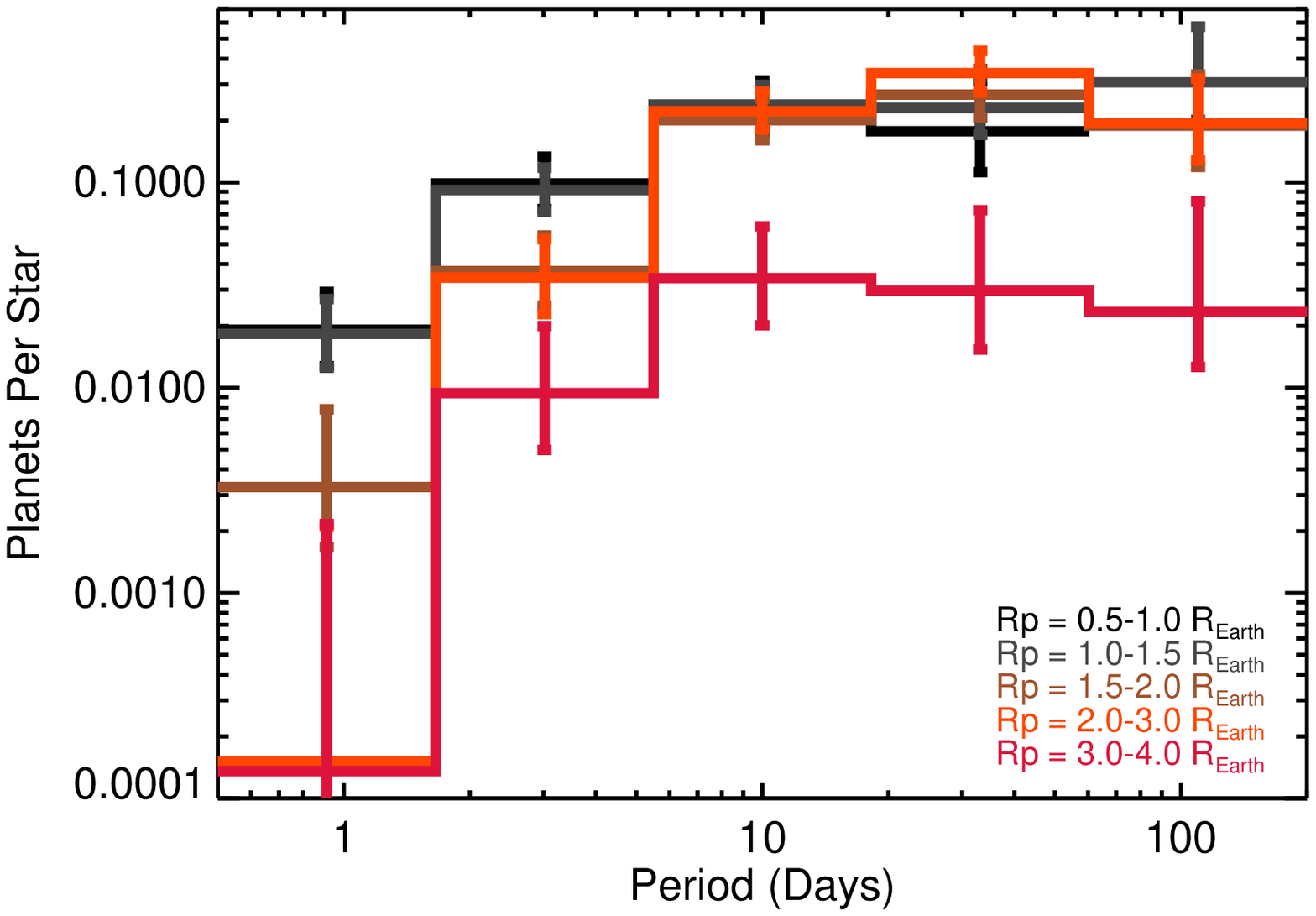}
\includegraphics[width=0.5\textwidth]{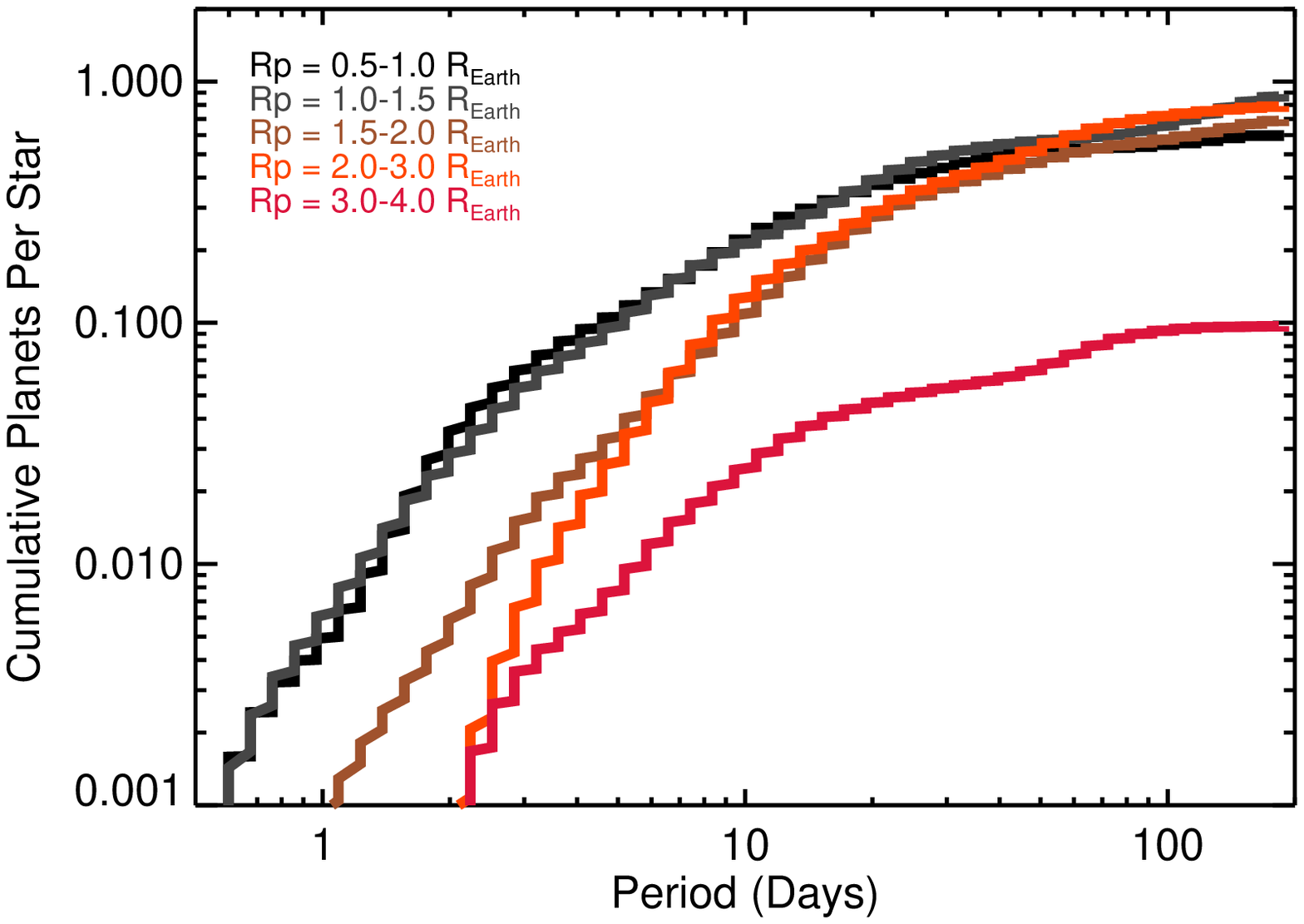}
\end{center}
\caption{Planet occurrence (top) and cumulative planet \mbox{occurrence} (bottom) versus orbital period for planets with radii of $0.5-1\rearth$ (black), $1-1.5\rearth$ (dark gray), $1.5-2.0\rearth$ (brown), $2-3\rearth$ (orange), and $3-4\rearth$ (red).}\label{fig:perdep}
\end{figure}

Figure~\ref{fig:rpdep}, Figure~\ref{fig:perdep}, and Table~\ref{tab:perocc} display the planet occurrence rate as a function of planet radius and orbital period. As in previous studies \citep{dressing+charbonneau2013, morton+swift2014}, we found that planets with radii $> 3\rearth$ are rare around small stars (at least out to orbital periods of 200~days). Concentrating on planets with periods shorter than 50~days, we observed a general trend of decreasing planet occurrence with increasing planet radius between $1\rearth$ and $4\rearth$. There is an indication in Figure~\ref{fig:perdep} that the planet occurrence rate may become flat in $\log_{\rm 10} P$ for periods longer than 10~days. However, the errors on the longest orbital bins are large enough that we cannot distinguish between a brief flattening between $10-100$~days and a plateau extending out to much longer orbital periods.

For orbital periods shorter than 50~days, we measure an occurrence rate of \occearthfifty Earth-size ($1-1.5\rearth$) planets and \occsefifty super-Earths ($1.5-2\rearth$) per small star. Extending the period range to 100~days, we estimate \occearthhundred Earths and \occsehundred super-Earths per star. Overall, we find \totalsmall planets per M~dwarf with radii of $1-4\rearth$ and periods shorter than 200~days. We provide cumulative planet occurrence rates for several additional choices of period and radius boundaries in Table~\ref{tab:cumulativep}.

\begin{deluxetable*}{crrrrr}
\tablecolumns{6}
\tabletypesize{\footnotesize}
\tablecaption{Number of Planets Per Star Versus Orbital Period (In Percentage)}
\tablehead{
\colhead{$R_p (\rearth)$} & 
\colhead{$0.5-1.7$ Days} & 
\colhead{$1.7-5.5$ Days} & 
\colhead{$5.5-18.2$ Days} &
\colhead{$18.2-60.3$ Days} &
\colhead{$60.3-200$ Days}
}
\startdata
$0.5-1.0$ & $1.92^{+1.01}_{-0.64} (70\%)$ & $9.88^{+3.45}_{-2.46} (54\%)$ & $23.06^{+8.29}_{-5.57} (33\%)$ & $17.75^{+13.34}_{-6.54} (17\%)$ & $(6\%)$ \\
$1.0-1.5$ & $1.83^{+0.87}_{-0.58} (87\%)$ & $9.18^{+2.62}_{-1.98} (84\%)$ & $<26.54 (79\%)\tablenotemark{a}$ & $23.08^{+9.38}_{-6.02} (64\%)$ & $30.70^{+26.67}_{-10.54} (41\%)$ \\
$1.5-2.0$ & $0.33^{+0.46}_{-0.16} (90\%)$ & $3.70^{+1.82}_{-1.18} (88\%)$ & $20.06^{+5.45}_{-4.04} (87\%)$ & $26.73^{+8.99}_{-6.08} (84\%)$ & $18.90^{+14.67}_{-6.99} (74\%)$ \\
$2.0-2.5$ & $<0.13 (90\%)$ & $2.25^{+1.57}_{-0.88} (90\%)$ & $15.69^{+4.91}_{-3.55} (89\%)$ & $23.65^{+8.81}_{-5.83} (88\%)$ & $14.12^{+11.55}_{-5.51} (84\%)$ \\
$2.5-3.0$ & $<0.11 (91\%)$ & $1.19^{+1.21}_{-0.54} (90\%)$ & $6.54^{+3.51}_{-2.18} (90\%)$ & $10.42^{+6.68}_{-3.74} (89\%)$ & $5.30^{+7.85}_{-2.52} (88\%)$ \\
$3.0-3.5$ & $<0.11 (91\%)$ & $0.58^{+0.91}_{-0.29} (91\%)$ & $2.55^{+2.41}_{-1.13} (91\%)$ & $2.44^{+4.14}_{-1.19} (90\%)$ & $1.83^{+5.32}_{-0.80} (89\%)$ \\
$3.5-4.0$ & $<0.11 (91\%)$ & $0.36^{+0.72}_{-0.18} (91\%)$ & $0.87^{+1.65}_{-0.43} (91\%)$ & $0.53^{+2.31}_{-0.17} (90\%)$ & $<2.71 (90\%)$ \\
\hline
\hline
$0.5-1.0$ & $1.38^{+0.93}_{-0.53} (66\%)$ & $8.42^{+3.53}_{-2.39} (44\%)$ & $20.59^{+8.70}_{-5.57} (26\%)$ & $(11\%)$ & $(3\%)$ \\
$1.0-1.5$ & $1.95^{+0.93}_{-0.61} (86\%)$ & $9.94^{+2.82}_{-2.13} (83\%)$ & $<26.63 (75\%)$ & $26.85^{+10.70}_{-6.79} (58\%)$ & $28.85^{+28.66}_{-10.34} (32\%)$ \\
$1.5-2.0$ & $0.41^{+0.51}_{-0.20} (89\%)$ & $4.15^{+1.94}_{-1.28} (88\%)$ & $<23.58 (86\%)$ & $24.59^{+9.08}_{-6.00} (82\%)$ & $19.98^{+16.07}_{-7.42} (67\%)$ \\
$2.0-2.5$ & $<0.13 (90\%)$ & $2.72^{+1.73}_{-1.01} (89\%)$ & $18.73^{+5.39}_{-3.95} (89\%)$ & $27.58^{+9.42}_{-6.31} (87\%)$ & $18.08^{+13.21}_{-6.57} (82\%)$ \\
$2.5-3.0$ & $<0.11 (90\%)$ & $1.59^{+1.39}_{-0.69} (90\%)$ & $8.29^{+3.96}_{-2.55} (90\%)$ & $14.51^{+7.71}_{-4.62} (89\%)$ & $8.61^{+9.78}_{-3.84} (87\%)$ \\
$3.0-3.5$ & $<0.12 (91\%)$ & $0.65^{+1.00}_{-0.32} (91\%)$ & $3.25^{+2.72}_{-1.37} (90\%)$ & $3.37^{+4.62}_{-1.62} (90\%)$ & $1.97^{+5.87}_{-0.85} (89\%)$ \\
$3.5-4.0$ & $<0.11 (91\%)$ & $0.38^{+0.77}_{-0.19} (91\%)$ & $1.05^{+1.82}_{-0.52} (91\%)$ & $0.56^{+2.32}_{-0.19} (90\%)$ & $<2.34 (89\%)$
\enddata
\tablecomments{In this table and all subsequent occurrence rate tables, the numbers in parentheses are the fraction of injected planets that were recovered within the given intervals. In addition, the first set of entries are our estimates of the planet occurrence rate when using the stellar properties in the \citet{huber_et_al2014} catalog. The second set of entries (below the double line) are alternative estimates constructed by revising the stellar radii to lie along an empirical temperature/radius relation from \citet[][see Section \ref{ssec:starbias} for details]{mann_et_al2013b}.}
\tablenotetext{a}{We provide one-sigma upper limits instead of two-sided errors for grid cells with large Poisson errors and very low occurrence rates.}
\label{tab:perocc}
\end{deluxetable*}

 \begin{deluxetable*}{cccccc}
\tablecolumns{6}
\tabletypesize{\footnotesize}
\tablecaption{Cumulative Number of Planets Per Star Versus Orbital Period (In Percentage)}
\tablehead{
\colhead{$R_p (\rearth)$} &
\colhead{$0.5-10$ Days} & 
\colhead{$0.5-50$ Days} & 
\colhead{$0.5-100$ Days} & 
\colhead{$0.5-150$ Days} & 
\colhead{$0.5-200$ Days}
}
\startdata
$0.5-1.0$ & $22.08^{+4.88}_{-3.79} (54\%)$ & $51.43^{+7.99}_{-6.02} (44\%)$ & $54.73^{+8.21}_{-6.13} (40\%)$ & $59.71^{+8.46}_{-6.21} (37\%)$ & $59.71^{+8.44}_{-6.21} (36\%)$ \\[1.2ex]
$1.0-1.5$ & $21.18^{+3.84}_{-3.11} (84\%)$ & $56.25^{+6.31}_{-5.01} (79\%)$ & $65.09^{+6.52}_{-5.05} (75\%)$ & $82.17^{+6.07}_{-4.24} (72\%)$ & $88.74^{+5.38}_{-3.42} (70\%)$ \\[1.2ex]
$1.5-2.0$ & $10.76^{+3.23}_{-2.40} (89\%)$ & $45.95^{+7.05}_{-5.47} (87\%)$ & $57.35^{+7.59}_{-5.76} (86\%)$ & $66.58^{+7.78}_{-5.72} (85\%)$ & $69.72^{+7.77}_{-5.63} (84\%)$ \\[1.2ex]
$2.0-2.5$ & $8.68^{+3.21}_{-2.26} (90\%)$ & $35.62^{+7.02}_{-5.35} (89\%)$ & $49.81^{+8.18}_{-6.13} (89\%)$ & $54.59^{+8.46}_{-6.26} (88\%)$ & $55.72^{+8.52}_{-6.29} (88\%)$ \\[1.2ex]
$2.5-3.0$ & $3.89^{+2.30}_{-1.38} (90\%)$ & $15.46^{+5.41}_{-3.79} (90\%)$ & $21.97^{+6.87}_{-4.86} (90\%)$ & $23.19^{+7.14}_{-5.05} (90\%)$ & $23.45^{+7.20}_{-5.09} (90\%)$ \\[1.2ex]
$3.0-3.5$ & $1.73^{+1.64}_{-0.77} (91\%)$ & $4.65^{+3.10}_{-1.75} (90\%)$ & $7.15^{+4.32}_{-2.53} (90\%)$ & $7.38^{+4.44}_{-2.60} (90\%)$ & $7.41^{+4.45}_{-2.61} (90\%)$ \\[1.2ex]
$3.5-4.0$ & $0.72^{+1.08}_{-0.36} (91\%)$ & $1.63^{+1.86}_{-0.77} (91\%)$ & $2.11^{+2.31}_{-0.98} (91\%)$ & $2.27^{+2.45}_{-1.05} (90\%)$ & $2.27^{+2.46}_{-1.05} (90\%)$ \\[1.2ex]
\hline
\hline
$0.5-1.0$ & $18.39^{+5.02}_{-3.74} (49\%)$ & $39.93^{+8.22}_{-6.08} (38\%)$ & $40.60^{+8.31}_{-6.14} (34\%)$ & $40.60^{+8.28}_{-6.12} (31\%)$ & $40.60^{+8.27}_{-6.12} (30\%)$ \\[1.2ex]
$1.0-1.5$ & $22.85^{+4.08}_{-3.31} (82\%)$ & $60.76^{+6.57}_{-5.13} (75\%)$ & $68.38^{+6.65}_{-5.06} (71\%)$ & $85.73^{+5.85}_{-3.89} (68\%)$ & $91.47 \pm 5.04 (66\%)$ \\[1.2ex]
$1.5-2.0$ & $11.96^{+3.42}_{-2.57} (88\%)$ & $45.87^{+7.07}_{-5.48} (86\%)$ & $57.26^{+7.65}_{-5.80} (85\%)$ & $66.74^{+7.86}_{-5.76} (83\%)$ & $70.23^{+7.85}_{-5.66} (82\%)$ \\[1.2ex]
$2.0-2.5$ & $9.98^{+3.46}_{-2.48} (90\%)$ & $42.48^{+7.42}_{-5.67} (89\%)$ & $58.49^{+8.27}_{-6.13} (88\%)$ & $65.45^{+8.39}_{-6.07} (88\%)$ & $67.12^{+8.40}_{-6.03} (87\%)$ \\[1.2ex]
$2.5-3.0$ & $4.94^{+2.58}_{-1.63} (90\%)$ & $20.60^{+6.17}_{-4.44} (90\%)$ & $29.85^{+7.79}_{-5.63} (90\%)$ & $32.40^{+8.21}_{-5.92} (89\%)$ & $33.01^{+8.31}_{-5.99} (89\%)$ \\[1.2ex]
$3.0-3.5$ & $2.09^{+1.84}_{-0.90} (91\%)$ & $6.32^{+3.73}_{-2.22} (90\%)$ & $8.67^{+4.74}_{-2.88} (90\%)$ & $9.13^{+4.94}_{-3.02} (90\%)$ & $9.25^{+4.99}_{-3.05} (90\%)$ \\[1.2ex]
$3.5-4.0$ & $0.83^{+1.19}_{-0.41} (91\%)$ & $1.92^{+2.08}_{-0.89} (91\%)$ & $2.14^{+2.27}_{-0.98} (90\%)$ & $2.20^{+2.33}_{-1.01} (90\%)$ & $2.21^{+2.34}_{-1.02} (90\%)$
\enddata
\tablecomments{As in Table~\ref{tab:perocc}, the entries below the double horizontal line are the estimates based on the revised stellar radii (see Section~\ref{ssec:starbias}).}
\label{tab:cumulativep}
\end{deluxetable*}

\begin{deluxetable*}{crrrrr}
\tablecolumns{6}
\tabletypesize{\footnotesize}
\tablecaption{Number of Planets Per Star Versus Insolation (In Percentage)}
\tablehead{
\colhead{$R_p (\rearth)$} & 
\colhead{$0.2-1.1 \fearth$} & 
\colhead{$1.1-6.3 \fearth$} & 
\colhead{$6.3-35.6 \fearth$} & 
\colhead{$35.6-200\fearth$}
}
\startdata
$0.5-1.0$ & $(11\%)$ & $22.63^{+11.24}_{-6.65} (24\%)$ & $23.48^{+6.45}_{-4.71} (44\%)$ & $3.52^{+1.67}_{-1.10} (60\%)$ \\
$1.0-1.5$ & $24.02^{+19.77}_{-8.69} (47\%)$ & $<35.71 (69\%)$ & $13.45^{+3.68}_{-2.78} (81\%)$ & $5.36^{+1.65}_{-1.24} (85\%)$ \\
$1.5-2.0$ & $17.70^{+11.70}_{-6.18} (75\%)$ & $<31.17 (85\%)$ & $10.59^{+3.41}_{-2.49} (88\%)$ & $1.38^{+0.93}_{-0.53} (89\%)$ \\
$2.0-2.5$ & $13.97^{+9.54}_{-5.07} (85\%)$ & $20.16^{+6.80}_{-4.73} (88\%)$ & $9.84^{+3.34}_{-2.40} (89\%)$ & $0.75^{+0.81}_{-0.35} (90\%)$ \\
$2.5-3.0$ & $5.47^{+6.67}_{-2.53} (88\%)$ & $8.90^{+5.04}_{-3.02} (90\%)$ & $3.94^{+2.28}_{-1.38} (90\%)$ & $0.42^{+0.62}_{-0.21} (90\%)$ \\
$3.0-3.5$ & $1.71^{+4.48}_{-0.78} (89\%)$ & $2.18^{+3.15}_{-1.06} (90\%)$ & $1.71^{+1.65}_{-0.77} (91\%)$ & $0.27^{+0.53}_{-0.13} (91\%)$ \\
$3.5-4.0$ & $<2.33 (90\%)$ & $0.80^{+2.28}_{-0.36} (91\%)$ & $0.57^{+1.14}_{-0.28} (91\%)$ & $0.19^{+0.44}_{-0.09} (91\%)$ \\
\hline
\hline
$0.2-1.1$ & 1.1-6.3 & 6.3-35.6 & 35.6-200.0 \\
$0.5-1.0$ & $(6\%)$ & $10.46^{+9.36}_{-4.32} (16\%)$ & $17.22^{+6.10}_{-4.22} (32\%)$ & $3.23^{+1.75}_{-1.09} (52\%)$ \\
$1.0-1.5$ & $26.87^{+23.57}_{-9.63} (37\%)$ & $29.27^{+9.12}_{-6.26} (63\%)$ & $13.19^{+3.79}_{-2.83} (78\%)$ & $5.84^{+1.75}_{-1.32} (84\%)$ \\
$1.5-2.0 $& $18.99^{+13.05}_{-6.69} (71\%)$ & $24.54^{+7.34}_{-5.21} (83\%)$ & $11.35^{+3.63}_{-2.64} (87\%)$ & $1.79^{+1.07}_{-0.64} (88\%)$ \\
$2.0-2.5$ & $15.29^{+10.41}_{-5.49} (83\%)$ & $22.53^{+7.21}_{-5.05} (88\%)$ & $10.93^{+3.59}_{-2.60} (89\%)$ & $1.05^{+0.94}_{-0.46} (90\%)$ \\
$2.5-3.0$ & $7.68^{+7.80}_{-3.36} (87\%)$ & $11.83^{+5.75}_{-3.63} (89\%)$ & $4.40^{+2.46}_{-1.51} (90\%)$ & $0.47^{+0.65}_{-0.23} (90\%)$ \\
$3.0-3.5$ & $1.90^{+4.89}_{-0.87} (89\%)$ & $2.70^{+3.34}_{-1.28} (90\%)$ & $2.11^{+1.83}_{-0.91} (90\%)$ & $0.20^{+0.47}_{-0.10} (91\%)$ \\
$3.5-4.0$ & $<1.81 (90\%)$ & $0.76^{+2.22}_{-0.33} (90\%)$ & $0.75^{+1.27}_{-0.37} (91\%)$ & $0.19^{+0.45}_{-0.09} (91\%)$
\enddata
\tablecomments{As in Table~\ref{tab:perocc}, the entries below the double horizontal line are the estimates based on the revised stellar radii (see Section~\ref{ssec:starbias}).}
\label{tab:fluxocc}
\end{deluxetable*}

\begin{deluxetable*}{crrrrrr}
\tablecolumns{7}
\tabletypesize{\footnotesize}
\tablecaption{Cumulative Number of Planets Per Star Versus Insolation (In Percentage)}
\tablehead{
\colhead{$R_p (\rearth)$} & 
\colhead{$0.2-200\fearth$} & 
\colhead{$1.0-200\fearth$} & 
\colhead{$10.0-200\fearth$} & 
\colhead{$50.0-200\fearth$} & 
\colhead{$100.0-200\fearth$} & 
\colhead{$150.0-200\fearth$}
}
\startdata
$0.5-1.0$ & $63.36^{+8.82}_{-6.34} (34\%)$ & $49.63^{+8.11}_{-6.09} (41\%)$ & $17.79^{+4.57}_{-3.46} (51\%)$ & $2.09^{+1.20}_{-0.74} (58\%)$ & $0.61^{+0.61}_{-0.28} (61\%)$ & $0.06^{+0.25}_{-0.02} (62\%)$ \\[1.2ex]
$1.0-1.5$ & $74.67^{+6.59}_{-4.85} (70\%)$ & $50.65^{+6.18}_{-4.94} (77\%)$ & $15.00^{+3.06}_{-2.46} (83\%)$ & $3.50^{+1.30}_{-0.93} (85\%)$ & $1.00^{+0.67}_{-0.39} (85\%)$ & $0.16^{+0.31}_{-0.08} (86\%)$ \\[1.2ex]
$1.5-2.0$ & $57.61^{+7.57}_{-5.75} (84\%)$ & $39.92^{+6.49}_{-5.09} (87\%)$ & $6.60^{+2.31}_{-1.66} (88\%)$ & $0.90^{+0.74}_{-0.38} (89\%)$ & $0.23^{+0.38}_{-0.11} (89\%)$ & $0.04^{+0.20}_{-0.01} (89\%)$ \\[1.2ex]
$2.0-2.5$ & $44.72^{+7.55}_{-5.76} (88\%)$ & $30.74^{+6.17}_{-4.77} (89\%)$ & $6.41^{+2.50}_{-1.74} (90\%)$ & $0.33^{+0.56}_{-0.17} (90\%)$ & $0.02^{+0.19}_{-0.00} (90\%)$ & $0.00^{+0.09}_{-0.00} (90\%)$ \\[1.2ex]
$2.5-3.0$ & $18.72^{+5.92}_{-4.22} (90\%)$ & $13.26^{+4.62}_{-3.26} (90\%)$ & $2.85^{+1.74}_{-1.03} (90\%)$ & $0.23^{+0.46}_{-0.11} (90\%)$ & $0.04^{+0.22}_{-0.01} (90\%)$ & $<0.00 (90\%)$ \\[1.2ex]
$3.0-3.5$ & $5.87^{+3.57}_{-2.10} (90\%)$ & $4.16^{+2.72}_{-1.56} (90\%)$ & $1.28^{+1.23}_{-0.57} (91\%)$ & $0.19^{+0.45}_{-0.09} (91\%)$ & $0.02^{+0.19}_{-0.00} (91\%)$ & $<0.00 (91\%)$ \\[1.2ex]
$3.5-4.0$ & $2.03^{+2.21}_{-0.94} (90\%)$ & $1.56^{+1.78}_{-0.73} (91\%)$ & $0.49^{+0.79}_{-0.24} (91\%)$ & $0.18^{+0.43}_{-0.09} (91\%)$ & $<0.00 (91\%)$ & $<0.00 (91\%)$ \\[1.2ex]
\hline
\hline
$0.5-1.0$ & $34.08^{+7.85}_{-5.78} (28\%)$ & $30.91^{+7.35}_{-5.43} (34\%)$ & $15.13^{+4.79}_{-3.46} (44\%)$ & $2.03^{+1.32}_{-0.77} (52\%)$ & $0.54^{+0.61}_{-0.25} (55\%)$ & $0.05^{+0.23}_{-0.01} (57\%)$ \\[1.2ex]
$1.0-1.5$ & $75.17^{+6.91}_{-5.00} (64\%)$ & $48.30^{+6.31}_{-5.02} (73\%)$ & $15.61^{+3.18}_{-2.55} (81\%)$ & $3.95^{+1.39}_{-1.01} (83\%)$ & $1.30^{+0.77}_{-0.47} (84\%)$ & $0.22^{+0.36}_{-0.11} (84\%)$ \\[1.2ex]
$1.5-2.0$ & $56.66^{+7.72}_{-5.84} (81\%)$ & $37.67^{+6.39}_{-5.00} (85\%)$ & $7.26^{+2.43}_{-1.77} (88\%)$ & $1.14^{+0.84}_{-0.46} (88\%)$ & $0.28^{+0.43}_{-0.14} (88\%)$ & $0.04^{+0.21}_{-0.01} (88\%)$ \\[1.2ex]
$2.0-2.5$ & $49.80^{+7.75}_{-5.89} (87\%)$ & $34.51^{+6.47}_{-5.01} (89\%)$ & $7.31^{+2.68}_{-1.90} (89\%)$ & $0.54^{+0.69}_{-0.26} (90\%)$ & $0.02^{+0.21}_{-0.00} (90\%)$ & $<0.00 (90\%)$ \\[1.2ex]
$2.5-3.0$ & $24.38^{+6.81}_{-4.94} (89\%)$ & $16.71^{+5.27}_{-3.79} (90\%)$ & $3.17^{+1.86}_{-1.13} (90\%)$ & $0.35^{+0.56}_{-0.17} (90\%)$ & $0.05^{+0.25}_{-0.01} (90\%)$ & $<0.00 (90\%)$ \\[1.2ex]
$3.0-3.5$ & $6.91^{+3.91}_{-2.36} (90\%)$ & $5.01^{+3.03}_{-1.79} (90\%)$ & $1.51^{+1.38}_{-0.67} (91\%)$ & $0.17^{+0.43}_{-0.08} (91\%)$ & $0.03^{+0.22}_{-0.00} (91\%)$ & $<0.00 (91\%)$ \\[1.2ex]
$3.5-4.0$ & $1.93^{+2.03}_{-0.89} (90\%)$ & $1.69^{+1.82}_{-0.78} (91\%)$ & $0.62^{+0.92}_{-0.31} (91\%)$ & $0.18^{+0.44}_{-0.09} (91\%)$ & $<0.01 (91\%)$ & $0.00^{+0.09}_{-0.00} (91\%)$ 
\enddata
\tablecomments{As in Table~\ref{tab:perocc}, the entries below the double horizontal line are the estimates based on the revised stellar radii (see Section~\ref{ssec:starbias}).}
\label{tab:cumulativef}
\end{deluxetable*}

\subsection{Dependence on Planet Radius \& Insolation}
\label{ssec:occrpflux}
In Figure~\ref{fig:fluxdep} and Tables~\ref{tab:fluxocc}-\ref{tab:cumulativef}, we present the planet occurrence rate as a function of stellar insolation and planet radius. In general, we find that planet occurrence increases both with decreasing planet radius (as discussed in Section~\ref{ssec:occrpper}) and with decreasing $\log_{10}F_P$. Intriguingly, we observed that the occurrence rates of Earths, Super-Earths, and mini-Neptunes are comparable for insolations below roughly $30\fearth$ but that planets larger than $1.5\rearth$ were less common than smaller planets at insolations above $30\fearth$. The error bars on the coolest insolation bin are rather large, but the divergence of the Earth and Neptune occurrence relations might be due to photo-evaporation at short orbital periods.

Across the size range we considered, the planet occurrence rate versus $\log_{\rm 10}F_P$ rises with decreasing insolation between $100 - 10\fearth$ and appears roughly flat in $\log_{\rm 10}F_P$ between $10 -0.2 \fearth$. Figure~\ref{fig:fluxdep} hints that the planet occurrence rate might increase again at cooler insolations, but a larger sample of long period planets will be required to test that hypothesis. 

\subsection{The Occurrence of Potentially Habitable Planets}
\label{ssec:hzocc}
\begin{figure}[btp] 
\begin{center}
\centering
\includegraphics[width=0.5\textwidth]{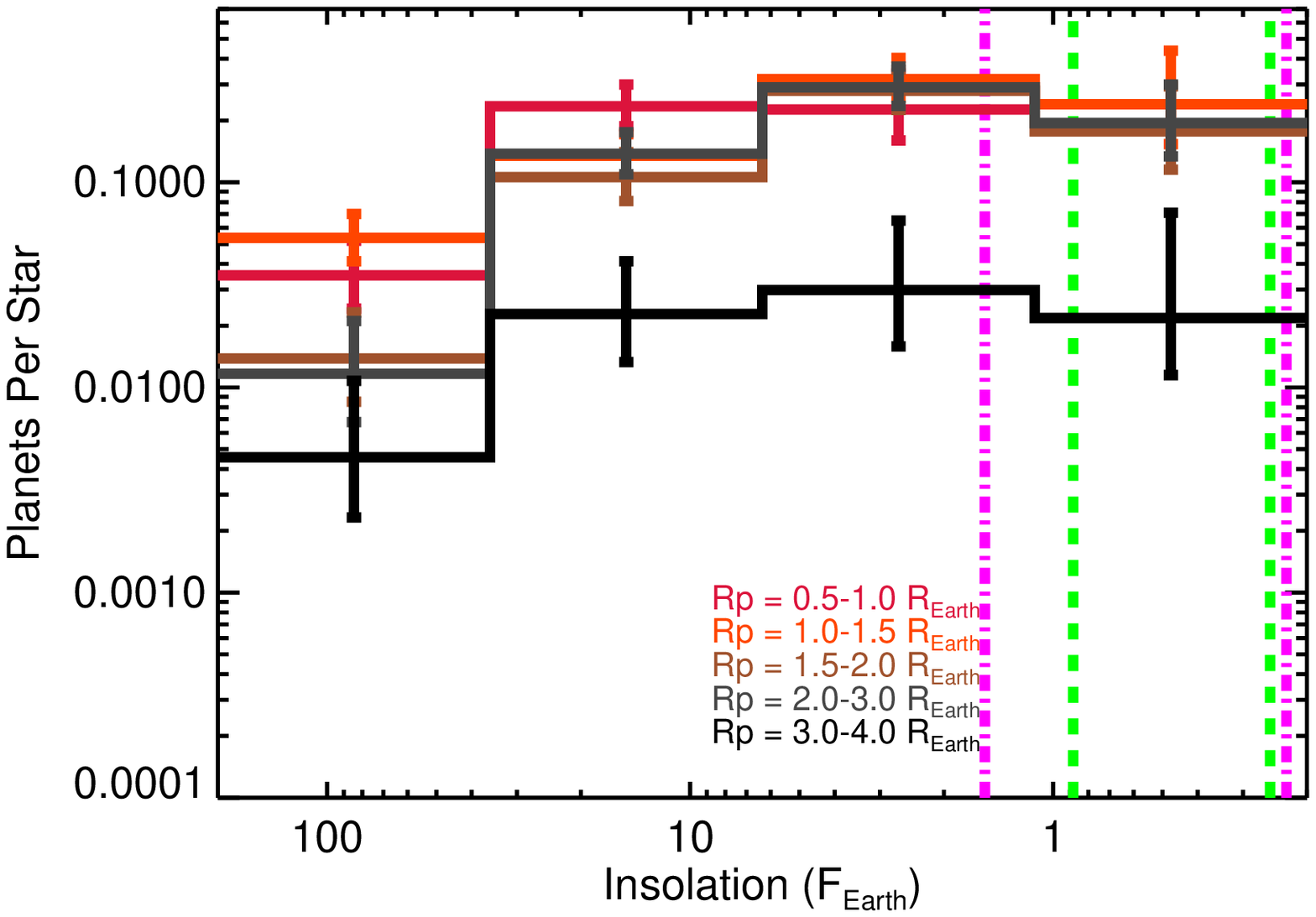}
\includegraphics[width=0.5\textwidth]{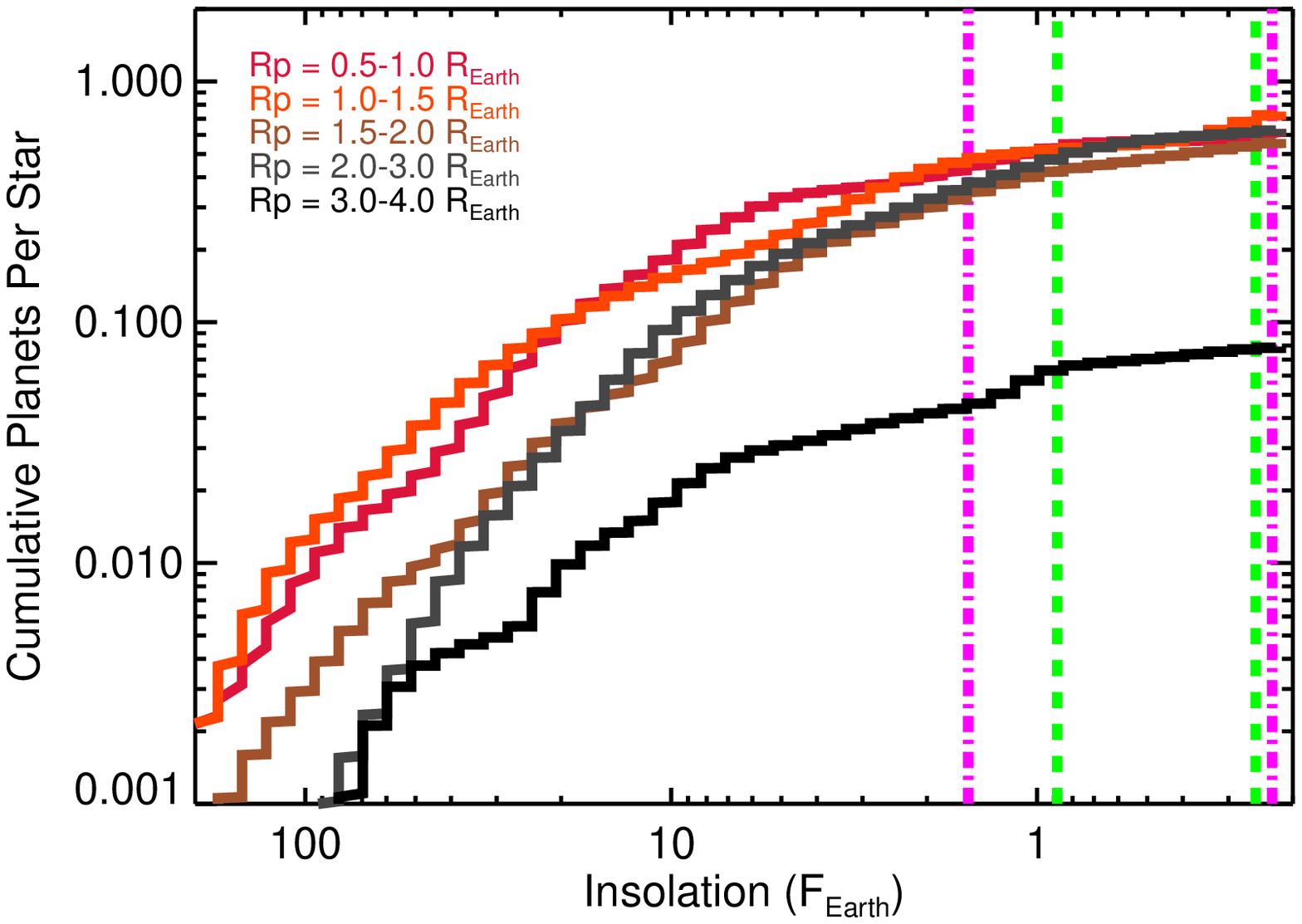}
\end{center}
\caption{Planet occurrence (top) and cumulative planet occurrence (bottom) versus insolation for planets with radii of $0.5-1\rearth$ (crimson), $1-1.5\rearth$ (orange), $1.5-2.0\rearth$ (brown), $2-3\rearth$ (dark gray), and $3-4\rearth$ (black).  As in Figure~\ref{fig:kidsearchmap}, the vertical lines mark two definitions of the habitable zone.}
\label{fig:fluxdep}
\end{figure}

The shaded regions in Figure~\ref{fig:fluxdep} display one choice of habitable zone boundaries, but the definition of the ``habitable zone'' (HZ) is still rather uncertain. Traditionally, astronomers have used the term to refer to the distance from the star at which liquid water could be present on the surface of a planet \citep{dole1964, hart1979, kasting_et_al1993}. In theory, there could also be water-based life on worlds with subsurface oceans or non-water-based life on worlds like Titan, but any associated biosignatures would be difficult to interpret remotely.  Accordingly, astronomers have concentrated thus far on the search for life as we know it, meaning surface-based lifeforms that depend on liquid water and might generate biosignatures that could alter the composition of their homeworld's atmosphere. 

Even within that rather narrow definition, there are many assumptions that can affect the choice of habitable zone boundaries. In particular, the assumed mass and composition of the planet's atmosphere affects the surface pressure and therefore the temperature range at which water would be liquid \citep[e.g.,][]{vladilo_et_al2013}. For instance, \citet{pierrehumbert+gaidos2011} showed that planets with thick hydrogen atmospheres would have sufficient surface pressure to retain surface liquid water out to distances of 2.4~AU. It is uncertain whether biosignatures would be detectable in such atmospheres \citep{seager_et_al2013, hu_et_al2013}, but those worlds could still be habitable. 

The presence of clouds adds an additional complication by both cooling and heating the planet. Clouds are particularly important in the case of tidally-locked planets, which might be a common fate for planets orbiting within the habitable zones of M~dwarfs. \citet{yang_et_al2013} demonstrated that a tidally-locked M~dwarf planet might develop a persistent cloud patch above the sub-stellar point. That cloud patch would have a higher albedo than the planetary surface and would allow the planet to be much cooler at a given separation than a cloud-free model would predict. As a result, the habitable zone for a cloudy, tidally-locked planet could extend to insolations as high as $F_P = 1.76\fearth$ for the moist greenhouse limit rather than the  limit of $F_P < 0.88\fearth$ calculated by  \citet{kopparapu_et_al2013b} for a cloud-free model. Even for non-tidally-locked planets, the presence of clouds can expand the distances corresponding to the boundaries of the habitable zone by roughly $40\%$ depending on the degree of cloud cover \citep{selsis_et_al2007}.

The orbital geometry of exoplanets is also important when assessing planetary habitability. For instance, planets with high obliquities or eccentric orbits might be partially habitable at certain latitudes or during certain times of year \citep{williams+kasting1997, williams+pollard2002, spiegel_et_al2008, spiegel_et_al2009, dressing_et_al2010, cowan_et_al2012, dobrovolskis2013, armstrong_et_al2014, linsenmeier_et_al2015}. Depending on the timescale for the temperature of the planet to change (which depends on factors such as the fraction of surface covered by ocean), such planets may undergo periodic global glaciations punctuated by short-lived epochs during which the surface is warm enough for liquid water \citep{pierrehumbert2005, spiegel_et_al2010}. In addition, the primordial obliquities of close-in planets orbiting M~dwarfs may be significantly eroded by tides \citep{heller_et_al2011a}.

Constructing a multi-dimensional habitable zone model for planets orbiting M~dwarfs is beyond the scope of this paper, but we aspire to provide enough information so that other researchers can assess the abundance of planets within their chosen habitable zone boundaries. We therefore provide occurrence rates for a few possible choices of habitable zone boundaries in Table~\ref{tab:hzocc}. 

\begin{deluxetable*}{crrrrrr}[tbp]
\tablecolumns{7}
\tabletypesize{\scriptsize}
\tablecaption{Habitable Zone Occurrence Rates (In Percentage)}
\tablehead{
\colhead{$F_P (\fearth)$} & 
\colhead{$ 0.25-0.88 $} & 
\colhead{$ 0.23-1.54 $} & 
\colhead{$ 0.25-4.00 $} & 
\colhead{$ 0.25-1.76 $} & 
\colhead{$ 0.25-2.78 $} & 
\colhead{$ 0.25-5.85 $} \\
\colhead{\emph{Outer HZ: }} & 
\colhead{Max GH\tablenotemark{a}} & 
\colhead{Early Mars\tablenotemark{a}} & 
\colhead{Fixed\tablenotemark{b}} & 
\colhead{Max GH\tablenotemark{a}} & 
\colhead{Max GH\tablenotemark{a}} & 
\colhead{Max GH\tablenotemark{a}} \\
\colhead{\emph{Inner HZ: }} & 
\colhead{Moist GH\tablenotemark{a}} & 
\colhead{Recent Venus\tablenotemark{a}} & 
\colhead{Fixed\tablenotemark{b}} & 
\colhead{Cloudy Moist GH\tablenotemark{c}} & 
\colhead{Desert (a=0.2)\tablenotemark{d}} & 
\colhead{Desert (a=0.8)\tablenotemark{d}}
}
\startdata
$0.5-1.0\rearth$ & $ (13\%)$ & $ (14\%)$ & $25.28^{+14.90}_{-7.96} (18\%)$ & $15.75^{+15.34}_{-6.41} (15\%)$ & $20.61^{+15.15}_{-7.37} (17\%)$ & $27.56^{+13.70}_{-7.86} (20\%)$ \\[1.2ex]
$0.8-1.0\rearth$ & $4.63^{+15.19}_{-1.78} (21\%)$ & $13.09^{+16.88}_{-5.73} (22\%)$ & $20.95^{+15.07}_{-7.41} (28\%)$ & $13.30^{+15.31}_{-5.72} (24\%)$ & $17.23^{+14.71}_{-6.66} (26\%)$ & $20.05^{+13.68}_{-6.99} (30\%)$ \\[1.2ex]
$1.0-1.5\rearth$ & $15.82^{+16.60}_{-6.54} (48\%)$ & $24.28^{+17.58}_{-8.39} (50\%)$ & $46.77^{+12.33}_{-8.12} (56\%)$ & $21.82^{+15.03}_{-7.54} (52\%)$ & $31.65^{+13.20}_{-8.00} (55\%)$ & $46.90^{+11.00}_{-7.54} (59\%)$ \\[1.2ex]
$1.5-2.0\rearth$ & $11.54^{+9.97}_{-4.67} (75\%)$ & $20.69^{+10.80}_{-6.32} (76\%)$ & $36.07^{+10.00}_{-6.90} (79\%)$ & $21.48^{+10.25}_{-6.22} (77\%)$ & $28.04^{+10.10}_{-6.62} (78\%)$ & $39.35^{+9.32}_{-6.64} (80\%)$ \\[1.2ex]
$2.0-2.5\rearth$ & $10.25^{+8.58}_{-4.14} (85\%)$ & $17.09^{+9.48}_{-5.50} (85\%)$ & $29.17^{+9.30}_{-6.33} (86\%)$ & $18.23^{+9.27}_{-5.56} (86\%)$ & $23.99^{+9.31}_{-6.06} (86\%)$ & $30.97^{+8.93}_{-6.23} (87\%)$ \\[1.2ex]
$2.5-4.0\rearth$ & $5.03^{+6.66}_{-2.37} (89\%)$ & $10.58^{+7.73}_{-4.04} (89\%)$ & $16.61^{+7.78}_{-4.87} (90\%)$ & $11.09^{+7.62}_{-4.10} (89\%)$ & $13.79^{+7.71}_{-4.53} (89\%)$ & $17.96^{+7.52}_{-4.89} (90\%)$ \\[1.2ex]
$1.0-2.0\rearth$ & $27.36^{+15.77}_{-8.40} (61\%)$ & $44.97^{+15.52}_{-9.29} (63\%)$ & $82.84^{+8.99}_{-5.33} (68\%)$ & $43.31^{+14.02}_{-8.74} (64\%)$ & $59.68^{+12.18}_{-7.96} (67\%)$ & $86.25^{+7.51}_{-4.45} (69\%)$ \\[1.2ex]
$2.0-3.0\rearth$ & $13.88^{+9.47}_{-5.03} (86\%)$ & $24.44^{+10.44}_{-6.53} (87\%)$ & $41.17^{+9.91}_{-6.96} (87\%)$ & $25.97^{+10.21}_{-6.55} (87\%)$ & $33.87^{+10.11}_{-6.88} (87\%)$ & $44.16^{+9.42}_{-6.76} (88\%)$ \\[1.2ex]
$3.0-4.0\rearth$ & $1.40^{+4.46}_{-0.58} (90\%)$ & $3.23^{+5.08}_{-1.57} (90\%)$ & $4.61^{+5.04}_{-2.10} (90\%)$ & $3.34^{+5.00}_{-1.62} (90\%)$ & $3.91^{+5.02}_{-1.85} (90\%)$ & $4.77^{+4.87}_{-2.13} (90\%)$ \\[1.2ex]
$2.0-4.0\rearth$ & $15.28^{+9.78}_{-5.33} (88\%)$ & $27.67^{+10.74}_{-6.86} (88\%)$ & $45.78^{+10.02}_{-7.07} (89\%)$ & $29.32^{+10.51}_{-6.86} (88\%)$ & $37.78^{+10.32}_{-7.09} (89\%)$ & $48.93^{+9.50}_{-6.83} (89\%)$ \\[1.2ex]
$0.5-1.4\rearth$ & $19.85^{+18.42}_{-7.70} (28\%)$ & $37.34^{+20.29}_{-10.30} (30\%)$ & $67.77^{+13.06}_{-7.97} (35\%)$ & $35.60^{+18.05}_{-9.67} (31\%)$ & $49.48^{+15.53}_{-9.34} (33\%)$ & $69.67^{+11.57}_{-7.32} (37\%)$ \\[1.2ex]
\hline
\hline
$0.5-1.0\rearth$ & $ (8\%)$ & $ (9\%)$ & $ (12\%)$ & $ (10\%)$ & $ (11\%)$ & $ (14\%)$ \\[1.2ex]
$0.8-1.0\rearth$ & $ (14\%)$ & $4.49^{+10.81}_{-2.03} (15\%)$ & $8.55^{+11.74}_{-3.93} (20\%)$ & $5.54^{+11.98}_{-2.54} (16\%)$ & $7.76^{+12.30}_{-3.61} (18\%)$ & $9.12^{+11.70}_{-4.14} (22\%)$ \\[1.2ex]
$1.0-1.5\rearth$ & $19.97^{+20.84}_{-7.94} (38\%)$ & $28.20^{+21.29}_{-9.56} (41\%)$ & $49.46^{+14.16}_{-8.85} (48\%)$ & $26.25^{+18.49}_{-8.84} (43\%)$ & $35.71^{+15.79}_{-9.06} (46\%)$ & $48.68^{+12.74}_{-8.30} (51\%)$ \\[1.2ex]
$1.5-2.0\rearth$ & $13.17^{+11.23}_{-5.25} (69\%)$ & $20.28^{+11.89}_{-6.59} (71\%)$ & $34.10^{+10.60}_{-7.11} (75\%)$ & $20.32^{+11.08}_{-6.36} (72\%)$ & $26.12^{+10.73}_{-6.76} (74\%)$ & $36.79^{+9.76}_{-6.80} (76\%)$ \\[1.2ex]
$2.0-2.5\rearth$ & $11.62^{+9.53}_{-4.61} (83\%)$ & $18.38^{+10.29}_{-5.90} (83\%)$ & $32.50^{+9.85}_{-6.71} (85\%)$ & $19.41^{+9.97}_{-5.92} (84\%)$ & $26.23^{+9.91}_{-6.44} (84\%)$ & $34.48^{+9.41}_{-6.58} (85\%)$ \\[1.2ex]
$2.5-4.0\rearth$ & $7.07^{+7.75}_{-3.17} (88\%)$ & $12.65^{+8.64}_{-4.62} (89\%)$ & $21.53^{+8.61}_{-5.61} (89\%)$ & $13.42^{+8.52}_{-4.72} (89\%)$ & $17.56^{+8.59}_{-5.26} (89\%)$ & $23.29^{+8.34}_{-5.61} (89\%)$ \\[1.2ex]
$1.0-2.0\rearth$ & $33.14^{+18.53}_{-9.60} (54\%)$ & $48.48^{+17.89}_{-10.06} (56\%)$ & $83.55^{+9.92}_{-5.51} (61\%)$ & $46.57^{+16.20}_{-9.54} (57\%)$ & $61.83^{+13.69}_{-8.47} (60\%)$ & $85.47^{+8.53}_{-4.86} (63\%)$ \\[1.2ex]
$2.0-3.0\rearth$ & $17.14^{+10.68}_{-5.84} (85\%)$ & $28.28^{+11.42}_{-7.16} (85\%)$ & $49.13^{+10.35}_{-7.24} (86\%)$ & $29.93^{+11.10}_{-7.13} (86\%)$ & $39.87^{+10.78}_{-7.35} (86\%)$ & $52.58^{+9.75}_{-6.95} (87\%)$ \\[1.2ex]
$3.0-4.0\rearth$ & $1.54^{+4.72}_{-0.66} (89\%)$ & $2.75^{+5.11}_{-1.33} (89\%)$ & $4.89^{+5.08}_{-2.20} (90\%)$ & $2.89^{+5.02}_{-1.41} (89\%)$ & $3.92^{+5.05}_{-1.86} (90\%)$ & $5.18^{+4.95}_{-2.26} (90\%)$ \\[1.2ex]
$2.0-4.0\rearth$ & $18.69^{+10.95}_{-6.13} (87\%)$ & $31.03^{+11.64}_{-7.40} (87\%)$ & $54.03^{+10.29}_{-7.20} (88\%)$ & $32.83^{+11.32}_{-7.35} (88\%)$ & $43.79^{+10.88}_{-7.46} (88\%)$ & $57.77^{+9.65}_{-6.85} (88\%)$ \\[1.2ex]
$0.5-1.4\rearth$ & $18.48^{+20.46}_{-7.54} (21\%)$ & $29.91^{+21.56}_{-9.83} (22\%)$ & $53.10^{+14.77}_{-9.06} (27\%)$ & $28.96^{+19.24}_{-9.33} (24\%)$ & $39.97^{+16.65}_{-9.52} (26\%)$ & $53.79^{+13.47}_{-8.57} (30\%)$
\enddata
\tablecomments{As in Table~\ref{tab:perocc}, the entries below the double horizontal line are the estimates based on the revised stellar radii (see Section~\ref{ssec:starbias}).}
\tablenotetext{a}{These habitable zone limits are from \citet{kopparapu_et_al2013}.}
\tablenotetext{b}{These are the ``simple'' habitable zone boundaries adopted by \citet{petigura_et_al2013b}.}
\tablenotetext{c}{This limit approximates the effect of clouds \citep{yang_et_al2014} by increasing the flux at the inner edge of the habitable zone a factor of two compared to the baseline calculation by \citet{kopparapu_et_al2013}.}
\tablenotetext{d}{For these calculations, the inner edge of the habitable zone was set to the values predicted by \citet{zsom_et_al2013} for hot ``desert'' worlds with low relative humidity.}
\label{tab:hzocc}
\end{deluxetable*}

In the most conservative case, we adopt the maximum greenhouse (Max GH) and moist greenhouse (Moist GH) insolation limits from \citet{kopparapu_et_al2013}. The Max GH limit is the insolation at which adding additional $\rm{CO_2}$ can no longer heat the surface of the planet because Rayleigh scattering begins to dominate over the greenhouse effect. At the inner edge, the Moist GH limit corresponds to the insolation at which the planet's stratosphere becomes dominated by water vapor. At that point, the planet's reservoir of hydrogen quickly escapes to space.     

Table~\ref{tab:hzocc} also provides estimates based on the assumption that Venus and Mars were habitable at earlier times in their histories. For a Sun-like star, those constraints correspond to insolation limits of $1.776\fearth$ and $0.321\fearth$, respectively \citep{kopparapu_et_al2013, kopparapu_et_al2013b}. The insolation boundaries are lower for planets orbiting M~dwarfs because the incoming radiation is redder. For a typical star in our sample ($T_{\rm eff} = 3748$K), the boundaries are $1.543\fearth$ and $0.228\fearth$, respectively. 

Even more optimistically, Table~\ref{tab:hzocc} includes HZ occurrence rates using the cloudy inner HZ from \citet{yang_et_al2014}, two choices of desert world albedos from \citet{zsom_et_al2013}, and the convenient limits of $0.25 - 4\fearth$ used by \citet{petigura_et_al2013b}. We do not provide an estimate based on the hydrogen atmosphere HZ of \citet{pierrehumbert+gaidos2011} because our search completeness is very low at the maximum allowed separation of 2.4~AU. 

The appropriate radius range to consider for a potentially habitable planet is more clearly defined than the appropriate insolation range. Based on radial velocity follow-up observations of \kepler planet candidates, \citet{rogers2015} argued that the majority of planets larger than $1.6\rearth$ contain too many volatiles to be rocky. This result agrees with previous fits to measured exoplanet masses and radii by \citet{weiss+marcy2014} and simulations by \citet{lopez+fortney2014}. Furthermore, \citet{dressing_et_al2015} found that all five exoplanets smaller than $1.6\rearth$ with masses and radii measured to a precision better than 20\% have densities consistent with an Earth-like mixture of iron and silicates. Like \citet{rogers2015}, they noted that planets larger than $1.6\rearth$ radii have densities inconsistent with rocky compositions. Due to observational constraints, the population of planets with well-constrained densities is strongly biased towards highly irradiated planets. We therefore include a broader range of radius choices in Table~\ref{tab:hzocc} to account for the possibility that the transition between rocky and gaseous planets might occur at a slightly different radius for less irradiated planets. For instance, the $2.35\rearth$ exoplanet Kepler-10c has a measured mass of $17.2\pm1.9\mearth$ and a bulk density of $7.1\pm1 \rm{g\,cm}^{-3}$, higher than the densities of most $2-3\rearth$ planets \citep{dumusque_et_al2014}. 

Adopting the most conservative assumptions ($1.0\rearth < R_P < 1.5\rearth$, outer HZ = Max GH, inner HZ = Moist GH), we estimate an occurrence rate of \occhzearth~potentially habitable $1-1.5\rearth$ planets per M~dwarf.  The predicted occurrence rates of super-Earths ($1.5-2.0\rearth$) and larger planets ($2\rearth < R_P < 4\rearth$) within the same habitable zone boundaries are \occhzse super-Earths and \occhzneps larger planets per M dwarf. Expanding the radius range to $1-2\rearth$ or increasing the habitable zone boundaries to the limits for recent Venus and early Mars increases the assumed occurrence rate to \occhzonetwo~and \occvenusmars, respectively. In the most optimistic case, we estimate an occurrence rate of \occzsom~desert worlds with albedos of 0.8 and radii of $1-2\rearth$ receiving insolations between the \citet{zsom_et_al2013} inner limit and the Max GH outer limit. The assumed occurrence rate of potentially habitable M~dwarf planets therefore varies by a factor of four depending on the specific choice of radius and insolation boundaries. 

The range of HZ possibilities in Table~\ref{tab:hzocc} is particularly useful for comparing our results to those of previous studies. For instance, \citet{petigura_et_al2013b} estimated that $22\%$ of FGK stars host $1-2\rearth$ planets receiving $0.25-4\fearth$. Within the same boundaries, we find an occurrence rate of \occpetigura~small planets per M~dwarf HZ. 

The difference in our estimates might suggest that habitable zone planets are more common around lower-mass stars, but the \citet{petigura_et_al2013b} prediction is based on an extrapolation of the occurrence rate for shorter period planets to longer period orbits assuming that planet occurrence is flat in $\log P$. If the planet occurrence rate actually increases with $\log P$ at longer periods, then perhaps the occurrence rates of potentially habitable planets orbiting FGK and M dwarfs are more similar. Such a change in the slope of the FGK star planet occurrence rate at longer periods could be explained by the radial- and temperature-dependence of the physics governing planet formation. 

\subsection{Implications of Systematic Biases in Modeled Stellar Radii}
\label{ssec:starbias}
The stellar parameters for the majority of the stars in our sample were estimated by fitting Dartmouth stellar models to photometric \citep{dressing+charbonneau2013, gaidos2013, huber_et_al2014} or spectroscopic \citep{mann_et_al2012, muirhead_et_al2012b} observations. The exceptions are one star with parameters from \citet{mann_et_al2013b} and two very-low mass stars with parameters from \citet{martin_et_al2013}. Both \citet{mann_et_al2013b} and \citet{martin_et_al2013} estimated stellar radii using empirical relations based on interferometric observations of low-mass stars \citep{boyajian_et_al2012}.

Several recent studies \citep[e.g.,][]{boyajian_et_al2012, mann_et_al2013c, newton_et_al2015} have demonstrated that theoretical stellar models do not accurately reproduce the observed radii of low-mass stars. As explained in \citet{newton_et_al2015}, there are two main issues: 
\begin{enumerate}
\item{The radii of model stars with $T_{\rm eff} < 4000{\rm K}$ are smaller than the interferometrically-measured radii by approximately $0.04-0.09\rsun$. }
\item{Variations in metallicity produce significant changes in the modeled radii of low-mass stars whereas observations reveal that metallicity actually has little influence on the radii of low-mass stars.}
\end{enumerate}
Although the Dartmouth stellar models perform better than many alternative models, both of these effects may have caused the radii of the stars in our sample to be systematically underestimated. In order to gauge the magnitude of this effect, we recalculated the radii for our stellar sample using an empirical temperature/radius relation for main sequence stars with $3300{\rm K} < T_{\rm eff}$ (Equation 6 in \citealp{mann_et_al2013c} with the additional significant figures reported by \citealp{newton_et_al2015}). We did not consider changes in the stellar temperatures, but \citet{newton_et_al2015} demonstrated that the temperatures we estimated in \citet{dressing+charbonneau2013} were consistent with predictions based on empirical observations (our values were lower by $40 \pm 110 {\rm K}$). Using the empirical temperature/radius relation to revise the radii of the 2437~stars in our sample with $T_{\rm eff}> 3300$, we found that the median change in radius ($\Delta R_*$) was an increase of 0.026$\rsun$ (6\%). The change was highly dependent on the assumed metallicity; stars with assigned [Fe/H]$ \leq -0.5$ displayed a median size increase of $0.05\rsun$ (11\%) while the estimated radii of stars with assigned [Fe/H] $\geq 0$ shrank by $0.016\rsun$ (5\%).

For the planet host stars in our sample, the increase in the stellar radii leads to larger predicted radii and increased insolation fluxes for the associated planet candidates. The median planet radius increase was 6.6\%, but the amplitude of the change varied considerably. The systems most strongly affected by the revision of the stellar radii were: KOI~3102~(+48\%), KOI~2650~(+26\%), KOI~2418~(+20\%), KOI~2006~(+17\%), and KOI~812~(+16\%). Two of the KOIs in these systems (KOI~3102.01 and KOI~2650.02) were missed by our planet detection pipeline so they did not enter into our calculation of the planet occurrence rate. 

\begin{figure*}[htbp] 
\begin{center}
\centering
\includegraphics[width=0.48\textwidth]{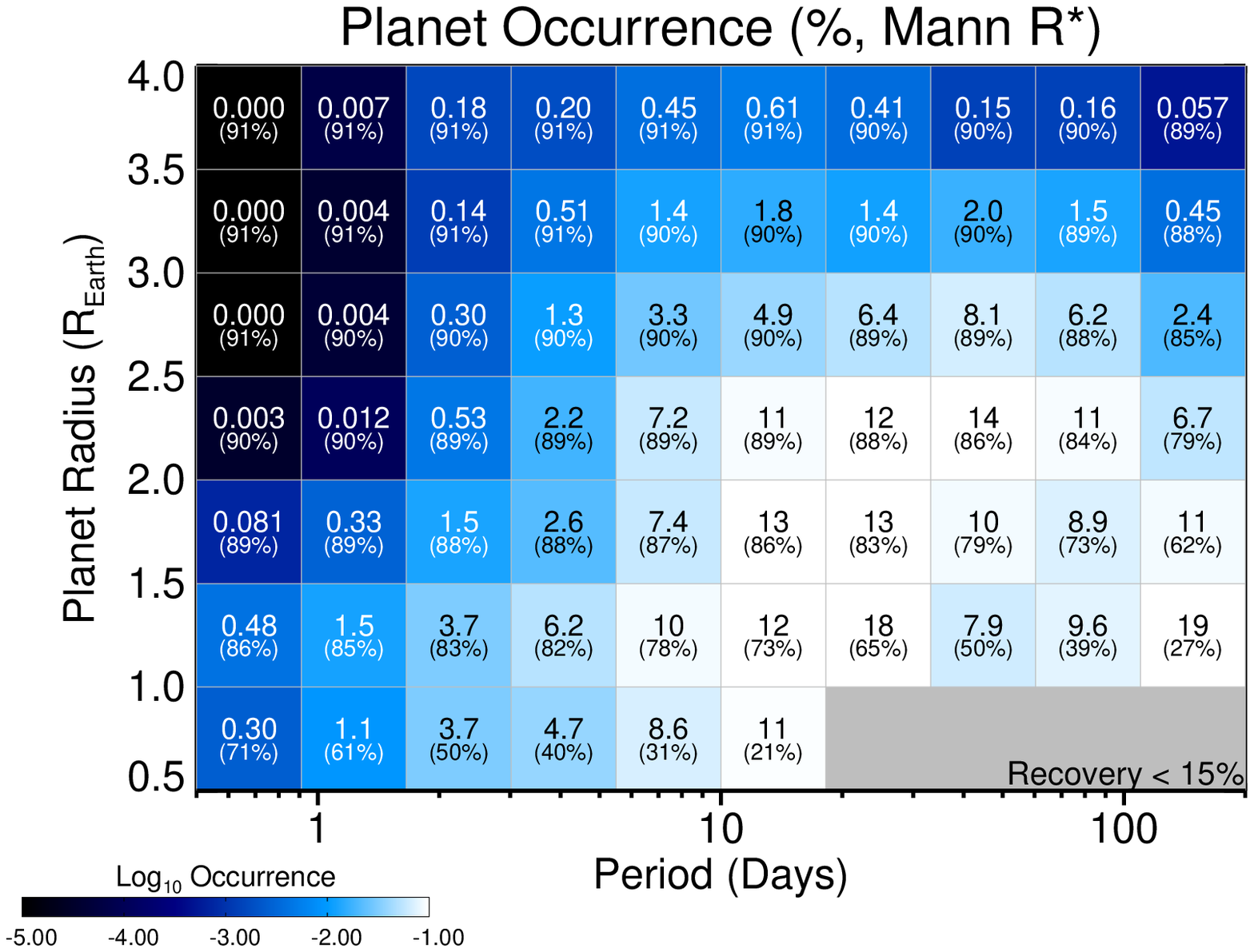}
\includegraphics[width=0.48\textwidth]{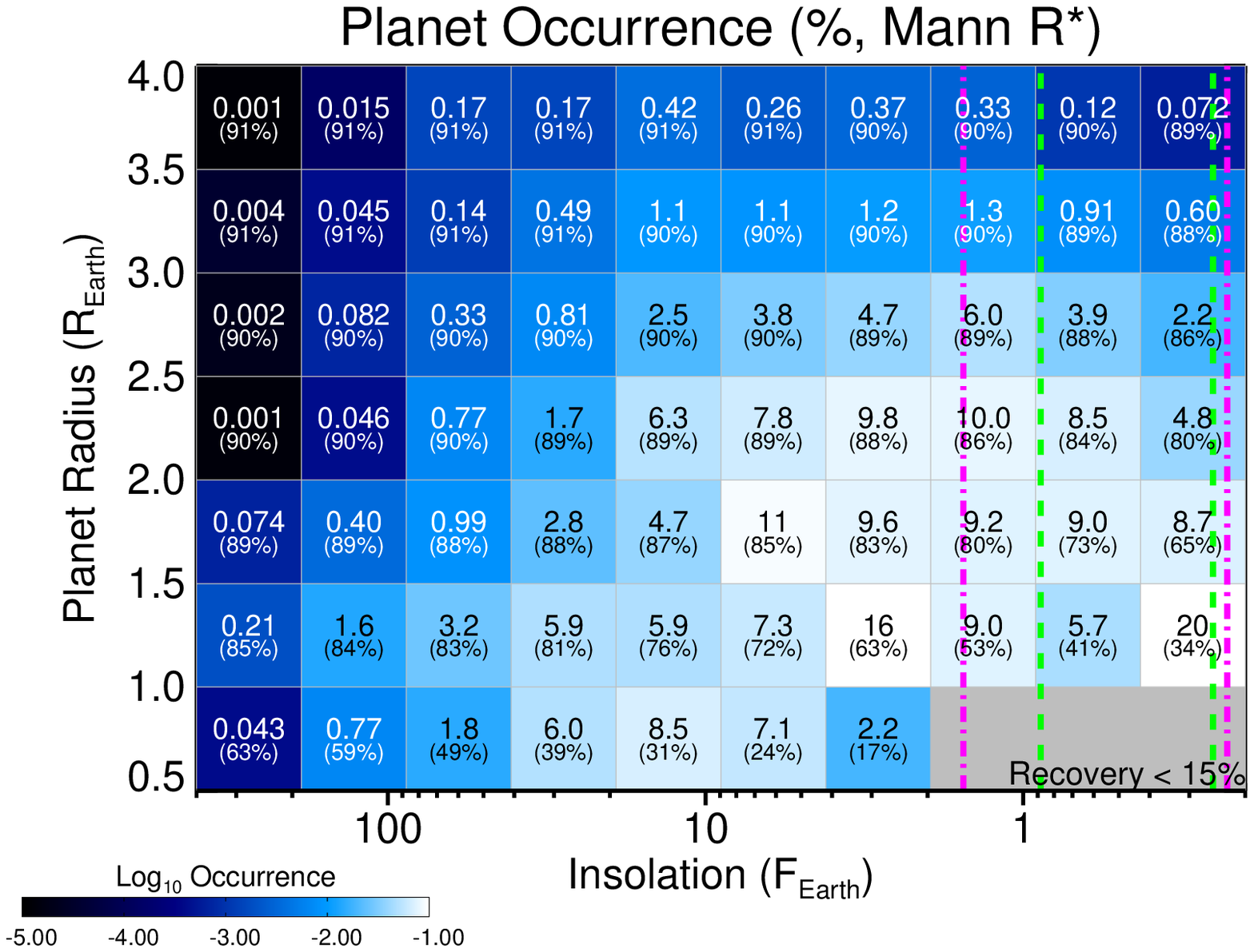}
\end{center}
\caption{Alternative calculation of the planet occurrence rate in period/planet radius space (\emph{Left}) and insolation/planet radius space (\emph{Right}) using the revised stellar radii (see Section~\ref{ssec:starbias}). The annotations are the same as in Figure~\ref{fig:occgrid}.  As in Figure~\ref{fig:kidsearchmap}, the vertical lines in the right panel mark two definitions of the habitable zone.}
\label{fig:empiricalocc}
\end{figure*}

In addition to altering the radius estimates for the detected planet candidates, the changes in the stellar radii affect the estimated survey completeness and, in turn, the derived occurrence rate. If the stellar radii are typically $6\%$ larger, then the search completeness we displayed in Figure~\ref{fig:searchmap} for planets between $0.50 - 4.00\rearth$ actually corresponds to $0.53-4.24\rearth$ planets. We accounted for this effect by generating new search completeness maps following the procedure outlined in Section~\ref{sec:injpipe} but correcting the radii and insolation flux environments of the injected planets to reflect the new radius estimates for each star. We then recalculated the planet occurrence rate using the updated search completeness maps and the revised planet properties. We present the corresponding planet occurrence maps in Figure~\ref{fig:empiricalocc} and include the resulting planet occurrence rates below the double horizontal lines in Tables~\ref{tab:perocc}--\ref{tab:hzocc}. 

As expected, the most noticeable difference between the occurrence maps displayed in Figures~\ref{fig:occgrid} and \ref{fig:empiricalocc} is that the ridge of high planet occurrence has moved upward to large radii. Similarly, the region of low search completeness now encompasses a slightly larger portion of our chosen parameter space. Using the revised stellar radii, we calculated occurrence rates of $0.61^{+0.07}_{-0.05}$~Earth-size planets and $0.46^{+0.07}_{-0.05}$~super-Earths per low-mass star with periods shorter than 50~days. These rates are nearly identical to the estimates presented in Section~\ref{sec:occrate}. Within the habitable zone, we estimate a frequency of $0.20^{+0.21}_{-0.08}$~Earths and $0.13^{+0.11}_{-0.05}$~super-Earths per star when adopting the moist GH inner limit and the maximum GH outer limit from \citet{kopparapu_et_al2013}. These estimates are 26\% and 14\% higher, respectively, than the rates of \occhzearth~Earths and \occhzse~super-Earths per HZ presented in Section~\ref{sec:occrate}. Although increasing the assumed stellar radii alters the inferred occurrence rates, the dominant source of error is the relatively small number of potentially habitable small planets.

\section{Summary \& Conclusions}
\label{sec:conc}
In this paper, we presented an updated estimate of the planet occurrence rate for early M~dwarfs based on the full four-year \kepler data set. We developed our own planet detection pipeline to search for transiting planets in the \kepler light curves. We then characterized the completeness of our pipeline by injecting simulated transiting planets into the \kepler light curves and attempting to recover them. Our search of the light curves of \nstars~small stars with at least $1000$~days of \kepler photometry revealed 3215~possible planetary transits. We thoroughly inspected all available follow-up observations for these objects and accounted for transit depth dilution for systems with close stellar companions. We accepted \ncois~planet candidates, one of which was not a previously known \kepler planet candidate. 

We then measured the occurrence rate of small planets around small stars by dividing smoothed maps of the detected planet population by maps of our pipeline search completeness in radius-period space and radius-insolation space. We found that Earth-sized planets ($1.0 -1.5\rearth$) are common and calculated an occurrence rate of  \occearthfifty Earth-sized planets with periods shorter than 50~days per early M~dwarf. We also found an occurrence of \occsefifty super-Earths ($1.5-2\rearth$) with periods shorter than 50~days per early M~dwarf. For orbital periods shorter than 200 days and planet radii of $1-4\rearth$, we estimated a cumulative planet occurrence rate of \totalsmall~planets per M dwarf.

Within a conservatively defined habitable zone based on the moist greenhouse and maximum greenhouse limits \citep{kopparapu_et_al2013, kopparapu_et_al2013b} we estimated occurrence rates of \occhzearth~Earth-size ($1.0-1.5\rearth$) planets and \occhzse ($1.5-2.0\rearth$) super-Earths per small star. Adopting a wider planet size range of $1-2\rearth$ and considering the effects of clouds \citep{yang_et_al2013} increased our estimate to \occhzcloudyonetwo~potentially habitable planets per star. Considering desert worlds \citep{zsom_et_al2013} would increase the measured occurrence rate to nearly one potentially habitable planet per M dwarf. These estimates span the range of previous estimates of the occurrence rate of potentially habitable M~dwarf planets. 

An order of magnitude calculation multiplying the occurrence rate of potentially habitable $1-1.5\rearth$ planets between the empirical early Mars outer boundary and the recent Venus inner boundary by an estimate of the number density of small stars in the galaxy from the RECONS survey \citep{henry_et_al2006, winters_et_al2015} therefore suggests that the nearest potentially habitable planet is most likely \dnearhz~pc away and is within \dnearhzuplim~pc with 95\% confidence. This estimate assumes that the occurrence rate of potentially habitable planets orbiting later M~dwarfs is identical to that for early M dwarfs, which is consistent with the results of \citet{berta_et_al2013}.

Correcting for the geometric probability of transit (assuming that 1.4\% of potentially habitable M~dwarf planets transit), the nearest transiting potentially habitable planet is likely to be \dtranshz~pc away and is within \dtranshzuplim~pc with 95\% confidence.\footnote{These distance estimates are based on the mean number of planets per star rather than the fraction of stars with planets. If potentially habitable planets are clustered such that M dwarfs hosting potentially habitable planets typically feature more than one potentially habitable planet, then the distance estimates will need to be increased to account for the relatively flat nature of multiplanet systems orbiting M~dwarfs \citep{ballard+johnson2014}.}  Early M~dwarfs at distances of \dnearhzuplim~pc and \dtranshzuplim~pc would have apparent K~band magnitudes of \knearhz~and \ktranshz, respectively, well within the magnitude range probed by current and upcoming planet surveys of nearby, bright stars such as CARMENES \citep{quirrenbach_et_al2010}, CHEOPS \citep{broeg_et_al2013}, ExoplanetSat \citep{smith_et_al2010}, ExTrA \citep{bonfils_et_al2014}, HPF \citep{mahadevan_et_al2010}, MEarth \citep{nutzman+charbonneau2008, berta_et_al2012}, PLATO \citep{rauer_et_al2014}, K2 \citep{howell_et_al2014}, SPECULOOS \citep{gillon_et_al2013}, SPIRou \citep{thibault_et_al2012}, and TESS \citep{ricker_et_al2014}. 

\LongTables
\begin{deluxetable*}{rrrrrrrrrrrc}
\tablecolumns{12}
\tabletypesize{\footnotesize}
\tablecaption{Candidates Accepted By Our Pipeline}
\tablehead{
\colhead{} & 
\colhead{} & 
\multicolumn{3}{c}{Period (Days)} & 
\multicolumn{3}{c}{$R_p$ ($\rearth$)} &
\multicolumn{3}{c}{$F_p$ ($\fearth$)} & 
\colhead{} \\
\colhead{KID} & 
\colhead{KOI} & 
\colhead{Value} & 
\colhead{- Err} & 
\colhead{+ Err} & 
\colhead{Value} & 
\colhead{- Err} & 
\colhead{+ Err} & 
\colhead{Value} & 
\colhead{- Err} & 
\colhead{+ Err} & 
\colhead{Provenance\tablenotemark{a}}
}
\startdata

2161536 & 2130.01 & 16.85595526 & 6.07E-05 & 5.96E-05 & 1.93 & 0.26 & 0.30 & 6.27 & 1.56 & 1.86 & 0 \\
2556650 & 2156.01 & 2.85234932 & 3.29E-06 & 3.38E-06 & 1.88 & 0.22 & 0.27 & 37.94 & 8.21 & 9.58 & 0 \\
2715135 & 1024.01 & 5.74773353 & 8.53E-06 & 8.58E-06 & 1.49 & 0.19 & 0.21 & 22.53 & 5.97 & 7.60 & 0 \\
2973386 & 3034.01 & 31.02089195 & 2.46E-04 & 2.30E-04 & 1.62 & 0.33 & 0.37 & 2.31 & 0.85 & 1.12 & 0 \\
3426367 & 2662.01 & 2.10434036 & 4.05E-06 & 4.03E-06 & 0.56 & 0.09 & 0.09 & 28.25 & 8.02 & 9.82 & 0 \\
3642335 & 3010.01 & 60.86661711 & 4.91E-04 & 5.80E-04 & 2.37 & 0.27 & 0.33 & 0.94 & 0.18 & 0.21 & 0 \\
3749365 & 1176.01 & 1.97376228 & 4.23E-07 & 4.24E-07 & 9.11 & 2.01 & 2.01 & 74.26 & 36.13 & 60.43 & 0 \\
4061149 & 1201.01 & 2.75759481 & 6.97E-06 & 6.65E-06 & 1.21 & 0.18 & 0.20 & 39.76 & 10.43 & 12.72 & 0 \\
4139816 & 812.01 & 3.34022436 & 8.45E-06 & 8.40E-06 & 2.11 & 0.32 & 0.33 & 38.49 & 12.17 & 16.10 & 1 \\
4139816 & 812.02 & 20.05993080 & 2.82E-04 & 3.03E-04 & 2.14 & 0.33 & 0.36 & 3.53 & 1.12 & 1.48 & 1 \\
4139816 & 812.03 & 46.18428829 & 6.06E-04 & 6.48E-04 & 1.90 & 0.29 & 0.31 & 1.16 & 0.37 & 0.49 & 1 \\
4139816 & 812.04 & 7.82527952 & 1.24E-04 & 1.37E-04 & 1.09 & 0.18 & 0.22 & 12.37 & 3.91 & 5.19 & 1 \\
4172805 & 4427.01 & 147.66174063 & 1.51E-03 & 2.16E-03 & 1.56 & 0.23 & 0.25 & 0.17 & 0.05 & 0.06 & 0 \\
4832837 & 605.01 & 2.62811645 & 2.03E-06 & 2.04E-06 & 2.52 & 0.41 & 0.43 & 69.25 & 28.58 & 43.81 & 0 \\
4832837 & 605.02 & 5.06549822 & 4.22E-05 & 4.37E-05 & 0.97 & 0.17 & 0.19 & 28.88 & 11.92 & 18.35 & 0 \\
4913852 & 818.01 & 8.11438395 & 1.25E-05 & 1.24E-05 & 2.14 & 0.26 & 0.27 & 11.23 & 2.81 & 3.47 & 0 \\
5364071 & 248.01 & 7.20387100 & 8.00E-06 & 8.00E-06 & 2.25 & 0.27 & 0.27 & 14.00 & 3.64 & 4.55 & 3 \\
5364071 & 248.02 & 10.91273200 & 2.10E-05 & 2.10E-05 & 2.74 & 0.36 & 0.45 & 8.04 & 2.08 & 2.58 & 3 \\
5364071 & 248.03 & 2.57654900 & 3.00E-06 & 3.00E-06 & 1.55 & 0.18 & 0.18 & 55.11 & 14.21 & 17.62 & 3 \\
5364071 & 248.04 & 18.59610800 & 7.90E-05 & 7.90E-05 & 1.31 & 0.17 & 0.17 & 3.94 & 1.01 & 1.28 & 3 \\
5384713 & 3444.02 & 60.32665084 & 4.47E-05 & 4.47E-05 & 2.98 & 0.34 & 0.36 & 1.08 & 0.18 & 0.21 & 4 \\
5531953 & 1681.00 & 21.91384343 & 1.83E-04 & 1.93E-04 & 1.03 & 0.15 & 0.17 & 1.85 & 0.47 & 0.57 & 0 \\
5531953 & 1681.01 & 6.93911381 & 2.60E-05 & 2.60E-05 & 0.99 & 0.13 & 0.13 & 8.55 & 2.07 & 2.38 & 2 \\
5531953 & 1681.02 & 1.99281275 & 7.47E-06 & 7.21E-06 & 0.72 & 0.10 & 0.11 & 45.27 & 11.46 & 13.98 & 0 \\
5531953 & 1681.03 & 3.53105829 & 2.39E-05 & 2.40E-05 & 0.68 & 0.10 & 0.12 & 21.12 & 5.34 & 6.52 & 0 \\
5617854 & 1588.01 & 3.51749675 & 3.96E-06 & 3.94E-06 & 1.21 & 0.15 & 0.16 & 40.95 & 9.81 & 11.76 & 0 \\
5640085 & 448.01 & 10.13962300 & 2.20E-05 & 2.20E-05 & 1.77 & 0.22 & 0.23 & 9.50 & 2.35 & 2.84 & 3 \\
5640085 & 448.02 & 43.59579200 & 1.25E-04 & 1.25E-04 & 2.48 & 0.31 & 0.32 & 1.36 & 0.33 & 0.41 & 3 \\
5794240 & 254.01 & 2.45524062 & 1.00E-07 & 1.00E-07 & 11.00 & 0.56 & 0.56 & 67.54 & 9.34 & 10.62 & 3 \\
5809954 & 1902.01 & 137.86453830 & 2.87E-04 & 2.87E-04 & 1.99 & 0.40 & 0.44 & 0.23 & 0.06 & 0.08 & 4 \\
6382217 & 2036.01 & 8.41102527 & 2.59E-05 & 2.61E-05 & 1.51 & 0.30 & 0.31 & 13.17 & 4.79 & 6.30 & 0 \\
6382217 & 2036.02 & 5.79529807 & 3.24E-05 & 3.52E-05 & 1.00 & 0.20 & 0.21 & 21.68 & 7.86 & 10.38 & 0 \\
6435936 & 854.01 & 56.05318853 & 1.31E-03 & 1.26E-03 & 2.09 & 0.29 & 0.32 & 0.65 & 0.16 & 0.20 & 1 \\
6497146 & 3284.01 & 35.23301880 & 2.24E-04 & 2.27E-04 & 1.01 & 0.16 & 0.17 & 1.31 & 0.38 & 0.47 & 0 \\
6666233 & 2306.01 & 0.51240811 & 5.59E-07 & 5.53E-07 & 0.94 & 0.13 & 0.14 & 541.12 & 143.41 & 173.03 & 0 \\
6679295 & 2862.01 & 24.57535492 & 1.37E-04 & 1.37E-04 & 1.60 & 0.18 & 0.20 & 2.51 & 0.56 & 0.67 & 0 \\
6773862 & 1868.01 & 17.76080479 & 2.70E-05 & 2.62E-05 & 2.13 & 0.21 & 0.24 & 5.69 & 1.12 & 1.31 & 0 \\
6867155 & 868.01 & 235.99802060 & 3.78E-04 & 3.78E-04 & 8.88 & 1.18 & 1.17 & 0.14 & 0.04 & 0.05 & 3 \\
7021681 & 255.01 & 27.52199799 & 4.98E-05 & 4.96E-05 & 2.56 & 0.31 & 0.31 & 2.27 & 0.54 & 0.65 & 1 \\
7021681 & 255.02 & 13.60335797 & 2.02E-04 & 2.01E-04 & 0.75 & 0.11 & 0.12 & 5.79 & 1.39 & 1.65 & 1 \\
7094486 & 1907.01 & 11.35011141 & 2.30E-05 & 2.33E-05 & 1.98 & 0.20 & 0.22 & 9.32 & 1.90 & 2.25 & 0 \\
7135852 & 875.01 & 4.22097130 & 3.02E-06 & 3.04E-06 & 2.58 & 0.45 & 0.46 & 29.90 & 18.13 & 36.70 & 0 \\
7287995 & 877.01 & 5.95489377 & 8.57E-06 & 8.63E-06 & 2.06 & 0.29 & 0.30 & 20.25 & 6.52 & 8.89 & 1 \\
7287995 & 877.02 & 12.03993346 & 2.85E-05 & 2.92E-05 & 1.87 & 0.26 & 0.26 & 7.92 & 2.56 & 3.48 & 1 \\
7287995 & 877.03 & 20.83776773 & 1.97E-04 & 1.94E-04 & 1.05 & 0.16 & 0.19 & 3.81 & 1.23 & 1.67 & 1 \\
7304449 & 1702.01 & 1.53818130 & 3.09E-06 & 3.12E-06 & 0.84 & 0.17 & 0.18 & 27.77 & 10.07 & 13.20 & 0 \\
7447200 & 676.01 & 7.97251347 & 1.82E-06 & 1.82E-06 & 2.88 & 0.33 & 0.32 & 14.41 & 3.22 & 3.70 & 6 \\
7447200 & 676.02 & 2.45323590 & 4.75E-07 & 4.75E-07 & 3.67 & 0.41 & 0.42 & 69.37 & 15.43 & 18.00 & 6 \\
7455287 & 886.01 & 8.01026000 & 3.00E-05 & 3.00E-05 & 2.39 & 0.35 & 0.43 & 10.04 & 2.56 & 3.09 & 3 \\
7455287 & 886.02 & 12.07238900 & 1.00E-04 & 1.00E-04 & 1.28 & 0.17 & 0.17 & 5.81 & 1.49 & 1.78 & 3 \\
7455287 & 886.03 & 20.99569400 & 1.43E-04 & 1.43E-04 & 1.40 & 0.18 & 0.19 & 2.77 & 0.70 & 0.85 & 3 \\
7603200 & 314.01 & 13.78113900 & 1.10E-05 & 1.10E-05 & 1.34 & 0.16 & 0.16 & 5.98 & 1.42 & 1.68 & 3 \\
7603200 & 314.02 & 23.08871300 & 3.10E-05 & 3.10E-05 & 1.30 & 0.16 & 0.18 & 3.01 & 0.71 & 0.84 & 3 \\
7603200 & 314.03 & 10.31236400 & 3.60E-05 & 3.60E-05 & 0.59 & 0.07 & 0.08 & 8.81 & 2.10 & 2.47 & 3 \\
7870390 & 898.01 & 9.77042372 & 3.02E-05 & 3.20E-05 & 2.38 & 0.29 & 0.31 & 10.52 & 2.56 & 3.13 & 1 \\
7870390 & 898.02 & 5.16981333 & 2.29E-05 & 2.19E-05 & 1.78 & 0.22 & 0.24 & 24.57 & 5.99 & 7.28 & 1 \\
7870390 & 898.03 & 20.09010000 & 1.33E-04 & 1.40E-04 & 2.04 & 0.25 & 0.27 & 4.02 & 0.98 & 1.20 & 1 \\
7871954 & 1515.01 & 1.93703537 & 2.77E-06 & 2.71E-06 & 0.95 & 0.12 & 0.13 & 94.54 & 22.56 & 26.94 & 0 \\
7871954 & 1515.02 & 7.06117534 & 1.34E-05 & 1.38E-05 & 1.16 & 0.15 & 0.17 & 16.86 & 4.04 & 4.81 & 0 \\
7907423 & 899.01 & 7.11369666 & 2.66E-05 & 2.68E-05 & 1.25 & 0.18 & 0.20 & 8.49 & 2.37 & 2.90 & 1 \\
7907423 & 899.02 & 3.30656751 & 1.39E-05 & 1.42E-05 & 0.99 & 0.15 & 0.16 & 23.55 & 6.58 & 8.03 & 1 \\
7907423 & 899.03 & 15.36834908 & 1.08E-04 & 1.12E-04 & 1.17 & 0.17 & 0.18 & 3.04 & 0.85 & 1.04 & 1 \\
8013419 & 901.01 & 12.73263426 & 5.97E-06 & 6.01E-06 & 5.02 & 0.86 & 0.87 & 7.98 & 5.02 & 10.44 & 0 \\
8018547 & 902.01 & 83.94017967 & 1.49E-03 & 1.38E-03 & 5.23 & 0.76 & 0.80 & 0.62 & 0.18 & 0.22 & 1 \\
8120608 & 571.01 & 7.26737224 & 1.93E-05 & 1.98E-05 & 1.26 & 0.17 & 0.17 & 11.52 & 3.00 & 3.63 & 1 \\
8120608 & 571.02 & 13.34295116 & 4.25E-05 & 4.45E-05 & 1.28 & 0.17 & 0.18 & 5.12 & 1.34 & 1.61 & 1 \\
8120608 & 571.03 & 3.88677934 & 9.86E-06 & 9.88E-06 & 1.06 & 0.14 & 0.15 & 26.51 & 6.90 & 8.35 & 1 \\
8120608 & 571.04 & 22.40788397 & 1.20E-04 & 1.27E-04 & 1.18 & 0.16 & 0.18 & 2.57 & 0.67 & 0.81 & 1 \\
8120608 & 571.05 & 129.94188177 & 2.32E-03 & 2.34E-03 & 1.02 & 0.14 & 0.16 & 0.25 & 0.06 & 0.08 & 0 \\
8167996 & 1867.01 & 2.54954597 & 1.25E-05 & 5.91E-06 & 1.12 & 0.12 & 0.14 & 53.87 & 11.96 & 14.43 & 1 \\
8167996 & 1867.02 & 13.96935372 & 9.90E-05 & 1.07E-04 & 1.81 & 0.21 & 0.26 & 5.58 & 1.24 & 1.49 & 1 \\
8167996 & 1867.03 & 5.21231974 & 5.15E-05 & 5.22E-05 & 1.16 & 0.13 & 0.15 & 20.77 & 4.61 & 5.56 & 1 \\
8189801 & 2480.01 & 0.66682583 & 9.30E-07 & 9.58E-07 & 1.31 & 0.23 & 0.25 & 466.69 & 145.16 & 180.18 & 0 \\
8229458 & 2238.01 & 1.64680052 & 2.72E-06 & 2.81E-06 & 0.93 & 0.10 & 0.12 & 118.41 & 22.88 & 26.30 & 0 \\
8235924 & 2347.01 & 0.58800191 & 7.34E-07 & 7.18E-07 & 1.02 & 0.10 & 0.13 & 550.07 & 101.56 & 116.51 & 0 \\
8351704 & 1146.01 & 7.09712741 & 2.98E-05 & 3.27E-05 & 0.82 & 0.13 & 0.14 & 7.97 & 2.35 & 2.94 & 0 \\
8367644 & 1879.01 & 22.08557431 & 5.05E-05 & 4.98E-05 & 2.42 & 0.37 & 0.39 & 1.96 & 0.57 & 0.71 & 0 \\
8509442 & 2992.01 & 82.65995306 & 6.70E-04 & 7.23E-04 & 2.23 & 0.35 & 0.37 & 0.70 & 0.20 & 0.25 & 0 \\
8547140 & 1266.01 & 11.41929554 & 2.42E-05 & 2.45E-05 & 1.54 & 0.26 & 0.27 & 7.27 & 2.55 & 3.52 & 0 \\
8561063 & 961.01 & 1.21377022 & 3.76E-07 & 3.76E-07 & 0.87 & 0.21 & 0.21 & 17.83 & 7.89 & 11.04 & 2 \\
8561063 & 961.02 & 0.45328734 & 8.60E-08 & 8.60E-08 & 0.90 & 0.21 & 0.21 & 67.88 & 31.62 & 54.63 & 3 \\
8561063 & 961.03 & 1.86511477 & 1.12E-06 & 1.12E-06 & 0.69 & 0.16 & 0.16 & 10.00 & 4.38 & 6.24 & 2 \\
8631751 & 2453.01 & 1.53051487 & 1.96E-06 & 1.88E-06 & 1.13 & 0.21 & 0.23 & 61.55 & 21.28 & 28.63 & 0 \\
8845205 & 463.01 & 18.47717612 & 1.92E-04 & 2.26E-04 & 1.67 & 0.28 & 0.28 & 1.27 & 0.41 & 0.51 & 1 \\
8874090 & 1404.01 & 13.32391953 & 5.44E-05 & 5.62E-05 & 1.35 & 0.27 & 0.28 & 4.79 & 2.11 & 3.29 & 0 \\
8874090 & 1404.02 & 18.90623146 & 2.81E-04 & 3.04E-04 & 0.96 & 0.20 & 0.21 & 3.01 & 1.32 & 2.06 & 0 \\
8890150 & 2650.01 & 34.98870981 & 2.25E-03 & 3.62E-03 & 0.96 & 0.15 & 0.17 & 1.17 & 0.38 & 0.62 & 1 \\
9214942 & 1403.01 & 18.75471931 & 5.57E-05 & 5.57E-05 & 1.78 & 0.28 & 0.29 & 5.03 & 1.59 & 2.11 & 0 \\
9388479 & 936.01 & 9.46787010 & 6.62E-05 & 6.29E-05 & 2.18 & 0.30 & 0.31 & 6.26 & 1.68 & 2.04 & 1 \\
9388479 & 936.02 & 0.89303356 & 4.90E-06 & 4.81E-06 & 1.26 & 0.18 & 0.18 & 145.89 & 39.21 & 47.53 & 1 \\
9390653 & 249.01 & 9.54930070 & 1.50E-05 & 1.42E-05 & 1.62 & 0.22 & 0.22 & 4.54 & 1.20 & 1.44 & 1 \\
9427402 & 1397.01 & 6.24703152 & 1.02E-05 & 1.03E-05 & 2.23 & 0.23 & 0.26 & 21.87 & 4.24 & 4.89 & 0 \\
9573685 & 2057.01 & 5.94565639 & 1.53E-05 & 1.52E-05 & 1.19 & 0.17 & 0.19 & 21.39 & 5.60 & 6.80 & 0 \\
9575728 & 5692.01 & 2.64180062 & 1.99E-05 & 2.22E-05 & 0.46 & 0.05 & 0.06 & 64.99 & 13.21 & 15.76 & 0 \\
9710326 & 947.01 & 28.59885823 & 2.34E-04 & 2.18E-04 & 1.92 & 0.26 & 0.27 & 1.81 & 0.47 & 0.57 & 1 \\
9757613 & 250.01 & 12.28293091 & 1.43E-05 & 1.41E-05 & 2.89 & 0.44 & 0.45 & 7.83 & 2.76 & 3.88 & 1 \\
9757613 & 250.02 & 17.25116937 & 3.21E-05 & 3.18E-05 & 2.51 & 0.40 & 0.45 & 4.98 & 1.75 & 2.46 & 1 \\
9757613 & 250.03 & 3.54391419 & 1.15E-05 & 1.15E-05 & 1.06 & 0.17 & 0.18 & 41.08 & 14.46 & 20.34 & 0 \\
9757613 & 250.04 & 46.82763492 & 1.60E-04 & 1.62E-04 & 2.10 & 0.33 & 0.36 & 1.31 & 0.46 & 0.65 & 1 \\
9787239 & 952.01 & 5.90129534 & 2.03E-05 & 1.94E-05 & 2.13 & 0.27 & 0.29 & 16.39 & 4.14 & 5.11 & 1 \\
9787239 & 952.02 & 8.75208025 & 2.04E-05 & 2.09E-05 & 1.98 & 0.27 & 0.34 & 9.70 & 2.46 & 3.02 & 0 \\
9787239 & 952.03 & 22.78068566 & 1.28E-04 & 1.31E-04 & 2.30 & 0.29 & 0.29 & 2.71 & 0.69 & 0.84 & 1 \\
9787239 & 952.04 & 2.89601166 & 8.92E-06 & 9.34E-06 & 1.09 & 0.14 & 0.14 & 42.36 & 10.69 & 13.21 & 0 \\
9787239 & 952.05 & 0.74295972 & 3.06E-06 & 3.65E-06 & 0.91 & 0.12 & 0.13 & 260.04 & 66.04 & 80.88 & 1 \\
10027247 & 2418.01 & 86.82818564 & 1.57E-03 & 1.27E-03 & 1.24 & 0.17 & 0.23 & 0.35 & 0.09 & 0.11 & 0 \\
10027323 & 1596.01 & 5.92367766 & 1.81E-05 & 1.89E-05 & 1.04 & 0.16 & 0.16 & 18.61 & 5.52 & 7.08 & 0 \\
10027323 & 1596.02 & 105.35789935 & 5.05E-04 & 5.36E-04 & 1.90 & 0.28 & 0.30 & 0.40 & 0.12 & 0.15 & 0 \\
10073672 & 2764.01 & 2.25297009 & 7.20E-06 & 7.26E-06 & 1.30 & 0.22 & 0.24 & 85.43 & 27.60 & 35.63 & 0 \\
10118816 & 1085.01 & 7.71790094 & 5.17E-05 & 4.78E-05 & 0.94 & 0.14 & 0.16 & 15.00 & 4.82 & 6.54 & 0 \\
10166274 & 1078.01 & 3.35372553 & 1.04E-05 & 1.06E-05 & 1.86 & 0.25 & 0.26 & 34.92 & 9.20 & 11.35 & 1 \\
10166274 & 1078.02 & 6.87749137 & 2.35E-05 & 2.40E-05 & 2.10 & 0.28 & 0.31 & 13.39 & 3.51 & 4.35 & 1 \\
10166274 & 1078.03 & 28.46416009 & 2.41E-04 & 2.27E-04 & 1.93 & 0.26 & 0.29 & 2.02 & 0.53 & 0.65 & 1 \\
10329835 & 2058.01 & 1.52372638 & 2.40E-06 & 2.46E-06 & 1.05 & 0.14 & 0.16 & 131.36 & 34.28 & 41.60 & 0 \\
10332883 & 1880.01 & 1.15116708 & 6.89E-07 & 6.97E-07 & 1.30 & 0.20 & 0.21 & 181.87 & 53.73 & 66.69 & 0 \\
10340423 & 736.01 & 18.79420980 & 5.75E-05 & 5.95E-05 & 1.82 & 0.30 & 0.31 & 2.81 & 0.89 & 1.14 & 0 \\
10340423 & 736.02 & 6.73899074 & 3.10E-05 & 3.13E-05 & 1.17 & 0.19 & 0.20 & 11.03 & 3.48 & 4.48 & 0 \\
10386984 & 739.01 & 1.28707954 & 4.42E-06 & 4.48E-06 & 1.45 & 0.18 & 0.19 & 129.24 & 30.84 & 36.80 & 1 \\
10388286 & 596.01 & 1.68269527 & 5.88E-06 & 5.95E-06 & 1.38 & 0.19 & 0.22 & 76.32 & 19.27 & 23.05 & 1 \\
10489206 & 251.01 & 4.16438060 & 2.78E-06 & 2.77E-06 & 2.61 & 0.31 & 0.31 & 30.10 & 6.98 & 8.25 & 1 \\
10489206 & 251.02 & 5.77446840 & 6.66E-05 & 6.90E-05 & 0.90 & 0.12 & 0.15 & 19.47 & 4.52 & 5.33 & 1 \\
10525027 & 2006.01 & 3.27346499 & 5.99E-06 & 6.07E-06 & 0.74 & 0.10 & 0.11 & 35.43 & 9.80 & 12.56 & 0 \\
10525049 & 4252.01 & 15.57135826 & 9.44E-05 & 9.36E-05 & 0.73 & 0.14 & 0.15 & 5.54 & 2.00 & 2.61 & 0 \\
10591855 & 2845.01 & 1.57408931 & 5.05E-06 & 5.08E-06 & 0.89 & 0.10 & 0.13 & 135.82 & 26.43 & 30.77 & 0 \\
10670119 & 2179.01 & 14.87155876 & 5.40E-05 & 5.01E-05 & 1.27 & 0.19 & 0.20 & 3.15 & 0.90 & 1.10 & 0 \\
10670119 & 2179.02 & 2.73277006 & 5.46E-06 & 5.49E-06 & 1.08 & 0.16 & 0.18 & 30.19 & 8.59 & 10.50 & 0 \\
10717241 & 430.01 & 12.37646610 & 1.30E-05 & 1.30E-05 & 2.04 & 0.26 & 0.27 & 6.15 & 1.57 & 1.92 & 3 \\
10717241 & 430.02 & 9.34052690 & 8.42E-05 & 8.42E-05 & 0.73 & 0.10 & 0.10 & 8.95 & 2.30 & 2.78 & 3 \\
10925104 & 156.01 & 8.04133834 & 9.01E-06 & 8.98E-06 & 1.39 & 0.18 & 0.20 & 15.72 & 3.58 & 4.24 & 1 \\
10925104 & 156.02 & 5.18856137 & 9.66E-06 & 9.67E-06 & 1.04 & 0.13 & 0.15 & 28.20 & 6.40 & 7.60 & 1 \\
10925104 & 156.03 & 11.77614238 & 7.55E-06 & 7.52E-06 & 2.02 & 0.23 & 0.23 & 9.45 & 2.15 & 2.55 & 0 \\
11129738 & 1427.01 & 2.61301749 & 5.76E-06 & 5.70E-06 & 1.36 & 0.20 & 0.22 & 58.27 & 17.40 & 22.67 & 0 \\
11187837 & 252.01 & 17.60462615 & 3.24E-05 & 3.21E-05 & 2.36 & 0.29 & 0.29 & 3.79 & 0.93 & 1.10 & 1 \\
11192235 & 2329.01 & 1.61535973 & 2.53E-06 & 2.53E-06 & 1.17 & 0.22 & 0.24 & 101.54 & 20.26 & 23.51 & 2 \\
11348997 & 2090.01 & 5.13248495 & 9.13E-06 & 9.09E-06 & 1.52 & 0.25 & 0.26 & 18.98 & 5.92 & 7.38 & 0 \\
11497958 & 1422.01 & 5.84164124 & 1.02E-05 & 1.02E-05 & 3.49 & 0.30 & 0.50 & 14.89 & 5.58 & 8.77 & 5 \\
11497958 & 1422.02 & 19.85028393 & 5.85E-05 & 5.85E-05 & 3.10 & 0.51 & 1.11 & 2.94 & 1.30 & 1.78 & 5 \\
11497958 & 1422.03 & 10.86443187 & 4.97E-05 & 4.97E-05 & 2.34 & 0.34 & 0.87 & 5.09 & 1.96 & 7.23 & 5 \\
11497958 & 1422.04 & 63.33666060 & 5.57E-04 & 5.57E-04 & 2.30 & 0.34 & 0.66 & 0.61 & 0.37 & 0.53 & 5 \\
11497958 & 1422.05 & 34.14189000 & 2.64E-04 & 2.64E-04 & 2.12 & 0.34 & 0.79 & 1.42 & 0.72 & 0.93 & 5 \\
11752906 & 253.01 & 6.38316009 & 9.25E-06 & 9.25E-06 & 2.68 & 0.39 & 0.40 & 17.61 & 6.43 & 9.30 & 1 \\
11752906 & 253.02 & 20.61727169 & 2.66E-04 & 2.85E-04 & 1.48 & 0.23 & 0.25 & 3.69 & 1.35 & 1.94 & 1 \\
11754553 & 775.01 & 16.38481298 & 7.05E-05 & 7.02E-05 & 1.90 & 0.26 & 0.27 & 5.56 & 1.61 & 2.08 & 1 \\
11754553 & 775.02 & 7.87740709 & 1.12E-05 & 1.12E-05 & 2.10 & 0.28 & 0.31 & 14.75 & 4.23 & 5.53 & 3 \\
11754553 & 775.03 & 36.44516433 & 2.60E-04 & 2.91E-04 & 1.96 & 0.27 & 0.29 & 1.91 & 0.55 & 0.72 & 1 \\
11768142 & 2626.01 & 38.09723780 & 2.86E-04 & 2.86E-04 & 2.36 & 0.53 & 0.44 & 0.92 & 0.46 & 0.57 & 5 \\
11852982 & 247.01 & 13.81496561 & 6.45E-05 & 6.47E-05 & 1.61 & 0.21 & 0.22 & 5.22 & 1.27 & 1.50 & 1 \\
11853130 & 3263.01 & 76.87935073 & 4.79E-05 & 4.79E-05 & 6.83 & 1.18 & 1.27 & 0.31 & 0.08 & 0.09 & 2 \\
11853255 & 778.01 & 2.24336795 & 1.13E-05 & 1.11E-05 & 1.37 & 0.19 & 0.20 & 52.26 & 14.75 & 18.85 & 1 \\
11923270 & 781.01 & 11.59822478 & 1.50E-05 & 1.49E-05 & 2.82 & 0.37 & 0.39 & 6.10 & 1.55 & 1.90 & 0 \\
12066335 & 784.01 & 19.27103179 & 3.52E-04 & 3.51E-04 & 1.81 & 0.25 & 0.28 & 3.29 & 0.88 & 1.10 & 1 \\
12066335 & 784.02 & 10.06525147 & 2.48E-05 & 2.46E-05 & 1.54 & 0.20 & 0.22 & 7.83 & 2.10 & 2.63 & 0 \\
12066569 & 3282.01 & 49.27623343 & 4.37E-04 & 5.12E-04 & 2.07 & 0.29 & 0.31 & 1.30 & 0.34 & 0.41 & 0 \\
12302530 & 438.01 & 5.93119294 & 5.02E-06 & 5.02E-06 & 1.80 & 0.24 & 0.25 & 23.47 & 7.53 & 10.42 & 3 \\
12302530 & 438.02 & 52.66160633 & 1.53E-04 & 1.53E-04 & 1.80 & 0.25 & 0.27 & 1.28 & 0.41 & 0.56 & 0 \\
12352520 & 3094.01 & 4.57700369 & 1.33E-05 & 1.34E-05 & 1.39 & 0.17 & 0.20 & 24.39 & 5.04 & 5.86 & 0 \\
12506770 & 1577.01 & 2.80624351 & 8.14E-06 & 8.33E-06 & 1.40 & 0.23 & 0.25 & 59.53 & 18.67 & 24.43 & 0
\enddata
\tablenotetext{a}{The planet parameters are from our fits to the long cadence data (provenance = 0), our fits to the short cadence data (provenance = 1), the NASA Exoplanet Archive (provenance = 2), \citet[][provenance = 3]{rowe_et_al2014}, \citet[][provenance = 4]{swift_et_al2015},  \citet[][provenance = 5]{cartier_et_al2014}, and \citet[][provenance = 6]{ioannidis_et_al2014}.}
\label{tab:cois}
\end{deluxetable*}

\acknowledgements
C.D. is supported by a National Science Foundation Graduate Research Fellowship. Support for this work was provided through the NASA Kepler Mission Participating Scientist Program grants NNX09AB53G and NNX12AC77G awarded to D.C.  This publication was made possible through the support of a grant from the John Templeton Foundation. The opinions expressed in this publication are those of the authors and do not necessarily reflect the views of the John Templeton Foundation. We thank the anonymous referee for providing feedback that improved the quality of this paper and the \kepler team for providing the community with a fantastic collection of data. We are grateful to Jonathan Irwin for sharing a fast implementation of the transit model \citep{mandel+agol2002} and for providing valuable advice. We thank Jessie Christiansen for answering questions about the \kepler pipeline completeness and providing helpful suggestions. Funding for the \kepler mission is provided by the NASA Science Mission directorate. This publication made use of the \kepler Community Follow-Up Observing Program website (\url{https://cfop.ipac.caltech.edu}), the NASA Exoplanet Archive, and the Mikulski Archive for Space Telescopes (MAST). The NASA Exoplanet Archive is operated by the California Institute of Technology, under contract with the National Aeronautics and Space Administration under the Exoplanet Exploration Program. STScI is operated by the Association of Universities for Research in Astronomy, Inc., under NASA contract NAS5-26555. Support for MAST for non-\emph{HST} data is provided by the NASA Office of Space Science via grant NNX09AF08G and by other grants and contracts. 
\bibliography{../mdwarf_biblio.bib}
\clearpage

\end{document}